\documentclass[journal]{IEEEtranTCOM}

\usepackage{color}
\usepackage{xcolor}
\usepackage{algorithmic}
\usepackage{setspace}
\usepackage{cite}
\usepackage{amssymb}
\usepackage[ruled,linesnumbered]{algorithm2e}

\usepackage{amsmath,amsthm}

\usepackage{stfloats}
\usepackage{cuted}
\usepackage{graphicx}
\usepackage{epsfig}
\usepackage{subfigure}
\usepackage{url}
\usepackage{multirow}
\usepackage{makecell}
\usepackage{enumerate}
\usepackage{cite}
\usepackage{multicol} % 调用跨栏长公式宏包
\usepackage{mathtools,cuted,lipsum}
\usepackage[numbers]{natbib}

\graphicspath{{pic/}}

\ifCLASSINFOpdf
\else
\fi
\begin{document}
\title{Fidelity-Guaranteed Entanglement Routing in Quantum Networks}

 \author{Jian~Li,~\IEEEmembership{Member,~IEEE,} Mingjun Wang, Kaiping Xue,~\IEEEmembership{Senior Member,~IEEE}, Nenghai Yu,~\IEEEmembership{Member,~IEEE,} Qibin Sun,~\IEEEmembership{Fellow,~IEEE,} Jun Lu% <-this % stops a space
\thanks{J. Li, M. Wang, K. Xue, N. Yu and Q. Sun are with the School of Cyber Science and Technology, University of Science and Technology of China, Hefei 230027, China.}% <-this % stops a space
\thanks{J. Lu is with the Department of Electronic Engineering and Information Science, Dalian Maritime University, University of Science and Technology of China, Hefei 230027, China.}
\thanks{Corresponding Author: K. Xue (kpxue@ustc.edu.cn)}
}


%\IEEEauthorblockA{\IEEEauthorrefmark{3}Starfleet Academy, San Francisco, Califo44rnia 96678-2391\\
%Telephone: (800) 555--1212, Fax: (888) 555--1212}


\maketitle

\begin{abstract}

Entanglement routing establishes remote entanglement connection between two arbitrary nodes, which is one of the most important functions in quantum networks. The existing routing mechanisms mainly improve the robustness and throughput facing the failure of entanglement generations, which, however, rarely include the considerations on the most important metric to evaluate the quality of connection, entanglement fidelity. To solve this problem, we propose purification-enabled entanglement routing designs to provide fidelity guarantee for multiple Source-Destination (SD) pairs in quantum networks. In our proposal, we first consider the single S-D pair scenario and design an iterative routing algorithm, Q-PATH, to find the optimal purification decisions along the routing path with minimum entangled pair cost. Further, a low-complexity routing algorithm using an extended Dijkstra algorithm, Q-LEAP, is designed to reduce the computational complexity by using a simple but effective purification decision method. Finally, we consider the common scenario with multiple S-D pairs and design a greedy-based algorithm considering resource allocation and re-routing process for multiple routing requests.
Simulation results show that the proposed algorithms not only can provide fidelity-guaranteed routing solutions, but also has superior performance in terms of throughput, fidelity of end-to-end entanglement connection, and resource utilization ratio, compared with the existing routing scheme.

%In this study we propose an effective routing scheme to enable automatic responses for multiple requests of entanglement generation between source-terminal stations on a quantum lattice network with finite edge capacities \cite{li2021effective}.
\end{abstract}

% \fi
\IEEEpeerreviewmaketitle

\section{Introduction}
In recent years, quantum information technologies have been widely developed and achieved remarkable breakthroughs especially in secure communications \cite{yin2017satellite}.
Along with the concept validation of quantum repeater and long-distance quantum communications \cite{kimble2008quantum, pu2017experimental}, %\cite{kimble2008quantum,muralidharan2016optimal,pu2017experimental,azuma2015all},
the quantum network, which is foreseen to be a ``game-changer'' to the classic network, is being developed at a rapid pace.

\begin{figure}[t]
  \centering
  \includegraphics[width=1\linewidth]{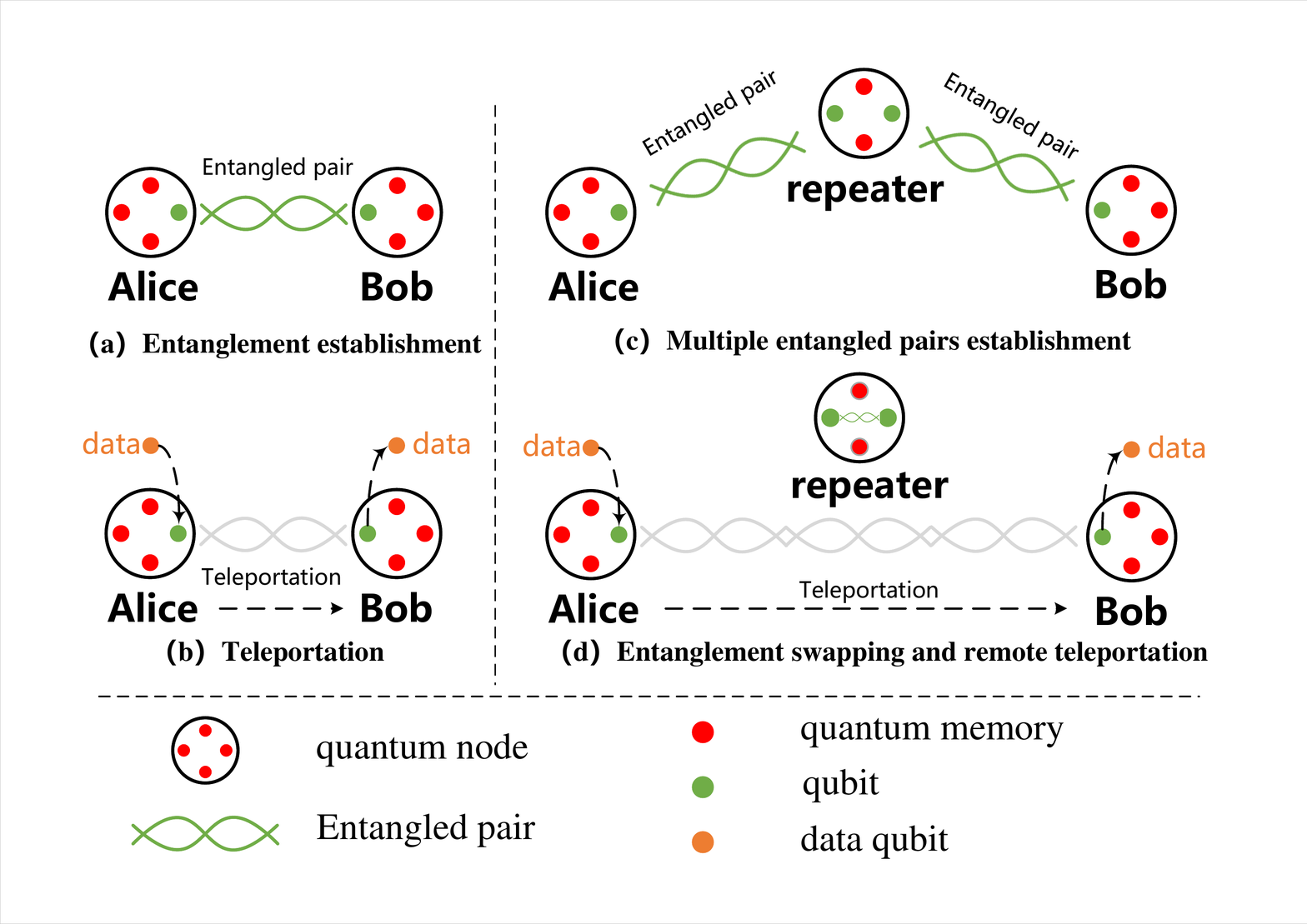}
  \setlength{\abovecaptionskip}{0.1cm}
  \caption{Illustration of information transmission in a quantum network, data qubit represents the state that Alice wished to teleport \cite[chapter 4]{van2014quantum}. (a)(b) an arbitrary single \textit{qubit} can be sent using short-distance quantum teleportation. (c)(d) establishment of remote entanglement connection through entanglement swapping, data transmission using long-distance quantum teleportation.}
  \label{entanglement}
\end{figure}

In a quantum network, quantum nodes (including quantum processors and repeaters) are interconnected via optical links, and they can generate, store, exchange, and process quantum information \cite{shi2020concurrent,li18building}. When two faraway quantum nodes, serving as source and destination, attempt to exchange information, the quantum network first establishes the entanglement connection between them, and then information is transmitted in the form of quantum bits (called \textit{qubits}) over entanglement connection experiencing noisy channel to the destination \cite{bugalho2021distributing}. As shown in Fig. \ref{entanglement}, to establish such end-to-end entanglement connection, entangled pairs between adjacent nodes are first generated. After that, quantum repeater connects quantum nodes over longer distances by performing entanglement swapping, i.e., joint Bell state measurements at the local repeater aided by classical communication \cite{li2021effective}.

To build up a large-scale functional quantum network with satisfying the dynamic requests from Source-Destination (S-D) pairs, the critical problem we have to face firstly is how to select routing path and utilize network resources efficiently (such as limited entangled pairs on each edge). Recently, some existing studies are dedicated to solve such problem \cite{zhao2021redundant}, and propose multiple entanglement routing designs to improve the robustness and throughput facing the failure of entanglement generations. Although the development of quantum network is currently at the primitive stage, these routing designs bring a good start to facilitate the process of quantum networks' construction in the future.

%To support various quantum applications, such as distributed quantum computing, sensing and metrology, clock synchronization, etc \cite{dahlberg2019link}, one main function of a quantum network is to distribute entangled pairs between two arbitrary quantum nodes. In a quantum network, information is represented by quantum bits  and transmitted between a pair of nodes over an established entanglement link or connection \cite{shi2020concurrent}.

%Facing the dynamic requests from multiple Source-Destination (S-D) pair, the entanglement routing becomes a critical problem especially in a large-scale quantum network. Considering the limited network resource such as quantum memory on quantum repeater and generated entanglement pairs on each link, how to efficiently select establishment path of entanglement connection and utilize the resources determines the performance in terms of throughput, latency and so on.
%Since each node has a small quantum memory to store a limited number of qubits and \textcolor{red}{each edge has a limited number of channels over which an entanglement link can be created}, how to efficiently select establishment path of entanglement connection and utilize the network resources in the quantum network becomes a critical problem.

%To facilitate the development of quantum networks, many researchers devote themselves to overcome the bottom-up challenges building a practical quantum network. The entanglement routing, as the core function provided by the quantum network, plays an important role in end-to-end transmission.
%After the first concept of quantum networks introduced by DARPA in the early 2000s, related explorations have begun by considering the unique characteristics of entanglement. The existing works focus on , such as \cite{shi2020concurrent}. However, due to the requirement of transmission quality, as the most important metric to evaluate the quality of entangled pairs or entanglement connection, the fidelity is vital to various quantum applications, such as distributed quantum computing, sensing and metrology, clock synchronization, etc \cite{dahlberg2019link}.

However, an important metric to evaluate the quality of remote entanglement connection, entanglement fidelity, is rarely considered in the exisitng entanglement routing designs. In practice, due to the noise in the system, quantum repeaters sometimes might not generate entangled pairs with a certain desired fidelity, which brings negative effects on various quantum applications \cite{cacciapuoti2019quantum}.
For example, in quantum cryptography protocols (e.g., BB84 protocol), an entanglement fidelity lower than the quantum bit error rate can reduce the security of key distribution \cite{jia2021improved}.
%an entanglement fidelity beyond the quantum bit error rate is required to ensure the security of key distribution \cite{li2021effective}.
%which is vital to various quantum applications (such as distributed quantum computing, sensing and metrology, clock synchronization, etc \cite{dahlberg2019link}),
To improve the fidelity of entanglement connection and satisfy the requirement of quantum applications, a technique called entanglement purification can be used to increase the fidelity of entangled pairs \cite{cacciapuoti2020entanglement}. It consumes shared lower-fidelity entangled pairs along the link between adjacent nodes to obtain one higher-fidelity entangled pair. By adopting purification technique, the entanglement routing can provide fidelity guarantee for end-to-end entanglement connection. Nevertheless, due to the nonlinear relationship between fidelity improvement and resource consumption in purification operation, the additional purification decision makes the entanglement routing problem more complicated. Thus, how to design such fidelity-guaranteed entanglement routing remains an unsolved problem.

%Therefore, in this paper, we explore the possibility of using such purification technology among adjacent nodes to enable the efficient fidelity guarantee service for end-to-end entanglement connection.

Based on such considerations, in this paper, we focus on purification-enabled entanglement routing design under the fidelity constraint in general quantum networks.
To address the complicated entanglement routing problem, we first study the entanglement routing problem in single S-D pair scenarios, and respectively propose an iterative routing algorithm to obtain the optimal solution and a low-complexity routing algorithm to obtain near-optimal but efficient solution. To obtain the optimal purification decisions, we also analyze the characteristic of purification operations and propose an optimal decision approach. After that, we further study the entanglement routing problem in multiple S-D pairs scenarios, and propose a greedy-based routing algorithm considering two resource allocation methods.
%For purification decision along a multi-hop path, we also design two efficient methods to obtain a near-optimal decision, rather than the optimal but high-complexity method.
We also conduct extensive simulations to show the superiority of the proposed algorithms compared with the existing ones.
Although the existing work \cite{chakraborty2020entanglement} has already proposed an entanglement distribution design and imposed a minimum end-to-end fidelity as a requirement, it does not take purification into consideration and then the fidelity of each Bell pair cannot be further improved. Thus, to the best of our knowledge, this is the first work that provides end-to-end fidelity-guaranteed entanglement routing with purification decision, which can fully leverage the advantages of purification operation and significantly improve the end-to-end fidelity with abundant low-fidelity Bell pairs.

The contributions of this paper can be summarized as follows:

\begin{itemize}
\item For the requirement of high-quality entanglement connections from various quantum applications, we propose the first entanglement routing and purification design that provides end-to-end fidelity-guaranteed connections for S-D pairs in ``advance generation'' model based quantum networks. For single S-D pair scenarios, we devise two novel entanglement routing algorithms, i.e., Q-PATH and Q-LEAP, respectively. The former one can obtain multiple routing paths for satisfying single S-D pair and provide the optimal routing solution with minimum entangled pair cost, and the latter one can efficiently provide the routing solution with minimum fidelity degradation and has the advantage of low computational complexity.
    %By analyzing the nature of the entanglement routing problem, we decompose the complex problem and start with the one in single S-D pair scenarios. We first propose an iterative routing design, Q-PATH, which may obtain multiple routing paths for satisfying single S-D pair, to provide the optimal routing solution with minimum entangled pair cost. Due to the high computational complexity, we further propose another low-complexity routing design, Q-LEAP, to provide the routing solution with minimum fidelity degradation. For purification decision along a multi-hop path, we also design a simple but efficient method to guarantee the fidelity constraint of end-to-end connection.}
%\item For the requirement of high-quality entanglement connections from various quantum applications, we propose the first entanglement routing design that provides end-to-end  fidelity-guaranteed connections for S-D pairs. By analyzing the nature of the entanglement routing problem, we decompose the complex problem and start with the one in single S-D pair scenarios. We first design an iterative routing design, Q-PATH, to obtain the optimal routing path and purification decision with minimum entangled pair cost. Due to the high computational complexity, we further propose another low-complexity routing design, Q-LEAP, to obtain the routing path with minimum fidelity degradation. For purification decision along a multi-hop path, we also design a simple but efficient method to guarantee the fidelity constraint of end-to-end connection.
\item Based on the routing solutions provided by algorithms designed for single S-D pair, we further consider the routing problem in multiple S-D pairs scenarios as a resource allocation problem, and propose a greedy-based routing design, which leverages two important factors of a given routing solution, i.e., resource consumption and degree of freedom, to globally allocate entanglement resources for various routing solutions and improve the efficiency of resource utilization.%Based on the paths provided by algorithms designed for single S-D pair, the routing problem in multiple S-D pairs scenarios is further considered as a resource allocation problem. We propose a greedy-based routing design for the requests of multiple S-D pairs. To allocate entanglement resources for various routing paths, we introduce a utility metric, which considers two important factors of a given routing solution, resource consumption and degree of freedom, to evaluate the degree of difficulty when satisfying each request. %By adopting the proposed utility metric, the throughput of the routing solution provided by the proposed routing design can be further improved.
\item To verify the effectiveness of the proposed algorithms, extensive simulations are conducted. Compared to the existing routing scheme with purification decisions, the proposed algorithms not only provide fidelity-guaranteed routing solutions, but also show the significant superiority in terms of throughput, the average fidelity of the end-to-end connections, and network resource utilization.
%We conduct simulations to verify the superiority of our scheme in terms of energy efficiency and traffic offloading. Simulation results demonstrate that,
\end{itemize}

The rest of this paper is organized as follows. Firstly, related work is discussed in Section II. Then, the motivation, the network model and the routing problem considered in this paper are given in Section III. After that, the entanglement routing designs for single S-D pair and multiple S-D pairs are given in Section IV and Section V, respectively.
Finally, the performance evaluation is conducted in Section VI and conclusions are drawn in Section VII.
%Although research has begun from 2010 after the concept of quantum networks introduced by the DARPA quantum network project, the existing works never consider fidelity as one routing metric.

\section{Related Work}
The interconnection of quantum devices forms a quantum network by enabling quantum communications among remote quantum nodes, and academic community believes the ultimate objective of the development of quantum networks is to build a global system, called quantum internet, with interconnected networks around the world that uses quantum internet protocol \cite{illiano2022quantum}, which is similar to the Internet. To achieve this ambition, Cacciapuoti \cite{cacciapuoti2019quantum}, Caleffi \cite{caleffi2020quantum} and others thoroughly survey the theoretical and practical problems of networking, and significantly push forward the development of quantum internet in terms of entanglement distribution, protocol design, optimization of physical devices and so on.
In quantum networks, long-distance entanglement connection is required by various quantum applications, such as distributed quantum computing, sensing and metrology and clock synchronization \cite{dahlberg2019link}. To establish a multi-hop quantum entanglement connection via quantum repeaters for multiple S-D pairs, an efficient entanglement routing solution is required.
In general, two kinds of quantum network models, i.e., \textit{``advance generation''} model (entanglement generation before routing decision) and \textit{``on-demand generation''} model (entanglement generation after routing decision), are concerned in existing work. The former basically isolates the functions between link layer and network layer, and only resource allocation and path selection should be considered in the routing design. The latter, however, tightly couples link layer and network layer, and not only path selection but also entanglement generation and potential failures should be considered in the routing design. Thus, in this paper, we focus on the entanglement routing problem based on \textit{``advance generation''} model.

%\textcolor{red}{Due to the decoherence of quantum memory, time-division model is adopted by all existing works.}

%For \textit{``advance generation''} model, %various routing designs are considered in different network topology.
%In existing works, the routing designs are different according to the network model
Most of the existing studies on quantum networks focus on a specific network topology, such as diamond \cite{pirandola2019end}, star \cite{vardoyan2019stochastic}, and square grid \cite{li2021effective,pant2019routing}.
%vardoyan2019stochastic,pant2019routing,gyongyosi2018decentralized,caleffi2017optimal,
Pirandola \cite{pirandola2019end}, Vardoyan {\textit{et al.}} \cite{vardoyan2019stochastic}, and Pant {\textit{et al.}} \cite{pant2019routing} considered the physical characteristics of quantum networks, such as quantum memory and decoherence time, and developed routing protocols and theoretical analysis about end-to-end capacities and expected number of stored \textit{qubits} for homogeneous systems.
Considering the routing problem with quantum memory failures, Gyongyosi. {\textit{et al.}} \cite{gyongyosi2019adaptive} proposed an efficient adaptive routing based on base-graph. Thus, when the failure happens, some entanglement connections can be destroyed but a seamless network transmission can still be provided since shortest replacement paths can be found by using the adaptive routing.
Caleffi \cite{caleffi2017optimal} considered stochastic framework that jointly accounts several physical-mechanisms such as decoherence time, atom–photon/photon–photon entanglement generation and entanglement swapping, and derived the closed-form expression of the end-to-end entanglement rate. Based on that, the authors further proposed an optimal routing protocol when using the proposed entanglement rate as routing metric.
After that, Hahn \textit{et al.} \cite{hahn2019quantum} utilized a graph state and proposed a general routing method in arbitrary networks.
To be noticed, the existing studies rarely consider fidelity as one of the metric in entanglement routing. One representative study was proposed by Li {\textit{et al.}} \cite{li2021effective}, who considered a lattice topology and proposed an effective routing scheme to enable automatic responses for multiple requests of S-D pairs. The authors considered the purification operation to ensure the fidelity of entanglement connection, however, the purification operation is performed before routing decision to satisfy the fidelity constraint. Due to the fidelity degradation of entanglement swapping, this simple purification decision cannot provide end-to-end fidelity guarantee. Thus, the lack of existing routing design involving fidelity encourages us to design a novel fidelity-guaranteed entanglement routing scheme in future quantum network.

\begin{table}[t]
  \centering
  \caption{Symbols used in the paper}
  \label{Parameters}
  \begin{tabular}{cp{6cm}}%p{4.8cm}
    \hline\hline
     \textbf{Symbol} & \textbf{Notation} \\\hline
     %Number of Tx antennas & 4 \\\hline
%     \multirowcell{4}[-.2ex][c]{\centering $(V,E,C)$} & Network topology with the set of nodes $V$, edges $E$, and capacity $C=\{c(u,v)|(u,v)\in E\}$, where $c(u,v)$ is the capacity on edge $(u,v)$. \\
%     $(u,v)$ & Edge between node $u$ and node $v$. \\
%     $\langle s_i,d_i\rangle$ & $i$-th source-destination pair. \\
     \multirowcell{2}[-.2ex][c]{$R_i$} & The number of routing requests from $i$-th S-D pair, and the unit is one end-to-end entanglement connection. \\
     $c(u,v)$ & Remaining capacity on edge $(u,v)$. \\
     $F^{th}_i$ & Fidelity threshold for $i$-th S-D pair.  \\
     %$M(u)$ & Quantum memory size at node $u$. \\
     $\mathcal{N}(u)$ & Number of neighbors of node $u$.  \\
     \multirowcell{2}[-.2ex][c]{\centering $F_{i,j}(s,d)$} & Fidelity of end-to-end entanglement for $j$-th routing path of $i$-th S-D pair.  \\
     \multirowcell{2}[-.2ex][c]{\centering $P_{i,j}(s,d)$} & Routing path $\{(s_i,u_1),...,(u_n,d_i)\}$ for $j$-th routing path of $i$-th S-D pair.\\
     \multirowcell{3}[-.2ex][c]{$N^{Pur}_{i,j}(u,v)$} & Number of purification rounds on edge $(u,v)$ for $j$-th routing path of $i$-th S-D pair. \\
     \multirowcell{3}[-.2ex][c]{\centering $D^{pur}_{i,j}$} & Purification decisions for $j$-th routing path of $i$-th S-D pair, $D^{pur}_{i,j}=[N^{pur}_{i,j}(s_i,u_1),...,N^{pur}_{i,j}(u_n,d_i)]$. \\
     \multirowcell{2}[-.2ex][c]{\centering $t_{i,j}^{EXT}(u,v)$} & Expected throughput on edge $(u,v)$ for $j$-th path of $i$-th S-D pair after purification.\\
     \multirowcell{2}[-.2ex][c]{\centering$T_{i,j}^{EXT}(s_i,d_i)$} & Expected throughput on routing path $P_{i,j}(s_i,d_i)$. \\\hline\hline
  \end{tabular}
\end{table}

\section{Network Model and Problem Description}
In this section, we first provide the motivation of our work, and then introduce the network model. Further, we define the entanglement routing problem in quantum networks and analyze its property. The notations used in this paper are summarized in Table \ref{Parameters}.

\subsection{Motivation}
The work proposed in this paper is motivated by the desire to provide fidelity guarantee for various quantum applications. Although several routing protocols and algorithms have already been proposed to handle the requests from multiple S-D pairs in existing work, the fidelity degradation during the entanglement swapping has not been considered and an end-to-end fidelity guarantee of entanglement connection\footnote{In this paper, we consider a entanglement connection as a end-to-end shared entangled pair between source node and destination node.} cannot be provided. Thus, we consider an entanglement generation before routing decision model, i.e., ``\textit{advance generation}'' model, and focus on satisfying end-to-end fidelity constraint with minimum resource consumption through optimizing path selection with considerations on purification operations. %To the best of our knowledge, this is the first work that studies the fidelity-guaranteed entanglement routing problem for end-to-end requirements of multiple S-D pairs.

\subsection{Network model}
%Most of the existing works on quantum networks focus on a specific network topology such as a chain, diamond, ring, star, square grid, etc \cite{pirandola2019end,vardoyan2019stochastic,pant2019routing,caleffi2017optimal}.
%To design a practical entanglement routing method, a general topology is considered in this paper.
A general quantum network is described by a graph $G=(V,E,C)$, where $V$ is the set of $|V|$ quantum nodes, $E$ is the set of $|E|$ edges, and $C$ is the set of edge capacity. An edge $(u,v)$ between two nodes means that two nodes share one or more quantum channels, and the capacity $c(u,v)$ determines the maximum number of the entangled pairs that can be provided.
%Capacity $C(u,v)$ on each edge is determined by the generation rate of entanglement pairs of EPR, and it can be calculated by $C(u,v)=r_{ent}t_{ent}$.
%(requires the rate of entanglement generation between nodes to exceed the decoherence (loss) rate of the entanglement).
For the quantum node $v\in V$, the quantum channel on edge $e\in E$, and the source-destination pair of a routing request, we give the definition as follows:

1) \textbf{\textit{Quantum Node:}} %An illustration of a quantum network is shown in Fig. \ref{topology}.
Quantum node is the device equipped with necessary hardware to perform quantum entanglement and teleportation on the \textit{qubits}.
%To support long-distance entanglement connection between two remote nodes via entanglement swapping, quantum nodes, which are equipped with necessary hardware to perform quantum entanglement and teleportation on the \textit{qubits}, are considered in the network.
Each quantum node holds the complete function of a quantum repeater\footnote{In this paper, we consider first-generation quantum repeaters, hence a finite number of qubit memories is considered, and entanglement generation and purification are applied on the repeaters, and quantum error correction is not available \cite{singh2021quantum,muralidharan2016optimal}.}.
Arbitrary quantum nodes are equipped with quantum processors and quantum applications can be deployed.
%Each node represents a quantum station with a finite number of qubit memories.

2) \textbf{\textit{Quantum Channel:}}
%A quantum channel can be established among quantum nodes through connecting them through polarization-maintaining optical fibers \cite{shi2020concurrent} with shared entanglement pairs to support \textit{qubits} transmissions.
A quantum channel is established between adjacent quantum nodes to support the transmission of \textit{qubits} via physical links (such as optical fibers \cite{shi2020concurrent} and free-space %[nature"mozi"]
\cite{ren2017ground})
with shared entanglement pairs and support qubit transmission. The encoding of \textit{qubits} may have multiple choices, e.g., polarzation-encoded \textit{qubits}, phase-encoded \textit{qubits}, etc.
%Quantum nodes in the network are connected through polarization-maintaining optical fibers \cite{shi2020concurrent} to establish quantum channels with shared entanglement pairs and support qubit transmission.
%A quantum link is successfully established with a shared entanglement pair between two nodes.
Here, a constant capacity of each quantum channel is considered, which means a constant number of the entangled pairs between two adjacent nodes are generated at the start of each time slot. %and the number of generated entanglement pairs is limited by maximum memory of quantum nodes, i.e., $M(u)=\sum_{v\in Neighbor(u)}C(u,v)$.
The process of entanglement generation can also be considered as a deterministic black box (Nitrogen-Vacancy platform is considered), and the fidelity of a generated entanglement pair on each quantum channel can be approximately calculated by a deterministic formula without consideration of noise \cite{humphreys2018deterministic,caleffi2020quantum}.
%Meanwhile, the process of entanglement generation can be considered as a fully deterministic black box (Nitrogen-Vacancy platform is considered), fidelity of generated entanglement pair on each quantum channel can be calculated by $F=1-\alpha$ \cite{humphreys2018deterministic,dahlberg2019link}, %the generation rate $r_{ent}$, and the fidelity of the entangled pair $F_{det}=p_{succ}F_{succ}+(1-p_{succ})F_{unent}$.
%where $\alpha$ is decided by the distance of the channel and physical media.

%Capacity $C(\eta)=-\log(1-\eta_{min})$ with transmissivity $\eta$.

3) \textbf{\textit{Source-Destination Pair:}} %In the network, the quantum processors, which are similar to the end hosts in classical networks, are equipped with quantum processors and running quantum applications. %Different from classical end hosts, each quantum processor is equipped with a limited number of $M(u)$ memory to store qubits, and
%Each quantum processor is equipped with necessary hardware to perform quantum entanglement and teleportation on the \textit{qubits}.
Due to the requirement of quantum applications, a quantum node may intend to establish entanglement connection with the other node. Herein, we name such pair of quantum nodes with the intention of entanglement connection establishment as a Source-Destination (S-D) pair. %such as $\langle s_1,d_1\rangle$ in Fig. \ref{iterative_routing_example1}.
%Note that the quantum processor includes the complete function of a repeater, thus, we name both quantum processors and repeaters as quantum nodes.

Based on the above definitions on quantum node, channel, and S-D pairs, we introduce the network management method.
In quantum networks, all quantum nodes are connected via classical networks, and each node has a certain level of classical computing and storage capacity.
Similar to the existing studies \cite{zhao2021redundant,shi2020concurrent}, we assume a time-synchronous network operating in time slots\footnote{The network is synchronized to a clock where each timestep is no longer than the memory decoherence time.}. To manage a quantum network, all quantum nodes are controlled by a centralized controller via classical networks. The controller holds all the basic information of the network, such as network topology and resources, which can be reported and updated by the quantum nodes.
For an entanglement routing process, it consists of three phases. At the beginning of each time slot, adjacent quantum nodes start to generate the entangled pairs, and the controller collects routing requests from quantum nodes. Then, the controller executes routing algorithm to determine the routing path of each S-D pair and resource allocation in the network. Note that part of routing requests might be denied due to the connectivity or resource limitation. Finally, according to the instructions from the controller, all quantum nodes perform purification and swapping to concatenate single-hop entangled pair and establish multi-hop entanglement connections for S-D pairs.

%Supported by deterministic state-delivery protocol \cite{humphreys2018deterministic}, which can be cast into a fully deterministic black box, the fidelity can be calculated by $F_{det}=p_{succ}F_{succ}+(1-p_{succ})F_{unent}$ with constant physical equipment.
%The entanglement generation adopts the state-of-the-art quantum processor platform ( Nitrogen-Vacancy (NV) centers in diamond) \cite{dahlberg2019link,humphreys2018deterministic}.

%\textcolor{red}{To ensure this fidelity is above a certain threshold Fth, the network is initialized by quantum entanglement purification on each edge, which results in a reduced number of shared entanglement pairs, thus a reduced edge capacity. We assume the fidelities of multiple entangled pairs along the same edge to be identical while the fidelities on different edges can vary, as influenced by various factors such as geographical constraints or human interventions. The average entanglement fidelity of a certain edge can be calculated by pre-characterization; here we assume the fidelities on all edges to follow a normal distribution N(Fmean, Fstd). Note we take the fidelity to be one if it exceeds unity.}

%\begin{figure}[t]
%  \centering
%  \includegraphics[width=0.5\linewidth]{purification_example2}
%  \caption{The process of 2-round purification operations.}
%  \label{fig:purification}
%\end{figure}
\begin{figure}[t]
  \centering
    \subfigure[The process of 2-round purification operations]{
    \label{purification_performance-0}
    \includegraphics[width=3.3in]{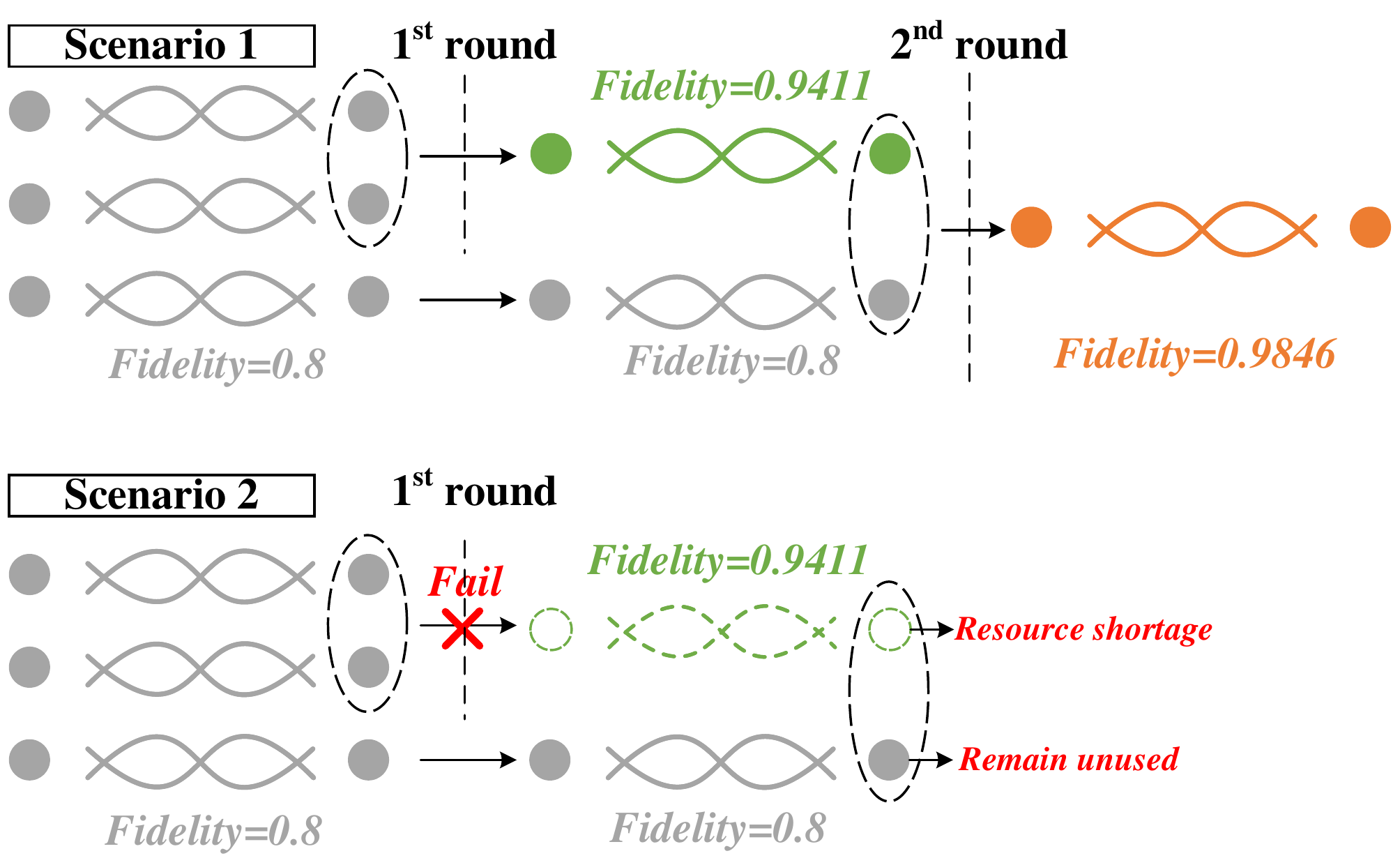}
 }
  \subfigure[success probability]{
    \label{purification_performance-1}
    \includegraphics[width=1.62in]{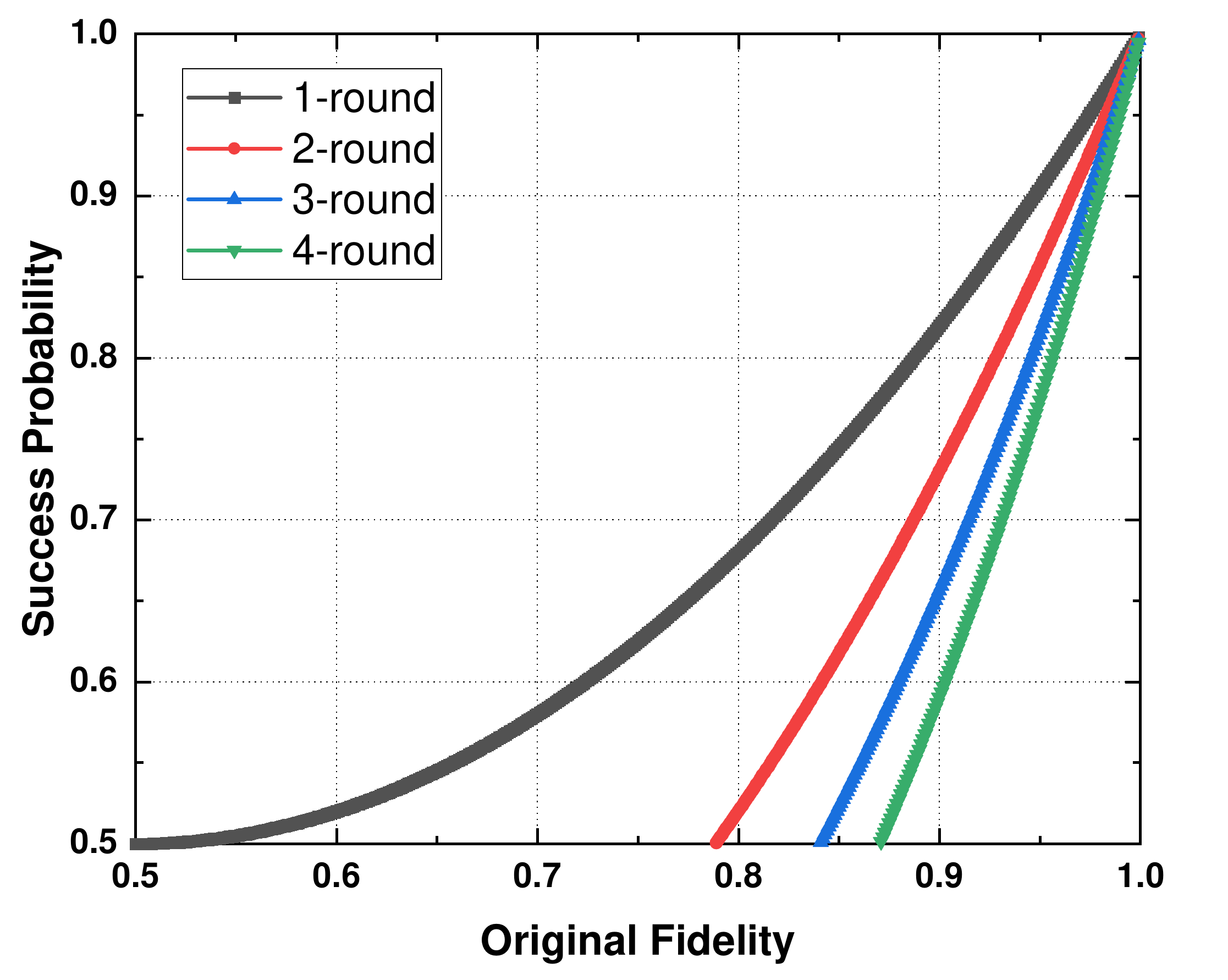}
 }
  \subfigure[resulting fidelity]{
    \label{purification_performance-2}
    \includegraphics[width=1.62in]{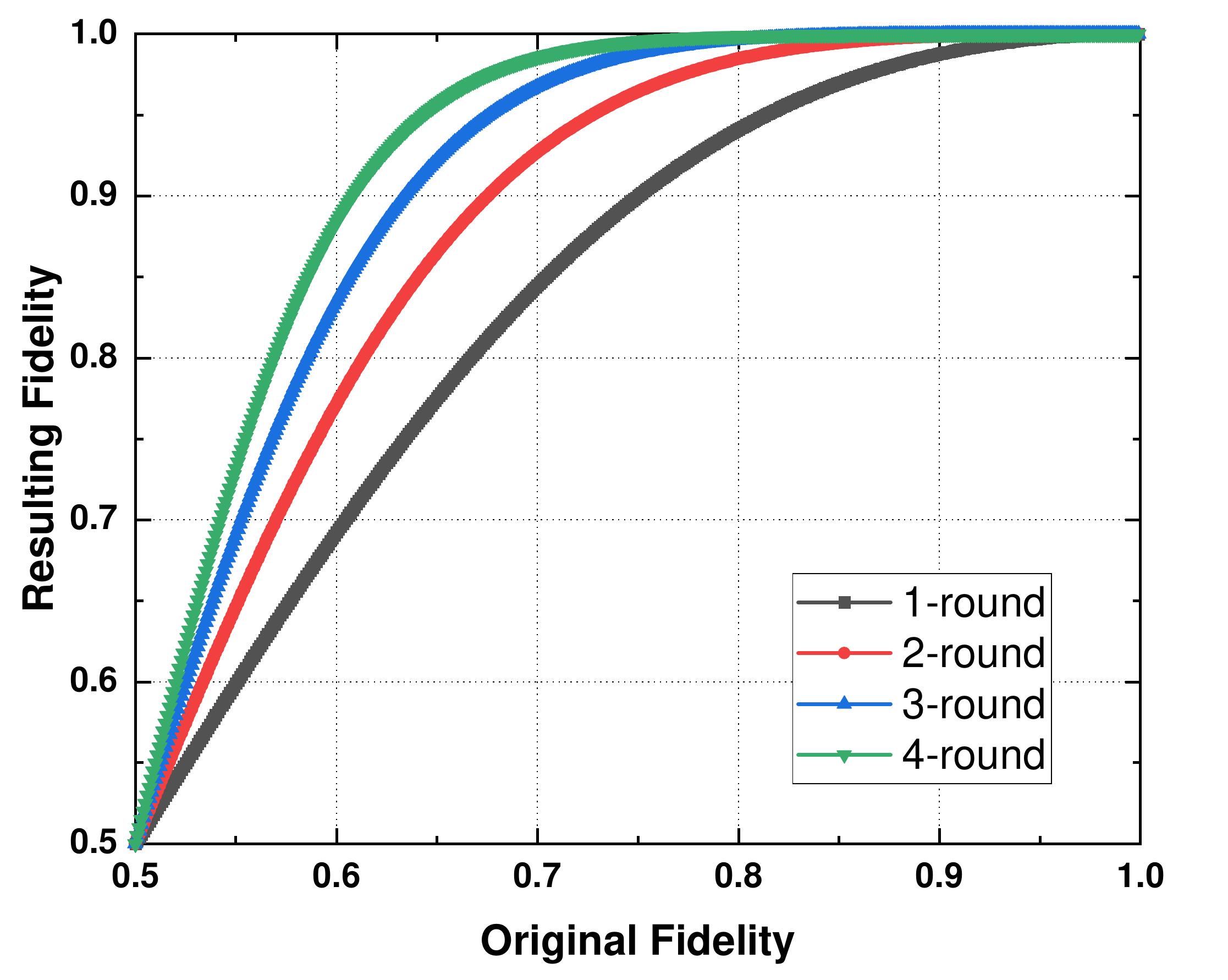}
    }
    \setlength{\abovecaptionskip}{0.1cm}
  \caption{Illustration of purification operations and performance for multi-round purification operations.}
  \label{fig:purification_performance}
\end{figure}

%Finally, the unique operations during the entanglement routing process are described as follows.
To establish end-to-end entanglement connection in quantum networks, three unique operations, i.e., entanglement generation, purification and swapping, which have no analogue in classical networking, should be considered: %In the following, we would introduce these concepts in details.

1) \textbf{\textit{Entanglement Generation:}}
Physical entanglement generation can be performed between two controllable quantum nodes, which connect to an intermediate station, called the heralding station, over optical fibers by using various hardware platform, such as Nitrogen-Vacancy centers in diamond \cite{dahlberg2019link}. After one success generation attempt, the entangled pair \footnote{In this paper, only Bell entangled state is considered.} can be stored in the memory of quantum nodes as the available resource to establish entanglement connection and transmit an arbitrary single \textit{qubit} state using teleportation.
%Due to the characteristic of physical equipment, the generation of entanglement pair is lossy and decoherent. The longest-distance record for producing entanglement keeps being broken. With the development of physical equipment, the capability of quantum nodes can be stronger.
%Although the success of one  generation attempt is probabilistic,

%Entanglement generation between the electron spins of two emitters separated with a certain distance. Although each entanglement is generated probabilistically, with the generation rate and generation time, the generated pairs can be regarded as a constant resource.
%The entanglement-generation rate across a link of transmissivity $\eta$, in the absence of any repeater mediation, is limited to $C=-\log(1-\eta)$.

2) \textbf{\textit{Entanglement Purification:}}
Entanglement purification enables two low-fidelity Bell pairs to be merged into single higher-fidelity one, which can be implemented using CNOT gates or optically using polarizing beamsplitters \cite{hu2021long}.
%The purification operation can be implemented using CNOT gates between adjacent nodes, both parties perform a measurement and  the resulting pair is determined to have a higher fidelity or be discarded \cite{victora2020purification}.
By considering bit flip errors, %and phase flip errors,
the resulting fidelity after purification operation can be calculated by \cite{van2014quantum}:
%\begin{equation}
%f(F_1,F_2)=\frac{F_1F_2+\frac{1}{9}(1-F_1)(1-F_2)}{F_1F_2+\frac{1}{3}F_1(1-F_2)+\frac{1}{3}F_2(1-F_1)+\frac{5}{9}(1-F_1)(1-F_2)}.
%\end{equation}
%\begin{equation}\label{fidelity_after_purification}
%f(x)=\frac{x^2+\frac{1}{9}(1-x)^2}{x^2+\frac{2}{3}x(1-x)+\frac{5}{9}(1-x)^2}.
%\end{equation}
%{\footnotesize
%\begin{align}\label{fidelity_after_purification}
%f(x_1,x_2)=\frac{x_1x_2+\frac{1}{9}(1-x_1)(1-x_2)}{x_1x_2+\frac{1}{3}x_1(1-x_2)+\frac{1}{3}x_2(1-x_1)+\frac{5}{9}(1-x_1)(1-x_2)}.
%\end{align}
%}
\begin{align}\label{fidelity_after_purification}
f(x_1,x_2)=\frac{x_1x_2}{x_1x_2+(1-x_1)(1-x_2)}.
\end{align}
where $x_1,x_2$ is the fidelity of two Bell pairs in the purification operation.
If we consider the fidelity of Bell pairs on the same edge are the same, then $x_1=x_2$, and the formula can be simplified as
\begin{align}\label{fidelity_after_purification2}
f(x)=\frac{x^2}{x^2+(1-x)(1-x)}.
\end{align}
This process can be applied recursively so as to in principle achieve arbitrarily high fidelities.
An example of multi-round purification operation on the same edge is shown in Fig. \ref{purification_performance-0}. Note that the dashed oval in the figure represents a purification operation, and the resulting fidelity is obtained by using Eq. (\ref{fidelity_after_purification}). Meanwhile, all entangled pairs before 1st round purification are generated on the same edge with the same fidelity (i.e., 0.8),  a \textit{pumping} purification scheme \cite{van2008system} is considered, which means that each round of purification operation consumes an extra entangled pair. In this example, we consider two scenarios, the first one is that all purification operations are implemented successfully, and thus the final fidelity after 2nd round purification is 0.9846. The second one is that 1st purification operation is failed, and two low-fidelity entangled pairs are thus broken. In the next, 2nd purification operation cannot be implemented, and the bottom entangled pair in scenario 2 remains unused and it can be used for other purification operations in the future.  %Due to the failure during the 1st round purification, only one higher-fidelity entangled pair is obtained from the 2nd round purification. In short, to apply $n$ rounds of purification operations, the required number of Bell-pair resource scales as $2^n$.
The specific performance for multi-round purification operation is shown in Fig. \ref{purification_performance-1}-\ref{purification_performance-2}. One challenge problem for designing a fidelity-guaranteed routing is to determine whether purification will be performed or not and the number of purification rounds for the intermediate nodes on the path between source and destination if necessary.

3) \textbf{\textit{Entanglement Swapping:}} To connect quantum nodes and establish long-distance entanglement connection, entanglement swapping can be regarded as an attractive approach. As shown in Fig. \ref{entanglement}, a quantum repeater that carries entangled pairs to both Alice and Bob can turn the two one-hop entanglements into one direct entanglement between Alice and Bob \cite{illiano2022quantum,van2008system}. By repeating swapping operations, multi-hop entanglement connection along the path of repeaters carrying entangled pairs can be established. To be noticed, due to the imperfect measurement (i.e., noisy operation \cite{childress2005fault,perseguers2013distribution}) on the repeater, the fidelity of multi-hop entanglement would degrade during entanglement swapping. Meanwhile, considering fidelity variance of entangled pairs on different quantum channel, different routing paths can lead to distinct fidelity results of end-to-end entanglement connection after swapping. It is another challenge issue for designing fidelity-guaranteed entanglement routing.

\subsection{Entanglement Routing Problem}
A fidelity-guaranteed entanglement routing problem can be described as follows: Given a quantum network with topology and edge capacity as $G=(V,E,C)$, finding routing solutions, including purification decision $D^{pur}_{i,j}$ and path selection $P_{i,j}(s_i,d_i)$, to enable the entanglement establishment between S-D pairs to satisfy fidelity constraint $F_{i,j}(s_i,d_i)>F^{th}_i, \forall i$.

The routing problem in traditional networks, as a classic one, has been studied for decades \cite{yen1971finding,szymanski2013max}. %and the efficient algorithms have been proposed to solve the basic routing problem such as shortest-path problem or max-flow problem \cite{van2013path,yen1971finding,li2019temporal,szymanski2013max}.
However, the routing problem for multiple S-D pairs, which belongs to multi-commodity flow problem \cite{ahuja1993network,zhang2018dynamic}, has been proven as a NP-hard problem. In quantum networks, due to the unique characteristics such as entanglement purification, entanglement routing problem becomes knotty since the special characteristics and constraints have to be considered. In specific, for the entanglement routing problem with fidelity guarantee, additional purification decision problem are coupled with path searching problem, which makes the entanglement routing problem more complicated.
%One of the earlier works is \cite{van2013path}, which adopted Dijkstra algorithm to find a single path for each S-D pair \cite{li2019temporal}....
%\textbf{To the best of our knowledge, this is the first work that studies the fidelity-guaranteed entanglement routing problem for multiple S-D pairs.}
%provide fidelity-guaranteed entanglement connection for as long as the. for increasing fidelity during the process of entanglement connection establishment.

\section{Routing Design for Single S-D Pair}
In this section, we focus on the routing problem for single S-D pair. At first, we propose Q-PATH, a \underline{P}urification-enabled iter\underline{A}tive rou\underline{T}ing algorit\underline{H}m,  to obtain the optimal routing path and purification decisions with minimum entangled pair cost. To further reduce the high computational complexity, we propose Q-LEAP, a \underline{L}ow-compl\underline{E}xity routing \underline{A}lgorithm from the perspective of ``multi\underline{P}licative'' routing metric of the fidelity degradation. %Based on these two routing designs for single S-D pair, in the next section, a greedy approach-based entanglement routing algorithm for multiple S-D pairs is further proposed.
%\textcolor{red}{pseudo-optimal}
%Min-Cut-Max-Flow (MCMF)

\begin{algorithm}[t]
\caption{Q-PATH: Iterative Routing}
\label{Iteration_routing}
\KwIn{$G=(V,E,C)$, $F^{th}_i$, request $R_i$ and $\langle s_i,d_i\rangle$;}
\KwOut{$P_{i,j}(s_i,d_i)$, $D^{pur}_{i,j}$, $F_j(s_i,d_i)$, $T^{EXT}_{i,j}$;}

\textit{\textbf{Step 1 Initialization:} \\}
Calculate \textit{Purification Cost Table} for $(u,v)\in E$;\label{Initialization1}\\
Delete all edges $(u,v)$ from $G$, if $F^{pur}_{max}(u,v)<F^{th}$;\\
Construct auxiliary graph $G^a=(V,E^a,C^a,Cost)$;\label{Initialization3}\\
$Q\leftarrow$ Priority queue according to value of $min\_cost$;\label{Initialization4}\\
Find shortest path on $G$ with $H^{min}$ by using BFS;\label{shortestPath}\\
%$q\leftarrow$ Priority queue saving edges along the routing path according to value of its fidelity;\\
\For{$min\_cost=H^{min}:|E|C_{max}$\label{mincost_loop}}{
\textit{\textbf{Step 2 Path Selection Procedure:} \\}
Multiple shortest paths set $P^{SPF}_{min\_cost}$ with the same cost $min\_cost$ $\leftarrow$ $K$-shortest path algorithm;\\
\If{no available path for $\langle s_i,d_i\rangle$}{
\textbf{break};\\
}
\textit{\textbf{Step 3 Edge Cost Update:} \\}
%Select $\mathcal{P}\subseteq P^{SPF}_{min\_cost}$ with minimum hops;\\
%Sort $P^{SPF}_{min\_cost}$ with ascending path length;\\
\For{$P_{i,j}(s_i,d_i)\in P^{SPF}_{min\_cost}$ with minimum hops}{
%\textcolor{red}{Construct $q$ for path $P_{i,j}(s_i,d_i)$;}\\
%$D^{pur}_{i,j}\leftarrow$ purification decision by Algorithm \ref{Purification_decision};
\While{$F_{i,j}(s_i,d_i)<F^{th}_i$\label{fidelity_check1}}{
Find $(u,v)\in P_{i,j}(s_i,d_i)$ with maximum fidelity improvement;\label{max_fidelity_improvement}\\
$N^{pur}_{i,j}(u,v)=N^{pur}_{i,j}(u,v)+1$;\\\label{fidelity_check2}
}
}
$Q\leftarrow$ $P_{i,j}(s_i,d_i)$, $cost(P_{i,j}(s_i,d_i))$, $D^{pur}_{i,j}$;\\
\textit{{\textbf{Step 4 Throughput Update:}} \\}
\While{$cost(Q.pop)\leq min\_cost+1$\label{mincost_satisfied}}{
Find $W^{min}_{i,j}$ along the path $P_{i,j}(s_i,d_i)$ in $G^a$;\\
%Calculate available width $W_{i,j}= \frac{C^{min}_{i,j}}{N_{i,j}^{pur}(u,v)+1}$;\\
\If{$W^{min}_{i,j} > 1$}{
Subtract $\min\{W_{i,j},R_i\}\times ({N_{i,j}^{pur}(u,v)}+1)$ on each $(u,v)\in P_{i,j}(s_i,d_i)$ from $C^a$ in $G^a$;\label{throughput_update_1}\\
}
%and delete $(u,v)\in E_r$ if $c(u,v)\leq 0 $;\\
$T^{EXT}_{i,j}(s_i,d_i)\leftarrow$ Calculate expected throughput of each edge $(u,v)\in P_{i,j}(s_i,d_i)$;\\
Output $P_{i,j}(s_i,d_i)$, $D^{pur}_{i,j}$, $\min\{W_{i,j},R_i\}$ $F_{i,j}(s_i,d_i)$ and delete this solution from $Q$;\\
\If{$\sum_j T^{EXT}_{i,j}(s_i,d_i)\geq R_i$}{
\textbf{terminate};\\
}\label{throughput_update_2}
}
}
\end{algorithm}

\subsection{Problem Definition and Design Overview}
\textbf{\textit{Problem Definition:}} Given a routing request from single S-D pair and a quantum network with topology and edge capacity as $G=(V,E,C)$, finding routing solutions, including purification decision $D^{pur}_{i,j}$ and path selection $P_{i,j}(s_i,d_i)$, to enable the entanglement establishment between S-D pairs to satisfy fidelity constraint $F_{i,j}(s_i,d_i)>F^{th}_i, \forall i$.

\textbf{\textit{Design Overview:}} In order to solve the entanglement routing problem in quantum networks, we start with the investigation on single S-D pair scenario, and first design a routing algorithm that can provide the optimal routing solution with both path selection and purification decisions. By utilizing such algorithm, we can obtain the upper bound of the routing performance and provide guidance for the optimal purification decisions. After that, due to the relatively high computational complexity of the optimal routing algorithm, we further design a heuristic routing algorithm which can efficiently find near-optimal routing solution with ``best quality'' path and fidelity guarantee.

\subsection{Iterative Routing Design for Single S-D Pair}
For a single S-D pair, the goal of routing design is to find the routing path satisfying fidelity constraint with minimum entangled pair cost.
To reach such goal, the proposed Q-PATH algorithm is described in Algorithm \ref{Iteration_routing}. The basic idea of Q-PATH can be described as follows. It searches all possible solutions (including routing path and purification decision) from the lowest routing cost (i.e., the number of consumed entangled pairs) and update purification decision in an iterative manner. In each iteration, the algorithm sets a certain ``expected cost value'' (i.e., $min\_cost$), and checks all possible routing solutions if they equals to ``expected cost value''. Multiple routing solutions may be found in each iteration by using Step 2 and Step 3, but all routing solutions output in the same iteration must have the same entangled pair cost, i.e., $min\_cost$. Once the actual entangled cost of the routing solution meets the ``expected cost value'', the routing solution can be output by the algorithm as the one with minimum cost.

It should be clarified that Q-PATH considers minimum entangled pair cost (i.e., minimum cost) as the metric to find the optimal routing solution rather than minimum number of hops in classic routing algorithm such as Dijkstra. For a given routing request $R_i$, the entangled pair cost means the total cumulation of resource consumption to establish one entanglement connection that satisfies corresponding fidelity threshold $F_i^{th}$ for $i$-th S-D pair, it includes the consumption for basic path establishment (consumes one entangled pair on each edge along the routing path $P_{i,j}(s,d)$) and extra purification operation (consumes $N_{i,j}^{pur}(u,v)$ entangled pairs on edge $(u,v)$).

In specific, Q-PATH contains 4 steps as follows:

1) \textit{\textbf{Initialization}}: Q-PATH first calculates a ``\textit{Purification Cost Table}'' for each edge $(u,v)\in E$. Since the fidelity constraint $F_{i,j}(s_i,d_i)\geq F^{th}$ must be satisfied for the routing path, if an edge $(u,v)$ cannot provide entangled pairs satisfying $F^{pur}_{max}(u,v)\geq F^{th}$ even after purifications, it should be deleted on graph $G$ for complexity reduction.
After that, Q-PATH constructs an update graph $G^a$ to record purification decision. Finally, Q-PATH finds the shortest path on graph $G$ by using Breadth-First-Search (BFS) to ensure the possible minimum cost $H^{min}$ used in step 2, and also constructs a priority queue to save potential routing paths during iterations.

2) \textit{\textbf{Path Selection Procedure}}:
%(Discussion for searching process in line 9)
To obtain the optimal routing path with minimum cost, we design an iterative method that utilizes the cost as the metric to continue each loop, which provides an indicator during the path searching. Here, $k$-shortest path algorithm \cite{yen1971finding} is used to obtain multiple shortest paths with the same cost $min\_cost$. %Here, BFS and $k$-shortest path algorithm are two options to obtain multiple shortest paths with the same cost $min\_cost$. The BFS algorithm can breadthwise search the path based on hop count, and the original $k$-shortest path algorithm \cite{yen1971finding}, which is based on Dijsktra algorithm, can output the shortest path one-by-one and also obtain multiple shortest paths with the same cost $min\_cost$.

%\begin{table}[t]
%  \centering
%  \caption{Purification Cost Table}
%  \label{purification_cost_table}
%  \begin{tabular}{|c|c|c|}%p{4.8cm}
%    \hline
%     \textbf{Fidelity} & \textbf{Residual Capacity} & \textbf{Cost}\\\hline
%     %Number of Tx antennas & 4 \\\hline
%     $F^{pur}_{min}(u,v)$ & $C(u,v)$ & $1$\\\hline
%     $F^{pur}_{1}(u,v)$ & $\lfloor C(u,v)/2\rfloor$ & $2^1$\\\hline
%     $F^{pur}_{2}(u,v)$ & $\lfloor C(u,v)/2^2 \rfloor$ & $2^2$\\\hline
%     ... & ... & ...\\\hline
%     $F^{pur}_{max}(u,v)$ & $1$ & $2^n$\\\hline
%  \end{tabular}
%\end{table}

To establish the end-to-end entanglement connection, entanglement swappings are required after the generation of entangled pairs. Considering the imperfect measurement on quantum repeaters, each swapping operation brings fidelity degradation. By adopting extra purification operations, the expected fidelity of the entanglement connection following routing path $P_{i,j}(s_i,d_i)$ can be calculated as:
\begin{align}
\label{fidelity_calculation}
F_{i,j}(s_i,d_i)=&\prod_{(u,v)\in P_{i,j}(s_i,d_i)} F^{pur}_{i,j}(u,v,N_{i,j}^{pur}),
\end{align}
where $F^{pur}_{i,j}(u,v,N_{i,j}^{pur})$ denotes the fidelity of quantum channel on edge $(u,v)$ after $N_{i,j}^{pur}$ round purification operations allocated for $j$-th routing path of $i$-th S-D pair.
Given that the purification protocol as shown in Fig. \ref{purification_performance-0}, the total number of entangled pairs consumed after $N^{pur}(u,v)$ round purification should be $N^{pur}(u,v)$. Thus, fidelity $F^{pur}_{i,j}(u,v,N_{i,j}^{pur})$ can be calculated by:
%\begin{align}
%\label{purification_calculation}
%F^{pur}_{i,j}(u,v,N_{i,j}^{pur})=
%\left\{
%\begin{aligned}
%   &F^{0}(u,v),N_{i,j}^{pur}=0,\\
%   &f\left(F^{0}(u,v)\right),N_{i,j}^{pur}=1,\\
%   &f\left(f\left(F^{0}\left(u,v\right))\right)\right),N_{i,j}^{pur}=2,\\
%   &...
% \end{aligned}
%\right.
%\end{align}
\begin{align}
\label{purification_calculation}
&F^{pur}_{i,j}(u,v,N_{i,j}^{pur})\notag\\&=
\left\{
\begin{aligned}
   &F^{0}(u,v),N_{i,j}^{pur}=0,\\
   &f\left(F^{0}(u,v),F^{pur}_{i,j}\left(u,v,N_{i,j}^{pur}-1\right)\right),N_{i,j}^{pur}\geq1,\\
 \end{aligned}
\right.
\end{align}
where $F^{0}(u,v)$ denotes the original fidelity of the generated entangled pair on $(u,v)$, and $f(\cdot)$ represents the resulting fidelity of quantum channel after purification in Eq. (\ref{fidelity_after_purification}).
%By only considering bit flip errors, the resulting fidelity after purification operation can be calculated by \cite{sheng2013hybrid}:
%\begin{equation}\label{fidelity_after_purification}
%f(x)=\frac{x^2}{x^2+(1-x)^2}.
%\end{equation}

%\begin{algorithm}[t]
%\caption{Purification Decision Process}
%\label{Purification_decision}
%%\KwIn{$G=(V,E,C)$, $F^{th}_i$, request $R_i$ and $\langle s_i,d_i\rangle$;}
%%\KwOut{$P_{i,j}(s_i,d_i)$, $D^{pur}_{i,j}$, $F_j(s_i,d_i)$, $T^{EXT}_{i,j}$;}
%\For{$\{(u,v),c^{pur}\}=q.pop()$}{
%\If{$c^{pur}>1$}{
%Find $c^{pur}$ edges with lowest fidelity that $N_{i,j}^{pur}(u',v')<N_{i,j}^{pur}(u,v)$;\\
%}
%\If{$\sum\limits_{1}^{c^{pur}}Improvement((u',v'),N_{i,j}^{pur}(u',v')+1)>Improvement((u,v),N_{i,j}^{pur}(u,v)+1)$\label{fidelity_improvement_comparison}}{
%$N_{i,j}^{pur}(u',v')=N_{i,j}^{pur}(u',v')+1$;\\
%\textbf{continue};
%}
%\Else{$N_{i,j}^{pur}(u,v)=N_{i,j}^{pur}(u,v)+1$;}
%\If{$F_{i,j}(s_i,d_i)\geq F^{th}_i$}{
%\textbf{break}\label{fidelity_check2};
%}
%}
%\end{algorithm}

3) \textit{\textbf{Edge Cost Update}}: %To ensure the fidelity of entanglement connection can satisfy the requirement of fidelity threshold $F^{th}_i$,
%Q-PATH must check if the fidelity of entanglement connection is satisfied for each path.
%To ensure the fidelity of entanglement connection can satisfy the requirement of fidelity threshold $F^{th}_i$, Algorithm \ref{Purification_decision} is designed to make the optimal purification decision and check if fidelity threshold is satisfied.
%line \ref{fidelity_check1}-\ref{fidelity_check2} check if fidelity threshold is satisfied, otherwise $N^{pur}_{i,j}(u,v)=N^{pur}_{i,j}(u,v)+1$ for the edge $(u,v)$ with minimum fidelity along the path.
For a given path $P_{i,j}(s_i,d_i)$, edge cost (i.e., number of consumed entangled pairs) equals to the throughput without purification. However, due to the fidelity constraint $F_i^{th}$, purification might be required on some edges along the path. To guarantee the end-to-end fidelity with minimum entangled pair cost, %Algorithm \ref{Purification_decision}
the purification decision process in line \ref{fidelity_check1}-\ref{fidelity_check2} is designed to check $F_{i,j}(s_i,d_i)$ and add the number of purification round $N^{pur}_i$ once a time. Note that $N^{pur}_{i,j}(u,v)=0$ for all $(u,v)\in E$ at the start of the algorithm.

\begin{table}[t]
  \centering
  \caption{Purification Cost Table}
  \label{purification_cost_table}
  \begin{tabular}{ccc}%p{4.8cm}
    \hline\hline
     \textbf{Round} & \textbf{Fidelity} & \textbf{Fidelity Improvement}\\\hline
     %Number of Tx antennas & 4 \\\hline
     0 & $F^{pur}_{min}(u,v)$ & $0$ \\
     1 & $F^{pur}_{1}(u,v)$ & $F^{pur}_{1}(u,v)-F^{pur}_{min}(u,v)$ \\
     2 & $F^{pur}_{2}(u,v)$ & $F^{pur}_{2}(u,v)-F^{pur}_{1}(u,v)$ \\
     ... & ... & ...\\
     $n$ & $F^{pur}_{max}(u,v)$ & $F^{pur}_{max}(u,v)-F^{pur}_{n-1}(u,v)$ \\
     \hline\hline
  \end{tabular}
\end{table}

To provide all possible purification options and corresponding cost, we design a ``\textit{Purification Cost Table}'' for each edge $(u,v)\in E$ as described in Table \ref{purification_cost_table}. ``\textit{Purification Cost Table}'' gives the guidance for the resource consumption of purification operation, and also gives the maximum and the minimum fidelity that each quantum channel can provide. For example, for an edge $(u,v)$ with capacity $c(u,v)=5$, the original fidelity $F^{pur}_{min}(u,v)=0.75$, then the maximum purification round is 4. According to Eq. (\ref{fidelity_after_purification}), the resulting fidelity after 1st round can be easily calculated as $F^{pur}_{1}(u,v)=0.9$, and the fidelity improvement is $F^{pur}_{1}(u,v)-F^{pur}_{min}(u,v)=0.15$. Similarly, the resulting fidelity after 2nd round can be easily calculated as $F^{pur}_{2}(u,v)=0.9642$, and the fidelity improvement is $F^{pur}_{2}(u,v)-F^{pur}_{1}(u,v)=0.0642$ and the following table entries are so on in the same manner. Note that the maximum fidelity and improvement are $F^{pur}_{max}(u,v)=0.9959$ and $F^{pur}_{max}(u,v)-F^{pur}_{3}(u,v)=0.0081$ after 4-round purification, thus we can tell that the fidelity improvement is decreasing along with the increase of purification round on the same edge.
%in line \ref{max_fidelity_improvement} of Q-PATH, for an edge

%For an edge $(u,v)$ with capacity $c(u,v)$, to generate one higher-fidelity entangled pair after $N^{pur}(u,v)$ round purification, the total number of entangled pairs consumed should be $N^{pur}(u,v)$.
%$2^{N^{pur}(u,v)}$.%

%$2^{N^{pur}(u,v)}$.%
%Thus, the residual capacity after  becomes $\frac{C(u,v)}{2^n}$.

In the following, we prove that the greedy approach used in \textit{\textbf{Step 3}} can find the optimal purification decision. The basic idea behind such greedy approach is that, on the same edge $(u,v)$, when the original fidelity of the entangled pair is low, e.g., $F^0(u,v)=0.75$, the fidelity improvement is high after one purification operation, i.e., 0.15. However, when the original fidelity is high, e.g., $F^0(u,v)=0.95$, the fidelity improvement obtained after one purification operation significantly decreases, i.e., 0.0472. In other words, the fidelity improvement brought by purification operation has monotonicity. Hence, the greedy approach\footnote{Note that the greedy approach used in Q-PATH requires sufficient entangled pairs on edges to execute the optimal purification operations, otherwise it cannot find any effective solutions.} becomes useful by leveraging the monotonicity property of purification operation. The specific conclusion and proof procedure are given in Theorem 1.
%\textcolor{red}{Since the purification brings more improvement when the original fidelity is lower, at each time, we apply $N^{pur}_{i}(u,v)=N^{pur}_{i,j}(u,v)+1$ for the edge currently has lowest fidelity.}

%\newtheorem*{Theorem 1}{Theorem 1}
%\begin{Theorem 1}
%\label{Theorem 1}
%The greedy approach used in \textit{\textbf{Step 3 }} of Q-PATH can find the optimal purification decision with minimum entangled cost when the original fidelity is above $x^*$.
%\end{Theorem 1}
%\begin{proof}
%At first, we prove that the purification operation to the entangled pair close to $x^*$ can bring the highest improvement.
%The resulting fidelity after purification operation can be calculated by eq. (\ref{fidelity_after_purification}).
%The second derivative of eq. (\ref{fidelity_after_purification}) can be obtained as
%\begin{equation}
%\frac{df(x)}{dx}=\frac{(-24x^2+84x-6)}{(8x^2-4x+5)^2}.\notag
%\end{equation}
%Let $\frac{df(x)}{dx}=1$, then we can obtain $x^*\approx0.77081$ when $x\in[0.5,1]$, and we can also obtain that $\frac{df(x)}{dx}>0$ and $\frac{d^2f(x)}{dx^2}<0$ when $x\in[0.54,1]$.
%The specific performance of purification operation is shown in Fig. \ref{fig:purification_performance}.
%In this case, if the original fidelity equals to $x^*$, then the highest fidelity improvement can be obtained. In other words, among multiple entangled pairs on different edges, the purification operation to the entangled pair with the lowest fidelity can bring the highest improvement when the original fidelity is above $x^*$.
%\end{proof}

\newtheorem*{Theorem 1}{Theorem 1}
\begin{Theorem 1}
\label{Theorem 1}
The greedy approach used in \textit{\textbf{Step 3 }} of Q-PATH can find the optimal purification decision with minimum entangled pair cost when the original fidelity is above $x^*$.
\end{Theorem 1}
\begin{proof}
Please see Appendix A for details.
\end{proof}

%Since the purification \textcolor{red}{requires/consumes} extra entangled pairs to increase the fidelity of target entangled pair, we attempt to find the optimal purification decision with minimum resource consumption. A typical purification process is shown as Fig. \ref{fig:purification}.
%At each purification step, either both pairs are discarded (if the purification was not successful) or one pair is discarded (if the purification was successful) . The leftover pairs are again used for purification at the next step \cite{dur1999quantum}.

%Thus, a routing path with minimum memory cost should be given. Before we propose our routing algorithm.
%For a single S-D pair in the quantum network, the optimal routing path can be obtained through

\begin{figure*}[!th]
  \centering
  \subfigure[$min\_cost$=2, actual cost=2+1 (purification on edge $(s_1,r_3)$).]{
    \label{iterative_routing_example2}
    \includegraphics[width=1.62in]{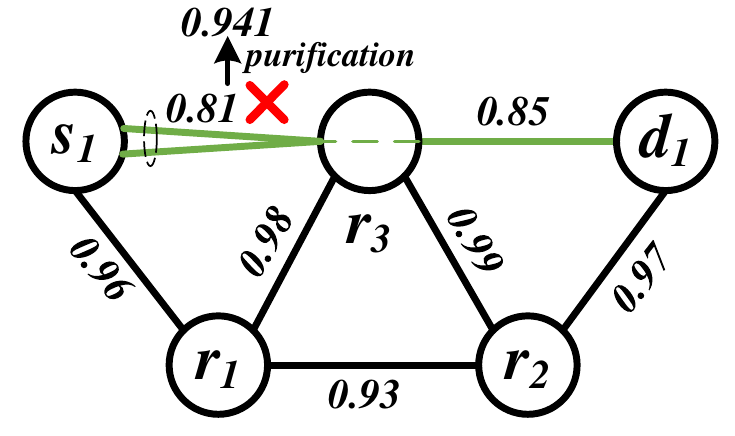}
 }
  \subfigure[$min\_cost$=3, actual cost=3$~~~$ (optimal path).]{
    \label{iterative_routing_example3}
    \includegraphics[width=1.62in]{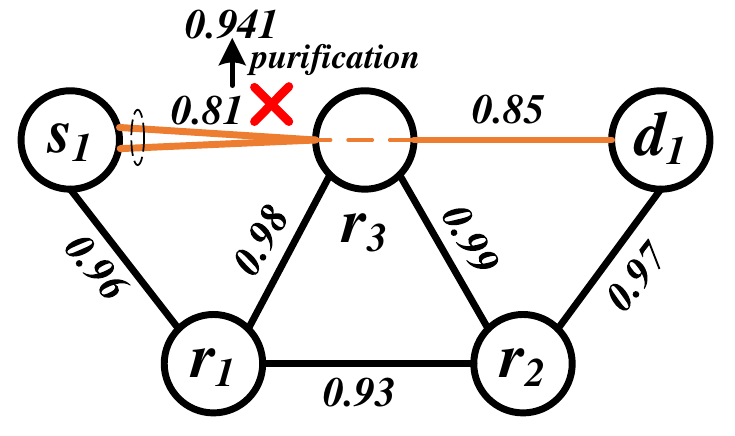}
 }
  \subfigure[$min\_cost$=3, actual cost=3$~~~$ (optimal path).]{
    \label{iterative_routing_example4}
    \includegraphics[width=1.62in]{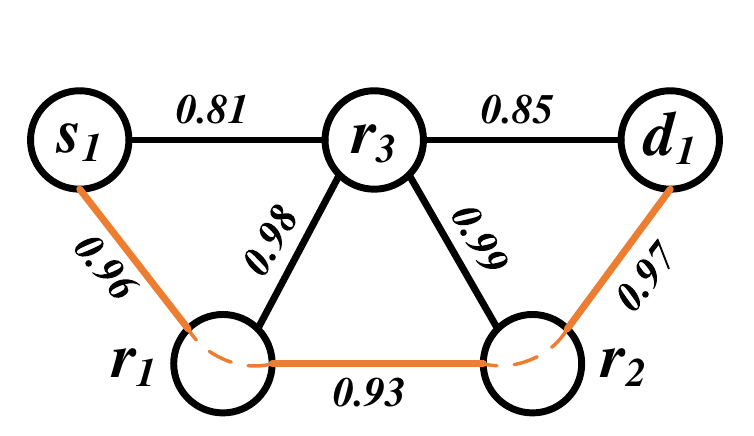}
 }
  \subfigure[$min\_cost$=3, actual cost=3+1 (purification on edge $(r_1,d_1)$).]{
    \label{iterative_routing_example5}
    \includegraphics[width=1.62in]{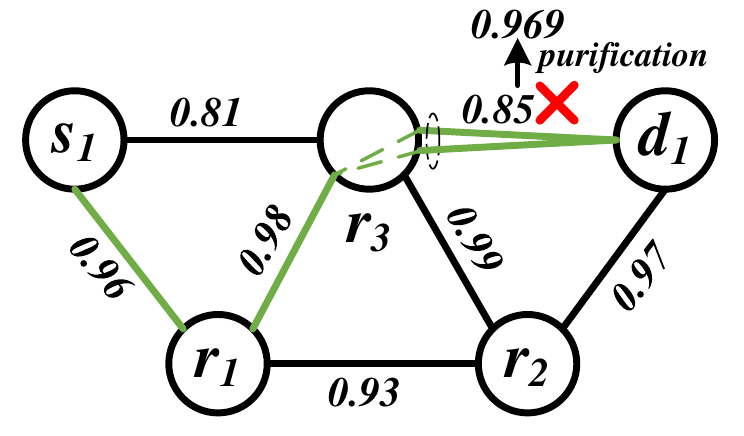}
 }
 \setlength{\abovecaptionskip}{0.1cm}
  \caption{A routing example for single S-D pair, and the fidelity threshold is $F^{th}_1=0.8$. (a)-(d) show the iterative searching procedure of Q-PATH.}
  \label{iterative_routing_example}
\end{figure*}

Due to the imperfect measurement on quantum repeaters, the purification is inherently probabilistic in nature \cite{van2014quantum,munro2015inside,bennett1996purification,dur2007entanglement}.
The probability of the successful one-round purification can be calculated by
%\begin{equation}\label{purification_equation1}
%P^{pur}(x, 1)=x^2+\frac{2}{3}x(1-x)+\frac{5}{9}(1-x)^2.
%\end{equation}
%\begin{align}\label{purification_equation1}
%P^{pur}(x_1, x_2, 1)=&x_1x_2+\frac{1}{3}x_1(1-x_2)+\notag\\
%&\frac{1}{3}x_2(1-x_1)+\frac{5}{9}(1-x_1)(1-x_2).
%\end{align}
$P^{pur}(x_1, x_2, 1)=x_1x_2+(1-x_1)(1-x_2)$.
For multi-round purification operation, the cumulative successful purification operation can be further calculated by:
%\begin{equation}\label{purification_equation2}
%P^{pur}\left(F^0,N^{Pur}_{i,j}\left(u,v\right)\right)=\prod^{N^{Pur}_{i,j}(u,v)}_{n=1}2^{n-1}\times P^{pur}(F^0,n).
%\end{equation}
{%\footnotesize
\begin{align}\label{purification_equation2}
P^{pur}&\left(F^0, N^{Pur}_{i,j}\left(u,v\right)\right)=\notag\\
&\prod^{N^{Pur}_{i,j}(u,v)}_{n=1}P^{pur}\left(F^0,F^{pur}_{i,j}(u,v,n),n\right).\notag
\end{align}
}

To verify the optimal routing solution including path selection and purification decision obtained by Q-PATH, a performance evaluation would be conducted in Section \ref{sectionVI-1} compared with brute-force method, and the results show that the routing solution between the one proposed in Q-PATH and the optimal one are the same.
%\textcolor{red}{Although the purification decision process in line \ref{fidelity_check1}-\ref{fidelity_check2} can provide the optimal solution for minimum purification rounds rather than the minimum entangled cost, considering the correlation between purification rounds and probability of successful purification operation, it is beneficial for maximizing throughput by minimizing purification rounds. On the other hand, the purification decision process for a given path can also be easily replaced by a brute-force method which can always find the purification decisions with minimum entangled cost, even though it takes non-polynomial time. In Section \ref{sectionVI-1}, a performance comparison would be conducted. Most of the time, the purification decisions between the one proposed in Q-PATH and the optimal one are the same, and the results also show that only negligible gap exists between these two methods.}

4) \textit{\textbf{Throughput Update}}: The algorithm calculates the maximum achievable throughput $W_{i,j}$ based on purification decision $D_{i,j}=[N^{pur}_{i,j}(s,u_1),...,N^{pur}_{i,j}(u_n,d)]$ for given path $P_{i,j}(s_i,d_i)$, and judges if the request of $i$-th S-D pair is satisfied. Note that available width $W_{i,j}$ means available number of end-to-end entanglement establishment for $j$-th routing path of $i$-th S-D pair based on given purification decision. If the request is satisfied, the algorithm terminates and outputs the routing path $P_{i,j}(s_i,d_i)$, purification decision $D_{i,j}$, and expected fidelity $F_{i,j}(s_i,d_i)$.
Due to the probability of failure situation of purification, the maximum throughput on a given path $P_{i,j}(s,d)$ would be lower than the one without purification operation. Similar to \cite{shi2020concurrent}, we define a metric called expected throughput (i.e., expected number of \textit{qubits}) to quantity an arbitrary end-to-end path $P_{i,j}(s,d)$:
%\begin{equation}
%t_{i,j}^{EXT}(u,v)=P_i^{pur}\left(F^0\left(u,v\right),N^{pur}\right)\times \lfloor \frac{c(u,v)}{2^{N_{i,j}^{pur}}} \rfloor.
%\end{equation}
\begin{equation}
t_{i,j}^{EXT}(u,v)=P^{pur}\left(F^0\left(u,v\right),N^{pur}_{i,j}(u,v)\right)\times W_{i,j}^{min}.\notag
\end{equation}
where $W^{min}_{i,j}=\min\limits_{(u,v)\in P_{i,j}(s_i,d_i)}\lfloor \frac{c(u,v)}{N_{i,j}^{pur}(u,v)+1}\rfloor$ represents the minimum width, i.e., the number of  entanglement connections, along the path $P_{i,j}(s_i,d_i)$.
Furthermore, the expected throughput on path $P_{i,j}(s_i,d_i)$ can be further calculated by:
\begin{equation}
T_{i,j}^{EXT}(s_i,d_i)=\min(t_{i,j}^{EXT}(u,v)|(u,v)\in E).\notag
\end{equation}

\subsection{Discussion and Complexity Analysis of Q-PATH}

To clearly explain the searching process of Q-PATH, a routing example is given in Fig. \ref{iterative_routing_example}. The fidelity value of entangled pairs is shown on each edge. Since $H^{min}=2$ (at least 2 hops are required to reach the destination) in this example, in the first iteration, the algorithm attempts to find the routing solution with $min\_cost = 2$. Only one potential path with length $l=2$ can be found as shown in Fig. \ref{iterative_routing_example2}. However, according to line \ref{fidelity_check1}, purification is required on edge $(s_1,r_1)$ to satisfy fidelity constraint. Thus, this routing solution with $cost = 3$ is enqueued into $Q$, and the cost condition in line \ref{mincost_satisfied} cannot be satisfied. Similarly, in the second iteration, the algorithm attempts to find the routing solution with $min\_cost = 3$. Two potential paths with length $l=3$ can be found as shown in Figs. \ref{iterative_routing_example4}-\ref{iterative_routing_example5}. After checking the fidelity constraint in line \ref{fidelity_check1}, these two routing solutions with $cost = 3$ and $cost = 4$ is enqueued into $Q$. Then, two routing solutions (illustrated in Figs. \ref{iterative_routing_example3}-\ref{iterative_routing_example4}) that satisfy the cost condition in line \ref{mincost_satisfied} can be found from $Q$ and they will be outputted. If request $R_i$ is satisfied with the expected throughput $t^{EXT}$ provided by the optimal paths, then the algorithm ends. Otherwise, the process repeats as above.

Based on Theorem 1, the following theorem theoretically prove that the proposed algorithm can find the routing path with minimum cost. The basic idea can be explained as follows. For a given routing path, Theorem 1 can guarantee that the optimal purification decision with minimum entangled pair cost. Meanwhile, the minimum entangled pair cost for a given routing path is decided by its path length. Thus, Q-PATH searches routing path according to its path length (i.e., minimum entangled pair cost), and then compare potential routing paths with their actually entangled pair cost after purification operations. If there exists an optimal routing path, it must close to its minimum entangled pair cost, and it can be found before other routing paths with larger minimum entangled pair cost. Then specific conclusion and proof procedure are given in Theorem 2.

\newtheorem*{Theorem 2}{Theorem 2}
\begin{Theorem 2}
\label{Theorem 2}
The Q-PATH can find the fidelity-guaranteed routing path with minimum entangled pair cost for arbitrary S-D pair in quantum networks.
%For a given path $P(s,d)$, the purification decision with minimum cost on $G^r$ can obtain the maximum throughput on path $P(s,d)$.
\end{Theorem 2}
\begin{proof}
Please see Appendix B for details.
\end{proof}

\begin{algorithm}[!t]
\caption{Q-LEAP: Low-complexity Routing}
\label{SPF_routing}
\KwIn{$G=(V,E,C)$, $F^{th}_i$, request $R_i$ and S-D pair;}
\KwOut{$P_{i,j}(s_i,d_i)$, $D^{pur}_{i,j}$, $F_j(s_i,d_i)$, and $T^{EXT}_{i,j}$;}

\textit{\textbf{Step 1 Initialization:} \\}
Same as line \ref{Initialization1}-\ref{Initialization4} in Q-PATH;\\
\For{$j=1: R_i$\label{loop1}}{
\textit{\textbf{Step 2 Best Quality Path Searching:} \\}
%Multiple path may be found if one flow cannot satisfy request $R_i$
$P_{i,j}(s_i,d_i),U(P_{i,j}(s_i,d_i))\leftarrow$ Using extended Dijkstra algorithm to search the path with minimum fidelity degradation according to fidelity multiplicative Eq. (\ref{fidelity_calculation});\\
\If{no available path for $\langle s_i,d_i\rangle$}{
\textbf{break};\\
}
\textit{\textbf{Step 3 Purification Decision:} \\}
Calculate path length $l$ and $F^{avg}_{i,j}=(F_i^{th})^{1/l}$;\\
%Let $N^{pur}_{i,j}(u,v)=1,\forall (u,v)\in P_{i,j}(s_i,d_i)$;\\
\For{$(u,v)\in P_{i,j}(s_i,d_i)$}{
\If{$F(u,v)<F^{avg}_{i,j}$}{
%\If{$F^{pur}_{max}(u,v)<F^{avg}_{i,j}$}{
%\textbf{continue};\\
%}
$N^{pur}_{i,j}(u,v)=\arg\min\limits_{N^{pur}_{i,j}(u,v)} F^{pur}_{i,j}(u,v)\geq F^{avg}_{i,j}$ according to \textit{Purification Cost Table};\\
%According to \textit{Purification Cost Table}, update purification operation with minimum cost $N^{pur}_{i,j}(u,v)$ to ensure $F^{pur}_{i,j}(u,v)\geq F^{avg}_{i,j}$;\\
}
}
$Q\leftarrow$ $P_{i,j}(s_i,d_i)$, $cost(P_{i,j}(s_i,d_i))$, $D^{pur}_{i,j}$;\\
\textit{{\textbf{Step 4 Throughput Update:}} \\}
%Same as line \ref{throughput_update_1}-\ref{throughput_update_2} in Q-PATH;\\
\While{$Q.pop!= null$}{
Find $W^{min}_{i,j}$ along the path $P_{i,j}(s_i,d_i)$ in $G^a$;\\
%Calculate available width $W_{i,j}= \frac{C^{min}_{i,j}}{N_{i,j}^{pur}(u,v)+1}$;\\
\If{$W_{i,j} > 1$}{
Subtract $\min\{W^{min}_{i,j},R_i\}\times ({N_{i,j}^{pur}(u,v)}+1)$ on each $(u,v)\in P_{i,j}(s_i,d_i)$ from $C^a$ in $G^a$;\\
}
%and delete $(u,v)\in E_r$ if $c(u,v)\leq 0 $;\\
$T^{EXT}_{i,j}(s_i,d_i)\leftarrow$ Calculate expected throughput of each edge $(u,v)\in P_{i,j}(s_i,d_i)$;\\
Output $P_{i,j}(s_i,d_i)$, $D^{pur}_{i,j}$, $\min\{W_{i,j},R_i\}$ $F_{i,j}(s_i,d_i)$ and delete this solution from $Q$;\\
\If{$\sum_j T^{EXT}_{i,j}(s_i,d_i)\geq R_i$}{
\textbf{terminate};\\
}
}

}
\end{algorithm}

%\textcolor{red}{
%Algorithm \ref{Iteration_routing} contains four main steps.
The computational complexity of Q-PATH can be analyzed as follows.
Let $|E|$ denote the number of edges in set $E$, $|V|$ denote the number of nodes in set $V$. $C_{max}$ denotes the maximum number of entangled pairs on an edge $(u,v)\in E$, %\textcolor{red}{and $\log_2C_{max}$ is the maximum purification rounds on a quantum channel.}
At \textit{\textbf{Step 1}}, the worst complexity of \textit{purification cost table} calculation and BFS can be calculated by $\mathcal{O}\left(|E|C_{max}\right)$ and $\mathcal{O}\left(|V|+|E|\right)$, respectively.
At \textit{\textbf{Step 2}}, $K$ paths would be obtained at most from $K$-shortest path algorithm, thus the worst complexity can be calculated by $\mathcal{O}\left(K|V|\left(|E|+|V|\log_2|V|\right)\right)$.
At \textit{\textbf{Step 3}}, to satisfy the fidelity threshold and make purification decisions, the worst complexity can be calculated by $\mathcal{O}\left(K|E|C_{max}\right)$, where $|E|C_{max}$ is the judgement times of line \ref{fidelity_check1} in the worst case.
At \textit{\textbf{Step 4}}, in the worst case, the throughput update procedure for each path obtained from \textit{\textbf{Step 2}} would be executed $R_i$ times in total, and the complexity of each throughput update procedure from line \ref{mincost_satisfied} to line \ref{throughput_update_2} is $\mathcal{O}\left(|E|\right)$. In worst case, the number of iterations in line \ref{mincost_loop} is $|E|C_{max}$. Thus, the worst computational complexity of Q-PATH is $\mathcal{O}\bigl(|V|+|E|+$ $|E|C_{max}\bigl(K|V||E|+|V|^2\log_2|V|+K|E|C_{max}+R_i|E|\bigr)\bigr)$.
%At \textit{\textbf{Step 4}}, in the worst case, the throughput update procedure for each path obtained from \textit{\textbf{Step 2}} would be executed $K|E|C_{max}$ times in total, and the complexity of each throughput update procedure from line \ref{throughput_update_1} to line \ref{throughput_update_2} is $\mathcal{O}\left(K|E|C_{max}\right)$.
%In worst case, the number of iterations in line \ref{mincost_loop} is $|E|C_{max}$.
%Thus, the worst computational complexity of Q-PATH is $\mathcal{O}\bigl(|V|+K|E|C_{max}\bigl(|V||E|+|V|^2\log_2|V|$
%$+|E|C_{max}\bigr)\bigr)$.

%At the first step, the worst complexity of BFS is $\mathcal{O}(|V|+|E|)$, thus the complexity can be calculated by $\mathcal{O}\left(|E|\left(\log_2 |C_{max}|+1\right)+|V|\right)$, where $|E|$ denotes the number of edges in set $E$, and $|C_{max}|$ denotes the maximum number of entangled pairs on an edge $(u,v)\in E$. At the second step, thus the worst complexity of BFS can be calculated by $\mathcal{O}\left(\left(|V|+|E|\right)\right)$. At the third step, to guarantee the fidelity of the path, the complexity is $\mathcal{O}\left(+(|E|^2\log_2|E|)\right)$. At the forth step, the worst complexity is $\mathcal{O}\left(+|E|\right)$. At last, due to the loop in line \ref{mincost_loop}, step 2-4 would be executed $|E|\log_2|C_{max}|$ times in the worst case.}
As a comparison, the complexity of brute-force approach is $\mathcal{O}\left(\left(C_{max})^{|E|}\right)\right)$.
Due to the existence of $|E|C_{max}|V|^2\log_2 V$ and $K\left(|E| C_{max}\right)^2$, the complexity of Q-PATH can rise quickly with the increase of edges and edge capacity. Hence, in the next, we further propose a low-complexity routing algorithm to overcome such problem.

%the worst complexity can be calculated by $\mathcal{O}(\frac{|R_i|}{|C_{min}|}\left(|V|\log_2 |V| + |E|\right))$, where $|R_i|$ denotes the number of requests of $i$-th S-D pair, and $|C_{max}|\geq 1$ denotes the minimum number of entangled pairs on an edge $(u,v)\in E$. At the third step, i.e., \textbf{\textit{Purification Decision}}, the worst complexity can be calculated by $\mathcal{O}(\frac{|R_i|}{|C_{min}|} \left(|E|\log_2 |C_{max}|\right))$. At the last step, i.e., \textbf{\textit{Throughput Update:}}, for each loop in line \ref{loop1}, it should be executed $\frac{|R_i|}{|C_{min}|}$ times to find minimum expected throughput, delete $t^{throughput}_{i,j}$ from $C$ in graph $G=(V,E,C)$ should be executed $|E|$ times in the worst case, the worst complexity can be calculated by $\mathcal{O}(\frac{|R_i|}{|C_{min}|}\left(\frac{|R_i|}{|C_{min}|}+|E|\right))$.

\begin{figure}[!t]
  \centering
  \subfigure[Network topology.]{
    \label{iterative_routing_example1}
    \includegraphics[width=1.62in]{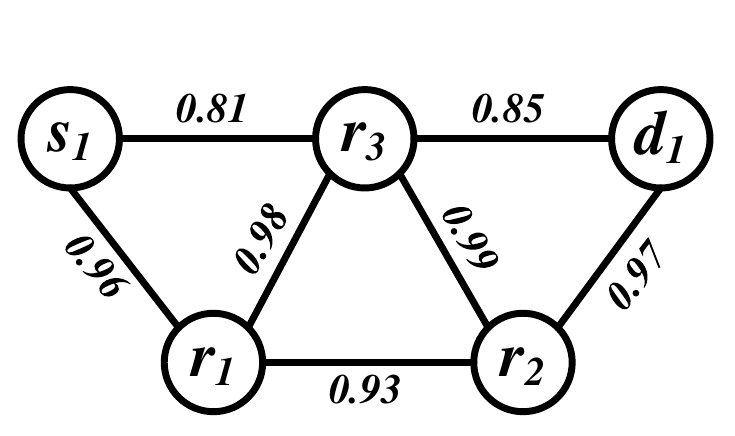}
 }
   \subfigure[$min\_cost$=4.]{
    \label{iterative_routing_example6}
    \includegraphics[width=1.62in]{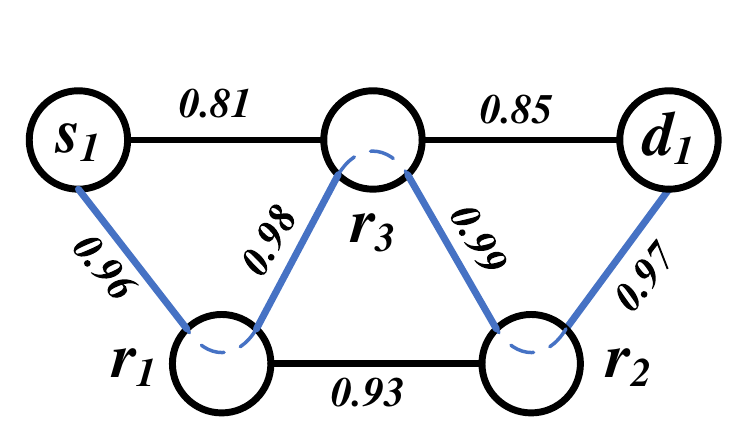}
 }
 \setlength{\abovecaptionskip}{0.1cm}
  \caption{A routing example for single S-D pair, and the fidelity threshold is $F^{th}_1=0.8$. (a) illustrates the network topology, (b) shows the path with the highest end-to-end fidelity found by extended Dijkstra in Q-LEAP.}
  \label{iterative_routing_example_extended}
\end{figure}

\subsection{Low-complexity routing design for single S-D pair}
Although Q-PATH provides the upper bound of the routing problem for single S-D pair, considering the high computational complexity, we further design a low-complexity routing algorithm, i.e., Q-LEAP, as described in Algorithm \ref{SPF_routing}. Similarly, Q-LEAP contains four steps, i.e., initialization, the path selection procedure, purification decision, and throughput update. The basic idea of Q-LEAP is that, for each iteration in Q-LEAP (line 3 in Algorithm 2), it searches one routing path with ``best quality'' (Step 2), and make purification decision to let it satisfy fidelity threshold (Step 3). The process repeats until the request $R_i$ is satisfied (see terminal condition in line 26). In the following, we discuss two main different parts between the low-complexity routing design in Q-LEAP and the iterative design in Q-PATH.

\textit{\textbf{Best Quality Path Searching}}:
Unlike the path searching process in Q-PATH, which searches all potential paths with the same cost, Q-LEAP attempts to find ``the optimal one'' with minimum fidelity degradation via an extended Dijkstra algorithm. As shown in Fig. \ref{iterative_routing_example_extended}, Q-LEAP finds the path with the highest end-to-end fidelity. Since the classical routing algorithm calculates the sum of the costs of all edges with ``additive'' routing metric, the original Dijkstra algorithm, which is based on greedy approach, cannot find the shortest path in quantum networks considering ``multiplicative'' routing metric for fidelity degradation in Eq. (\ref{fidelity_calculation}). Hence, we adopt non-additive but monotonic routing metric \cite{shi2020concurrent}, and design an extended Dijkstra algorithm to find the shortest path with minimum fidelity degradation between a given S-D pair.

%during the path selection procedure, we attempt to find the path with minimum fidelity degradation via Shortest-Path-First (SPF) algorithm, which represents the path with higher quality.
%The classical SPF algorithm such as Dijkstra calculates the sum of the costs of all edges with ``additive'' routing metric \cite{shi2020concurrent}. Thus, the algorithm can fail to find the shortest path in quantum networks considering ``multiplicative'' routing metric for fidelity degradation in (\ref{fidelity_calculation}). Hence, we modify the classic SPF algorithm as ``\textit{Path Selection Procedure}'' in Algorithm \ref{SPF_routing}, which designs for non-additive but monotonic routing metric, to find the shortest path with minimum fidelity degradation between a given S-D pair.

\textit{\textbf{Purification Decision}}:
%\textcolor{red}{As we mentioned before, finding the optimal purification decision on a given path is a NP-Hard problem.}
To make purification decisions $D^{pur}_{i,j}$ with lower computational complexity, we adopt the average fidelity to satisfy the requirement of fidelity threshold $F^{th}_i$. For a given path $P_{i,j}(s_i,d_i)$, each hop on the path should satisfy
$F^{avg}_{i}(l)=(F^{th}_i)^{\frac{1}{l}}$,
where $l$ is the length of routing path. By doing this, the worest complexity of purification decision can be significantly reduced to $\mathcal{O}\left(|E|\right)$.

\subsection{Discussion and Complexity Analysis of Q-LEAP}
Q-LEAP contains four main steps. At \textit{\textbf{Step 1}}, the complexity of \textit{purification cost table} can be calculated by $\mathcal{O}(|E|C_{max})$.
At \textit{\textbf{Step 2}}, the worst complexity for path selection can be calculated by $\mathcal{O}\left(|V|\log_2 |V| + |E|\right)$. At \textit{\textbf{Step 3}}, the worst complexity can be calculated by $\mathcal{O}\left(|E|\right)$. For the loop in line \ref{loop1}, it should be executed $R_i$ times. The complexity of \textit{\textbf{Step 4}} is $\mathcal{O}(|E|)$. Thus, the overall complexity is
$\mathcal{O}\Bigl(|E|C_{max}+R_i (|V|\log_2 |V| +|E|)\Bigr)$.

%to find minimum expected throughput, delete $t^{throughput}_{i,j}$ from $C$ in graph $G=(V,E,C)$ should be executed $|E|$ times in the worst case, the worst complexity can be calculated by $\mathcal{O}(\frac{|R_i|}{|C_{min}|}\left(\frac{|R_i|}{|C_{min}|}+|E|\right))$.

%with $\frac{|R_i|}{|C_{min}|}$ times, and $|C_{min}|\geq 1$ denotes the minimum number of entangled pairs on an edge $(u,v)\in E$.

%\begin{algorithm}[!htb]
%\caption{MCMF Routing Algorithm}
%\label{MCMF_routing}
%\KwIn{$G=(V,E,C)$, fidelity threshold $F^{th}_i$, S-D pair $<s,d>$, routing path $P(s,d)$;}
%\KwOut{Fidelity $F(s,d)=F[d]$,$D_i^j$,$fidelity\_label$;}
%
%{\textbf{Step 1 Initialization:} \\}
%Calculate \textit{Purification Cost Table} for $(u,v)\in E$;\\
%Delete all edges $(u,v)$ from $G$, if $F^{pur}_{max}(u,v)<F^{th}$;\\
%Construct $G_v=(V,E_v,C_v)$ with virtual edges;\\
%{\textbf{Step 2 Augmenting Path Searching:} \\}
%Dijkstra search augmenting path according to the hop count;\\
%function $SPF(s,d)$:\\
%
%{\textbf{Step 3 Max Flow:} \\}
%\While{$SPF(s,d)!=null$}{
%$P_{i,j}(s,d)=SPF(s,d)$;\\
%$Flow_j=DFS(s)$;\\
%}
%\end{algorithm}

\section{Routing design for Multiple S-D Pairs}
%As mentioned previously, the routing problem for multiple S-D pairs in classic networks can be regarded as a multi-commodity flow problem, which is NP-hard. In quantum networks, considering the extra purification decisions for fidelity constraint, the problem becomes more complicated.
%By exploring the routing design for single S-D pair in the last section, the routing problem for multiple S-D pairs can be easily divided into multiple single S-D pair scenarios and allocate the entanglement resource for multiple routing solutions.
%In general cases, requests from multiple S-D pairs are considered in a time slot especially for a large-scale quantum networks in the future.
%However, the routing problem for multiple S-D pairs in classic networks can be regarded as a multi-commodity flow problem, which is a NP-hard problem. In quantum networks, considering the extra purification decisions for fidelity constraint, the problem becomes more complicated.
%When the entanglement routing for multiple S-D pairs with purification operation is considered, the problem becomes more complicated.
In this section, based on aforementioned routing designs for single S-D pair, we further propose a greedy-based entanglement routing design to obtain fidelity-guaranteed routing paths for multiple S-D pairs, and design a utility metric with two important factors for resource allocation.

\subsection{Problem Definition and Design Overview}
\textbf{\textit{Problem Definition:}} Given multiple routing requests from multiple S-D pairs and a quantum network with topology and edge capacity as $G=(V,E,C)$, finding routing solutions, including purification decision $D^{pur}_{i,j}$ and path selection $P_{i,j}(s_i,d_i)$, to enable the entanglement establishment between S-D pairs to satisfy fidelity constraint $F_{i,j}(s_i,d_i)>F^{th}_i, \forall i$.

\textbf{\textit{Design Overview:}}
%Our final objective is to provide efficient entanglement routing solution to satisfy as many requests from multiple S-D pairs as possible with fidelity guarantee.
Based on the routing solutions obtained by the proposed two algorithms for single S-D pair, the routing problem in multiple S-D pairs scenario can be further regarded as a resource allocation problem. When sufficient resource (i.e.,entangled pairs) is provided in the network, the final solution is the combination of the optimal/near-optimal routing solution for single S-D pair. Otherwise, the requests which can hardly be satisfied should be denied. Thus, we consider two important allocation metrics, degree of freedom and resource consumption, to evaluate the performance of each routing solution for single S-D pair, and design a greedy-based routing algorithm to achieve efficient resource allocation for requests from multiple S-D pairs.

\begin{algorithm}[t]
\caption{Greedy-based Routing Design}
\label{Greedy_Routing}
\KwIn{$G=(V,E,C)$, $F^{th}_i$, requests $R_i$ and S-D pairs;}
\KwOut{$P_{i,j}(s_i,d_i)$, $D_{i,j}$, $F_j(s_i,d_i)$, $T^{EXT}_{i,j}$;}
\textit{\textbf{Step 1 Initialization:} \\}
Same as line \ref{Initialization1}-\ref{Initialization4} in Q-PATH;\\
Construct residual graph $G^r=(V,E^r,C^r)$;\\
\textbf{\textit{Step 2 Routing Path Predetermination:}} \\
%Search $K$-shortest-path $p$ for S-D pair;
\For{all S-D pairs} {
%Path searching
Find routing solutions $\leftarrow$Q-PATH or Q-LEAP;\\
$Q\leftarrow P_{i,j}(s_i,d_i)$ according to utility $U_{i,j}$;\label{utilitySort}\\
}
\While{$Q$ is not empty\label{resource_allocation}}{
\textit{\textbf{Step 3 Resource Allocation:} \\}
$P_{i,j}(s_i,d_i)\leftarrow Q$ and let $W^{min}_{i,j}=\rm{INT\_MAX}$;\\
\For{$(u,v)\in P_{i,j}(s_i,d_i) $}{
\If{$\lfloor \frac{c(u,v)}{N_{i,j}^{pur}(u,v)+1}\rfloor\leq W^{min}_{i,j}$}{
$W^{min}_{i,j}=\lfloor \frac{c(u,v)}{N_{i,j}^{pur}(u,v)+1}\rfloor$;\\
}
}
\If{$W^{min}_{i,j}\geq 1$}{
Subtract $\min\{W^{min}_{i,j},R_i\}\times (N_{i,j}^{pur}(u,v)+1)$ on each $(u,v)\in P_{i,j}(s_i,d_i)$ from $C^r$ in $G^r$;\\
%and delete $(u,v)\in E_r$ if $c(u,v)\leq 0 $;\\
Output $P_{i,j}(s_i,d_i)$, $D^{pur}_{i,j}$, $F_j(s_i,d_i)$, $T^{EXT}_{i,j}$;\label{throughput_update_end}\\
\If{$\sum_j T^{EXT}_{i,j}(s_i,d_i)\geq R_i$}{
Remove all paths $P_{i,j}(s_i,d_i),\forall j$ from $Q$;\\
}
}
\textit{\textbf{Step 4 Re-routing Process:} \\}
\Else{
Find routing solutions $\leftarrow$Q-PATH or Q-LEAP;\\
%Find the shortest path $P(s_i,d_i)$ for $i$-th S-D pair from $G$;\\
\If{$P_{i,j+1}(s_i,d_i)!=P_{i,j}(s_i,d_i)$}{
$Q\leftarrow P_{i,j+1}(s_i,d_i)$ according to utility $U_{i,j}$;\\
}
Delete $P_{i,j}(s_i,d_i)$ from $Q$;\\
}
}
%Notation:The following design is for one flow volume (one memory cost) each S-D pair
%$fail\_count=0$;\\
%\For{$i=1:(M_u-cost(u))+fail\_count$}{
%Find the shortest path $P(s_i,d_i)$ from $G'$;\\
%\If{$P(s_i,d_i)==null$}{
%$fail\_count=fail\_count+1$;\\
%}
%}
\end{algorithm}
\subsection{Algorithm Procedure}
The greedy-based routing design is described in Algorithm \ref{Greedy_Routing}.
The basic idea of the algorithm can be described as follows: it first calculates the routing path for single S-D pair by using the algorithms proposed in Section IV. To satisfy the requests from multiple S-D pairs as many as possible, we then allocate the network resource for each obtained routing solutions (path selection and purification decision) one by one. For each loop in line \ref{resource_allocation}, the corresponding resource $\min\{W^{min}_{i,j},R_i\}\times (N_{i,j}^{pur}(u,v)+1)$ consumed on each edge $(u,v)\in P_{i,j}(s_i,d_i)$ would be deleted on the residual graph $G^r$. If the resource of some edges on the routing path $P_{i,j}(s_i,d_i)$ has been exhausted, leading to an invalid path on $G^r$, a re-routing process will be performed on the residual graph $G^r$, and the re-routing solution for $i$-th S-D pair will be enqueued into priority queue $Q$ again and continue loop in line \ref{resource_allocation}.

\begin{figure*}[tp]
  \centering
  \subfigure[Network topology.]{
    \label{routing_example1}
    \includegraphics[width=1.27in]{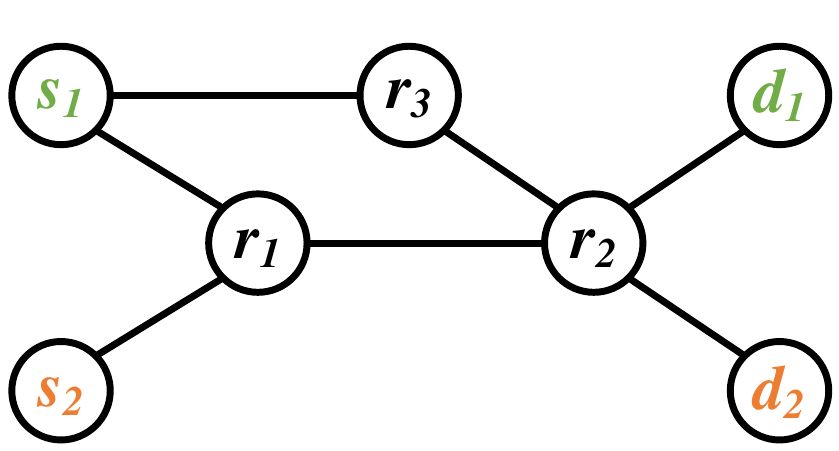}
 }
  \subfigure[Throughput=1.]{
    \label{routing_example2}
    \includegraphics[width=1.27in]{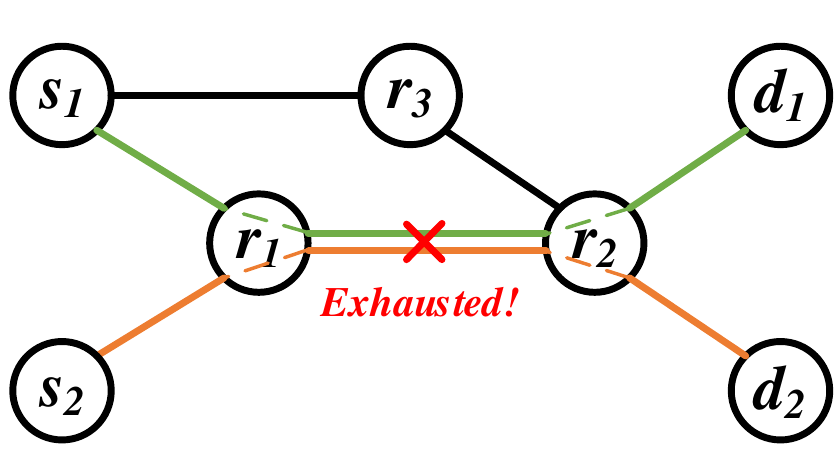}
    }
    \subfigure[Throughput=2.]{
    \label{routing_example3}
    \includegraphics[width=1.27in]{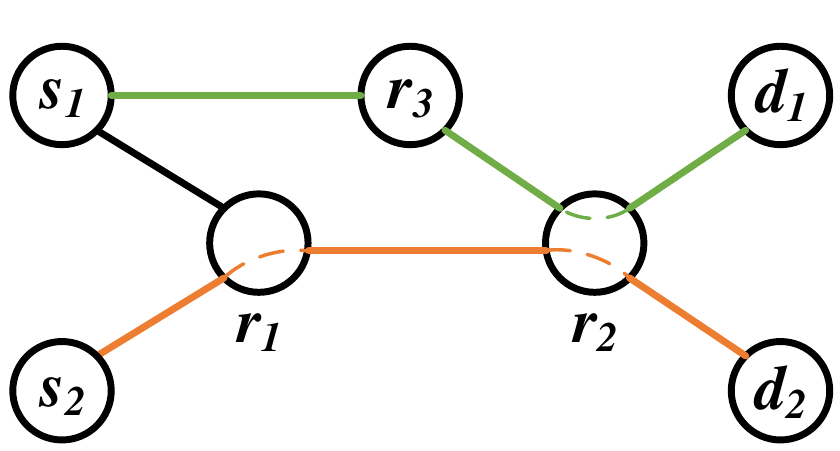}
    }
    \subfigure[Throughput=1.]{
    \label{routing_example4}
    \includegraphics[width=1.27in]{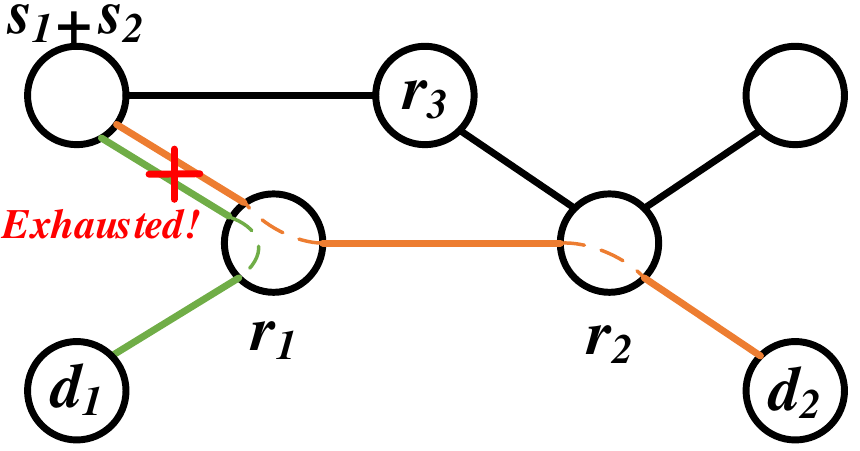}
    }
    \subfigure[Throughput=2.]{
    \label{routing_example5}
    \includegraphics[width=1.27in]{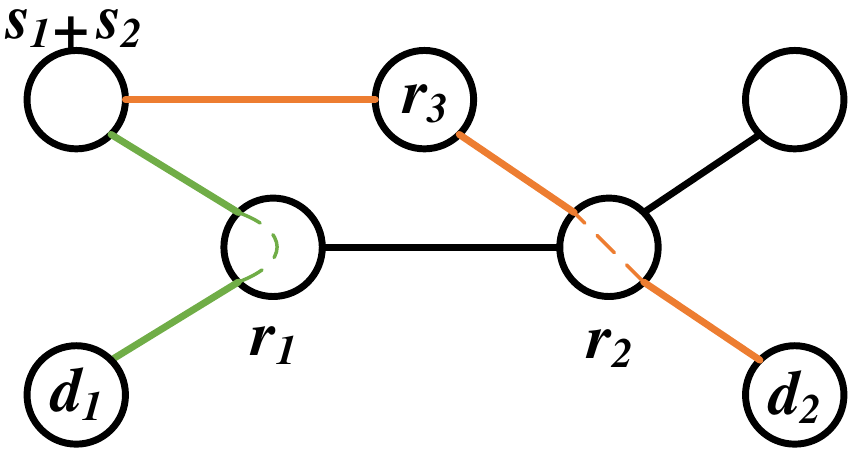}
    }
    \setlength{\abovecaptionskip}{0.1cm}
  \caption{Routing example for multiple S-D pairs, only one entangled pair is considered on each edge. }
  \label{routing_example}
\end{figure*}

In Algorithm \ref{Greedy_Routing}, the most important procedure, which has the most significant impact on the performance, is the order that which routing solution in $Q$ should be allocated first in Resource Allocation process. For example, in Figs. \ref{routing_example1}-\ref{routing_example2}, there are two requests for $\langle s_1,d_1\rangle$ and $\langle s_2,d_2\rangle$. By executing Q-PATH or Q-LEAP, we obtain two routing paths $P_{1,1}(s_1,d_1)$=$\{(s_1,r_1),(r_1,r_2),(r_2,d_1)\}$ and $P_{2,1}(s_2,d_2)$=$\{(s_2,r_1),(r_1,r_2),(r_2,d_2)\}$. If we first allocate the network resource for $P_{1,1}(s_1,d_1)$, then the resource on edge $(r_1,r_2)$ would be exhausted. After that, the request for $\langle s_2,d_2\rangle$ has to be denied since $P_{2,1}(s_2,d_2)$ becomes invalid and there is no other available routing solution for $\langle s_2,d_2\rangle$, which leads to a poor performance in terms of throughput.

\subsection{Resource Allocation}

To address the resource allocation problem, we introduce a utility metric to evaluate the degree of difficulty when the algorithm attempts to satisfy a given request, two important factors of a given routing solution is considered in the utility, i.e., resource consumption and degree of freedom:
\begin{equation}
U_{i,j}=\alpha\cdot G\left (P_{i,j}(s_i,d_i)\right)+\beta\cdot S\left(P_{i,j}(s_i,d_i),D_{i,j}^{pur}\right),
\label{utility}
\end{equation}
where $\alpha$ and $\beta$ are weight coefficient, $S\left(\cdot\right)$ and $G\left(\cdot\right)$ denotes the total resource consumption and the degree of freedom of a given routing path, respectively.

1) For degree of freedom, it is defined as the sum of routing options on each hop, which can be calculated by:
%The degree of freedom of the path $P_{i,j}(s_i,d_i)$ can be calculated by:
\begin{equation}
G\left(P_{i,j}(s_i,d_i)\right)=\sum_{{\rm{nodes~on~}}P_{i,j}(s_i,d_i)}\mathcal{N}(u).\notag
\end{equation}
The degree of freedom describes the successful re-routing possibility. For example, Figs. \ref{routing_example2}-\ref{routing_example3} show the advantage by considering degree of freedom as a factor when the algorithm decides the order of resource allocation, routing paths $P_{1,1}(s_1,d_1)$ and $P_{2,1}(s_2,d_2)$ are found by the Q-PATH or Q-LEAP as shown in Fig. \ref{routing_example2}, and $G\left(P_{1,1}\left(s_1,d_1\right)\right)=10$ and $G\left(P_{2,1}\left(s_2,d_2\right)\right)=9$ in Fig. \ref{routing_example2}. If we check $P_{1,1}(s_1,d_1)$ first, the resource on edge $(r_1,r_2)$ would be exhausted. Then $P_{2,1}(s_2,d_2)$ becomes invalid and there is no other available routing path for $\langle s_2,d_2\rangle$. If we check all routing solutions according to the value of degree of freedom, then the resource would be allocated to $P_{1,1}(s_1,d_1)$ first since $G\left(P_{2,1}(s_2,d_2)\right)<G\left(P_{1,1}(s_1,d_1)\right)$. In this case, maximum throughput is reached as shown in Fig. \ref{routing_example3}.

2) For resource consumption, it is defined as the sum of consumed entangled pairs for a given routing solution, which can be calculated by:
\begin{equation}
S\left(P_{i,j}(s_i,d_i),D_{i,j}^{pur}\right)=\sum_{(u,v)\in P_{i,j}(s_i,d_i)}N_{i,j}^{pur}(u,v).\notag
\end{equation}
Considering the limited resource of entangled pairs on edges, a routing path with lower resource consumption should be allocated earlier, in case the routing path with higher resource consumption consumes too much entangled pairs and makes the requests from other S-D pairs be denied. An example is shown in Figs. \ref{routing_example4}-\ref{routing_example5}, which show the advantage by considering resource consumption as a factor when the algorithm decides the order of resource allocation. $P_{1,1}(s_1,d_1)$ is a shorter path and consumes less entangled pairs, and $S\left(P_{1,1}(s_1,d_1),D_{1,1}^{pur}\right)=2$ and $S\left(P_{2,1}(s_2,d_2),D_{2,1}^{pur}\right)=3$ in \ref{routing_example4}. If we check $P_{2,1}(s_2,d_2)$ first, the resource on edge $(s_2,r_1)$ would be exhausted, then $P_{1,1}(s_1,d_1)$ becomes invalid and there is no other available routing solutions for $\langle s_2,d_2\rangle$. In this case, as shown in Fig. \ref{routing_example5}, if we check all routing solutions according to the value of resource consumption $S(\cdot)$, maximum throughput can be achieved.

\subsection{Computational Complexity Analysis}
The computational complexity of \textit{Algorithm \ref{Greedy_Routing}} can be analyzed as follows.
%The algorithm consists four main steps. %At \textit{\textbf{Step 1}}, i.e., ``\textbf{\textit{Initialization:}}'', the complexity can be calculated by $\mathcal{O}\left(N\left(|V|\log |V|+|E|\right)\right)$.
At \textit{\textbf{Step 2}}, the number of iterations is decided by the number of S-D pairs $R_i$, and the complexity of each path selection is the same as Q-PATH or Q-LEAP. At \textit{\textbf{Step 3}}, $K$ paths would be obtained at most from Q-PATH or Q-LEAP, the complexity can be calculated by $\mathcal{O}(K|E|)$. At \textit{\textbf{Step 4}}, re-routing procedure can be executed $\sum_iR_i$ times at most.

As a comparison, the complexity of brute-force searching approach is $\mathcal{O}\left(\sum_iR_i\left(C_{max}\right)^{|E|}\right)$.

%\subsection{Upper Bound of the Achievable Throughput/Analysis from the perspective of network flow}
%The constraints represent the limitation

%We propose an extended max-flow min-cost algorithm.
%algorithm reference:https://zh.wikipedia.org/wiki/%E6%9C%80%E5%A4%A7%E6%B5%81%E9%97%AE%E9%A2%98

%\subsection{Upper Bound with fidelity constraint}

\section{Performance Analysis}
%
%\begin{table}[t]
%  \centering
%  \caption{Simulation Parameters}
%  \label{SimulationParameters}
%  \begin{tabular}{p{6cm}c}%p{4.8cm}
%    \hline
%     \textbf{Parameter} & \textbf{Value} \\\hline
%     Network size & 10-500 \\
%     Edge capacity & 10-100 \\
%     Number of S-D pairs & 1-6 \\
%     Requests for each S-D pair & 100\\
%     Fidelity threshold & 0.55-0.9 \\
%     $\kappa$ in Waxmax model & 0.05-0.07\\
%     $\gamma$ in Waxmax model & $1$\\
%     Fidelity distribution of quantum channels & $\mathcal{N}[0.8, 0.1]$ \\\hline
%  \end{tabular}
%\end{table}

%\begin{figure}[t]
%  \centering
%  \includegraphics[width=1.0\linewidth]{topology.png}
%  \caption{Illustration of US backbone network topology that used in the simulation.}
%  \label{network_topology}
%\end{figure}

\subsection{Evaluation Setup}

%Since the value of fidelity threshold may vary in practice for different network functionalities.
To implement the proposed routing algorithms in quantum networks, we conduct a series of numerical evaluations. The simulation code is available in \cite{Quantum_Simulator}. %called QuantNetSim.
The software and hardware configuration of simulation platform are: AMD Ryzen 7 3700X 3.6GHz CPU, 32GB RAM, O.S. Windows 10 64bits.
Along with the development of quantum networks, it can be foreseen that quantum networks will become a critical infrastructure for security communications and quantum applications in the future, which is similar to the current Internet backbone. Thus, the US backbone network \cite{orlowski2010sndlib} % \footnote{The data of US backbone network is provided by SNDlib project \cite{orlowski2010sndlib}.}
is adopted as the examined network topology in our simulation.
For each set of parameter settings, simulations are run 1000 trials and the averaged results are given, the original fidelity of entangled pairs follows $\mathcal{N}[0.8, 0.1]$.
According to \cite{dahlberg2019link}, the typical qubit lifetime is 1.46s, thus we adopt the synchronization timestep as 500ms.
%\textcolor{red}{The key simulation parameters are summarized in Table \ref{SimulationParameters}.} %and the comparison scheme is given as follows.
\textbf{The performance of routing scheme proposed in \cite{li2021effective} is adopted as the baseline} since it is the only existing routing design that considers purification operation, and proportional share is adopted as its resource allocation scheme. %Note that only pre-purification is adopted before path selection, i.e., repeat purification operation on each edge $(u,v)\in E$ until $F(u,v)>F^{th}$, and proportional share is adopted as the resource allocation scheme.
In specific, US backbone network topology is used in Figs. 7-11.

%\begin{itemize}
%\item \textbf{\textit{Baseline 1:}} The classical Dijkstra algorithm is adopted as baseline 1. Note that the algorithm cannot make purification decision, i.e., $N^{pur}_{i,j}(u,v)=0,$ $\forall i,$ $j,$ $(u,v)$.
%\item \textbf{\textit{Baseline:}} The routing scheme proposed in \cite{li2021effective} is adopted as the baseline. Note that only pre-purification is adopted before path selection, i.e., repeat purification operation on each edge $(u,v)\in E$ until $F(u,v)>F^{th}$, and Proportional Share (PS) is adopted as the resource allocation scheme.
%%\item \textbf{\textit{Brute-force Routing:}} Brute-force searching is adopted to find the optimal routing path and purification decision.
%\end{itemize}

To evaluate the efficiency of the proposed algorithms, random network topologies are also generated in the simulations following the Waxman model \cite{shi2020concurrent,zhao2021redundant,waxman1988routing}, i.e., the probability that there is an edge between node $u$ and $v$ can be formulated as $p(u,v)=\kappa\exp\frac{-d(u,v)}{L\gamma}$ \citep[Eq.(4)]{waxman1988routing}, where $d(u,v)$ is the Euclidean distance (measured in kilometers) between node $u$ and $v$, $L$ is the largest distance between two arbitrary nodes. In specific, random network topologies are used in Table III.

%The network topology is randomly generated for simulations. We set a $10^5km\times10^5km$ to place quantum nodes. Edges are determined following the Waxman model \cite{waxman1988routing}, i.e., the probability that there is an edge between node $u$ and $v$ can be formulated as $p(u,v)=\gamma e^{-l(u,v)/\beta L}$, where is the Euclidean distance (measured in kilometers) between node $u$ and $v$, $L$ is the largest distance between two arbitrary nodes.

\begin{figure*}[!htp]
  \centering
  \subfigure[Throughput vs. Fidelity Threshold]{
    \label{simulation1-1}
    \includegraphics[width=1.9in]{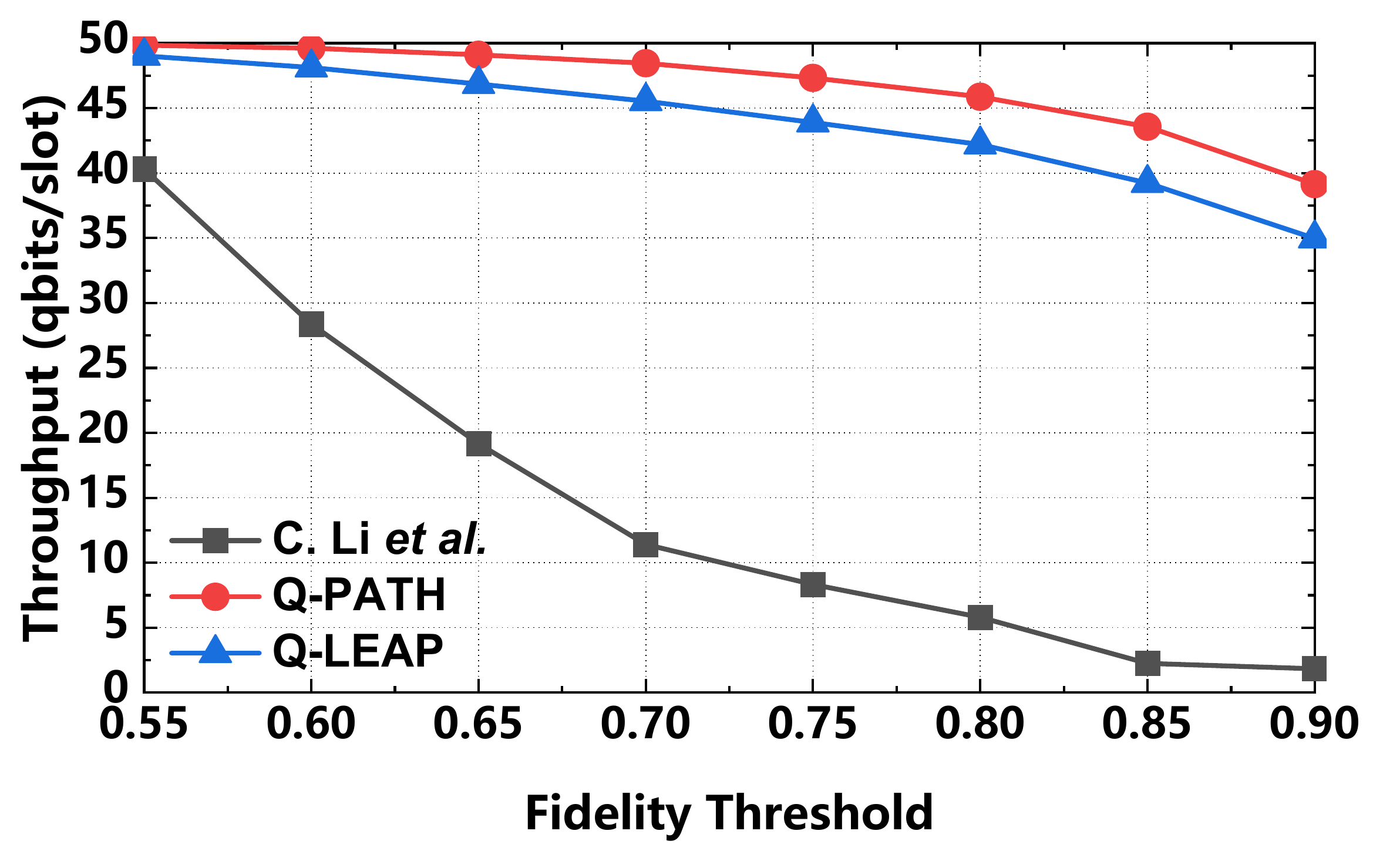}
 }
  \subfigure[Fidelity vs. Fidelity Threshold]{
    \label{simulation1-2}
    \includegraphics[width=1.95in]{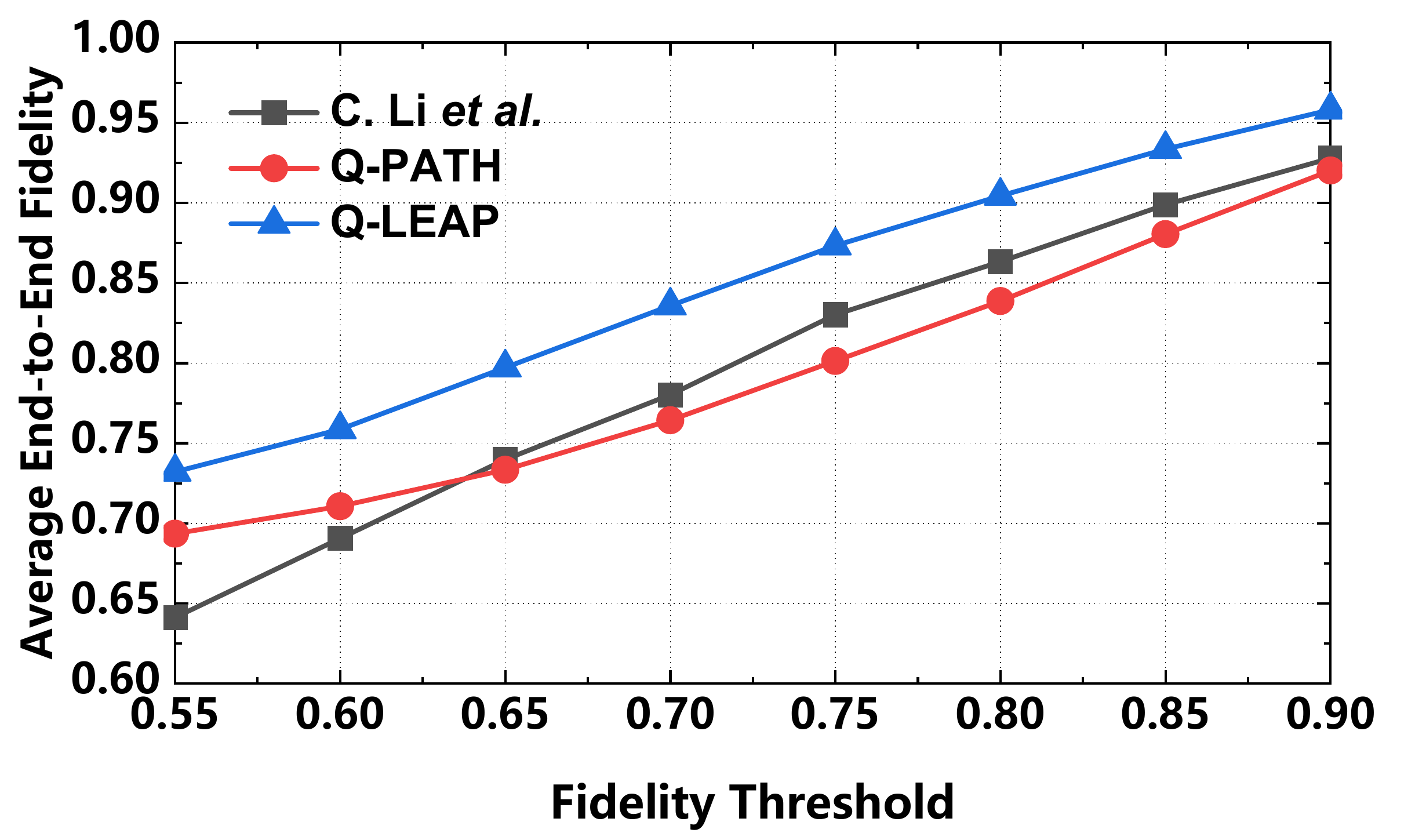}
    }
      \subfigure[Utilization vs. Fidelity Threshold]{
    \label{simulation1-3}
    \includegraphics[width=1.92in]{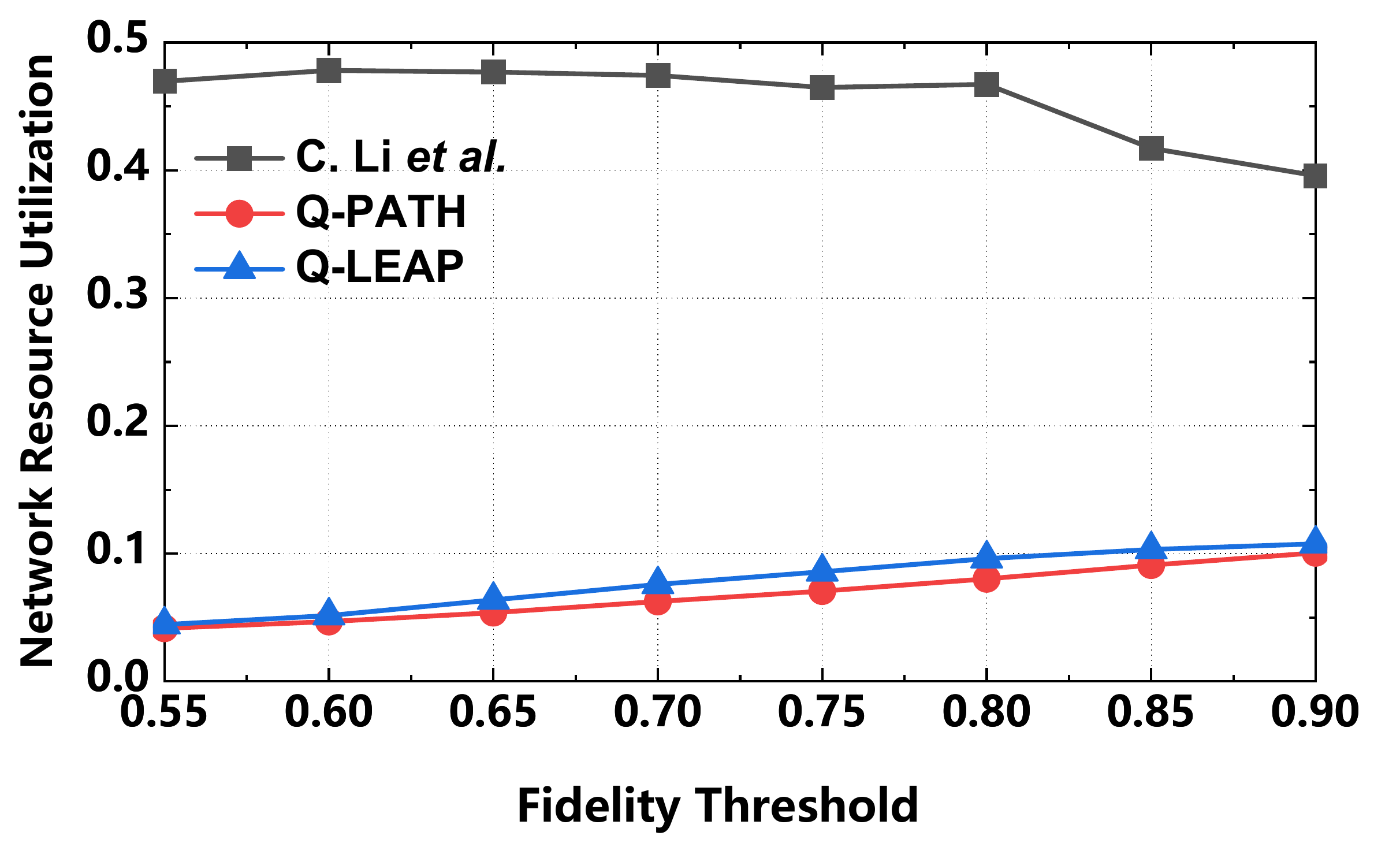}
    }
\setlength{\abovecaptionskip}{0.1cm}
  \caption{Performance comparison for single S-D pair in terms of throughput, average fidelity, and network resource utilization (channel capacity $= 50$).}
  \label{simulation1}
\end{figure*}
\begin{figure*}[t]
  \centering
    \subfigure[Throughput vs. Channel Capacity]{
    \label{simulation2-1}
    \includegraphics[width=1.9in]{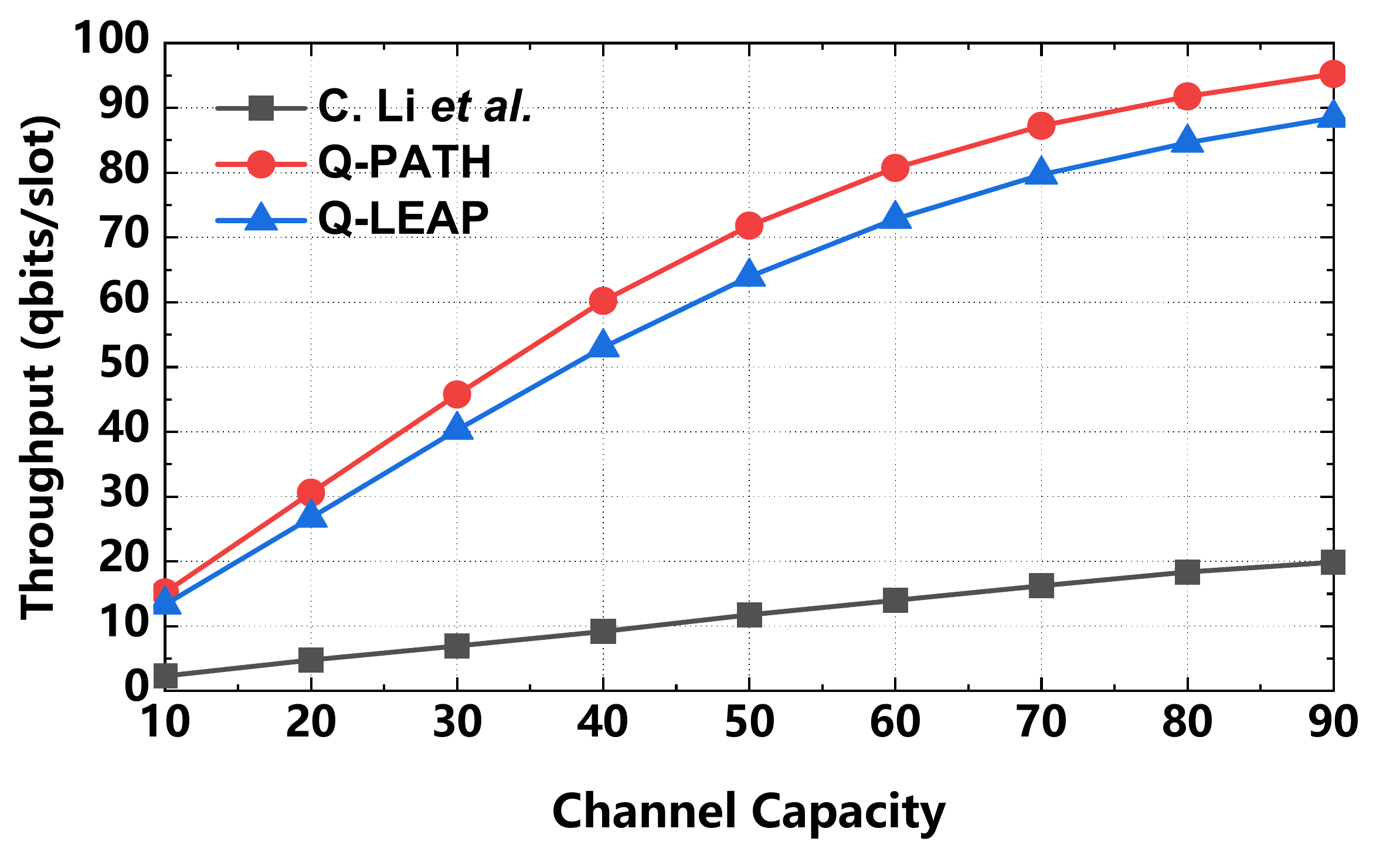}
 }
  \subfigure[Fidelity vs. Channel Capacity]{
    \label{simulation2-2}
    \includegraphics[width=1.9in]{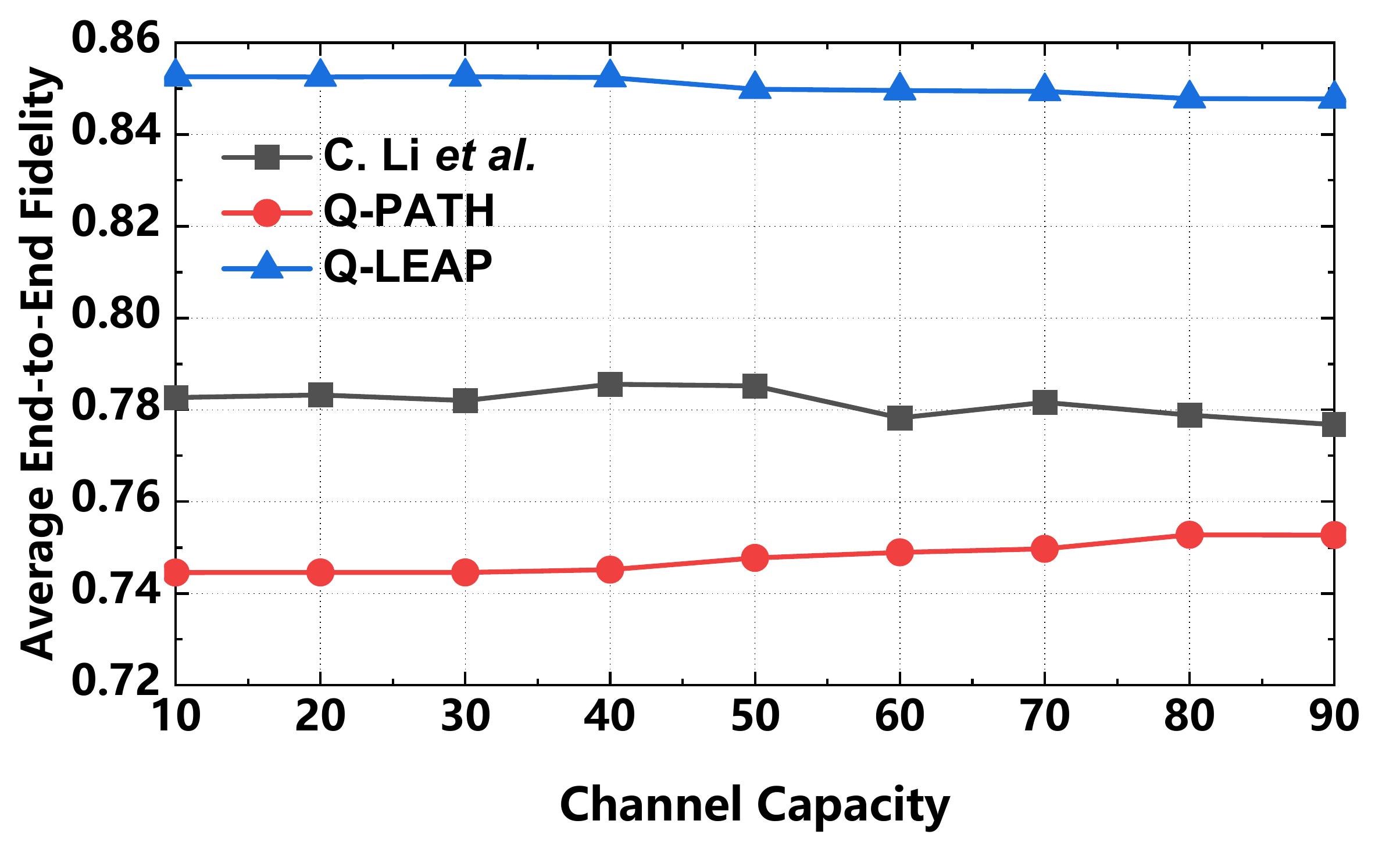}
    }
      \subfigure[Utilization vs. Channel Capacity]{
    \label{simulation2-3}
    \includegraphics[width=1.95in]{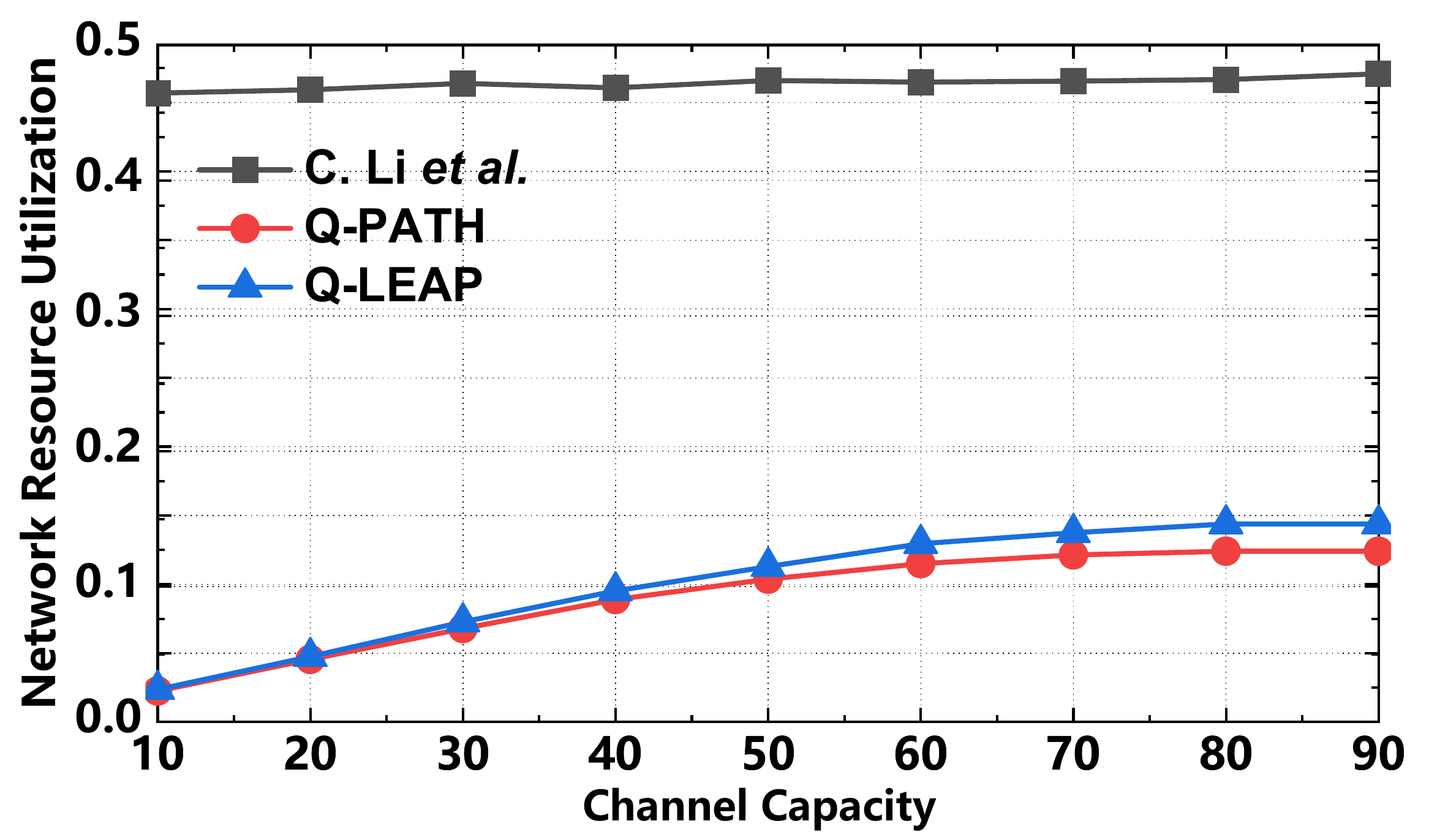}
    }
    \setlength{\abovecaptionskip}{0.1cm}
  \caption{Performance comparison for single S-D pair in terms of throughput, average fidelity, and network resource utilization (fidelity threshold $= 0.7$).}
  \label{simulation2}
\end{figure*}

\subsection{Results under Single S-D Pair Scenarios}
\label{sectionVI-1}

%\textbf{\textit{Performance Comparison with Brute-Force Method:}}
%In Fig. \ref{simulation0}, the performance of the proposed Q-PATH and brute-force searching method is evaluated. Note that $\kappa$=0.05-0.07, $\gamma$=1 in Waxman model, the objective of brute-force searching method is to find the optimal solution with minimum entangled pair cost, and the path found by two methods are the same. Due to the non-polynomial complexity of brute-force searching method, the evaluation can only be conducted on a small-scale network. For the results in Fig. \ref{simulation0-1}-Fig. \ref{simulation0-3}, the routing solutions obtained by Q-PATH and brute-force searching method are exactly the same, which indicates that Q-PATH can achieve the optimal routing performance in terms of throughput and average fidelity compared with brute-force searching method.

\begin{table}[t]
  \centering
  \caption{Algorithm Running Time (fidelity threshold = 0.6, channel capacity = 10)}
  \label{Algorithm_Running_Time1}
  \begin{tabular}{cccc}%p{4.8cm}
    \hline\hline
     \textbf{Network Scale} & \textbf{Q-PATH} & \textbf{Q-LEAP} & \textbf{Baseline}\\\hline
     %39 &116.88ms & 0.36ms & 11.87ms\\\hline
     100 &282.25ms & 0.47ms & 46.56ms\\
     200 &307.81ms & 1.4ms & 105.78ms\\
     300 &381.52ms & 2.55ms & 180.78ms\\
     400 &1056.11ms & 3.87ms & 275.46ms\\
     500 &3644.32ms & 6.76ms & 376.56ms\\
     \hline\hline
  \end{tabular}
\end{table}
\textbf{\textit{Computational Complexity Comparison:}}
The running time of three routing algorithms is shown in Table \ref{Algorithm_Running_Time1}. As we analyzed before, although Q-PATH can achieve the best routing performance, the complexity of Q-PATH is relatively high, which makes it time-consuming. As the results in Table \ref{Algorithm_Running_Time1}, with the increase of network scale, the running time of Q-PATH is 5x-10x higher than the baseline, but the running time of Q-LEAP is 50x lower than the baseline.
%In practical, the optimal routing algorithm can be applied in small-scale quantum networks,...

\textbf{\textit{Performance Comparison vs. Fidelity Threshold:}}
As shown in Fig. \ref{simulation1}, the performance of the proposed Q-PATH and Q-LEAP is evaluated compared with the baseline. In Fig. \ref{simulation1-1}, Q-PATH obtains the highest throughput, and the gap between Q-PATH and Q-LEAP reaches the peak at fidelity threshold of 0.85. The reason can be explained as follows. With the increase of fidelity threshold, multi-round purification operations are required to satisfy end-to-end fidelity requirement, then available entangled pairs on each quantum channel are reducing from default 50 pairs to several pairs after necessary purification. Thus, when the fidelity threshold is small, the gap between Q-PATH and Q-LEAP is negligible since multi-round purification operations are unnecessary. Once the number of available entangled pairs is limited, the solution space is reducing, which leads to similar routing solution obtained by Q-PATH and Q-LEAP. %and the gap between these two schemes is shrinking when the fidelity threshold is large enough.
For the baseline, due to the purification is performed before path selection, end-to-end fidelity can hardly be guaranteed. Thus, a poor performance in terms of throughput is obtained.

In Fig. \ref{simulation1-2}, Q-PATH also achieves the minimum fidelity but above fidelity threshold. This phenomenon shows that the purification decision and path selection obtained from Q-PATH can achieve the minimum cost to provide end-to-end fidelity guarantee. %Note that a cross point is shown in the figure between fidelity threshold 0.9 and 0.95,
In Fig. \ref{simulation1-3}, the network resource utilization is calculated as the ratio of the consumed entanglement pairs and the total entanglement pairs in the network. Although Q-PATH can achieve the highest throughput, the resource utilization of Q-PATH is lower than Q-LEAP.

\begin{figure*}[!ht]
  \centering
  \subfigure[Throughput vs. Fidelity Threshold]{
    \label{simulation3-1}
    \includegraphics[width=1.9in]{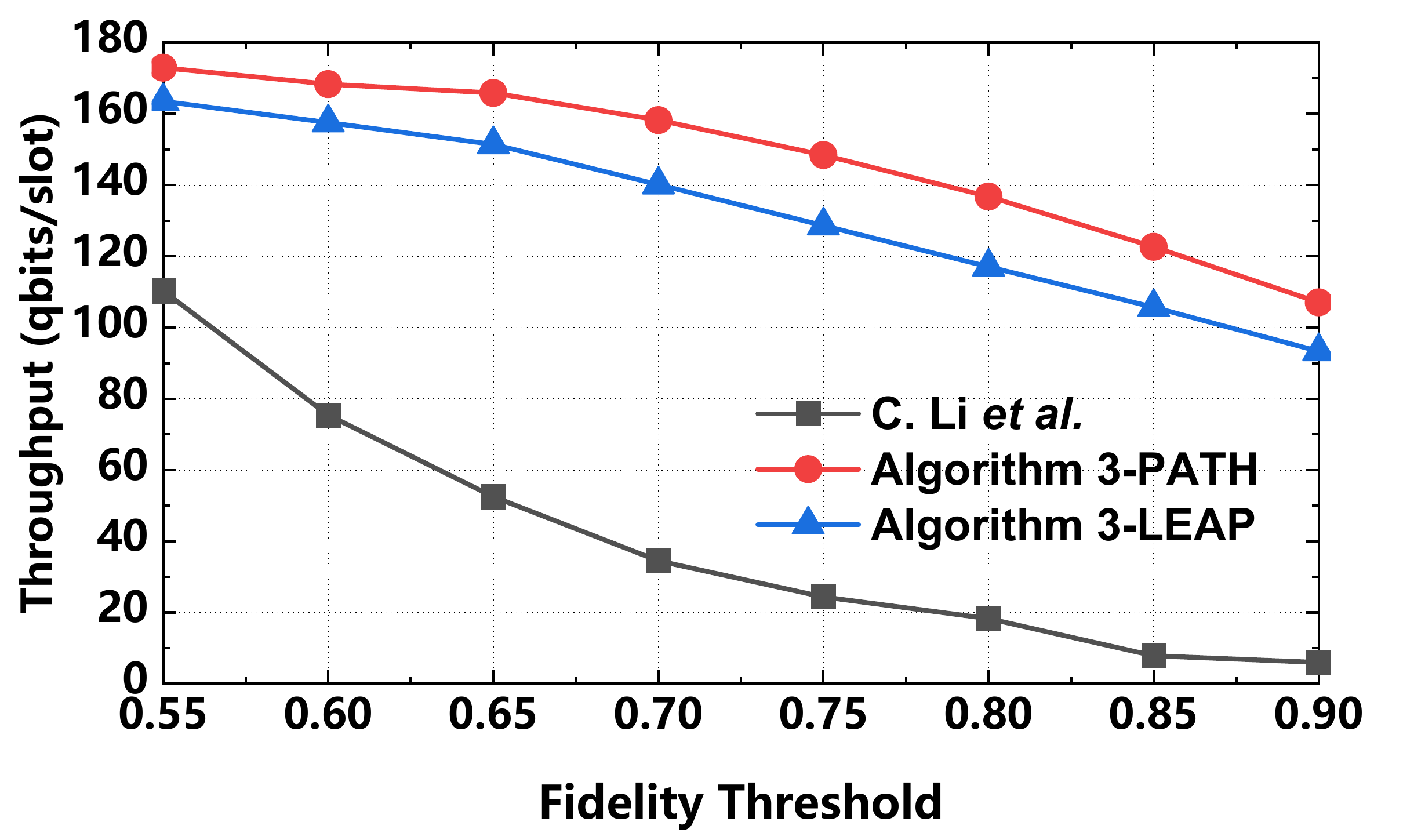}
 }
  \subfigure[Fidelity vs. Fidelity Threshold]{
    \label{simulation3-2}
    \includegraphics[width=1.9in]{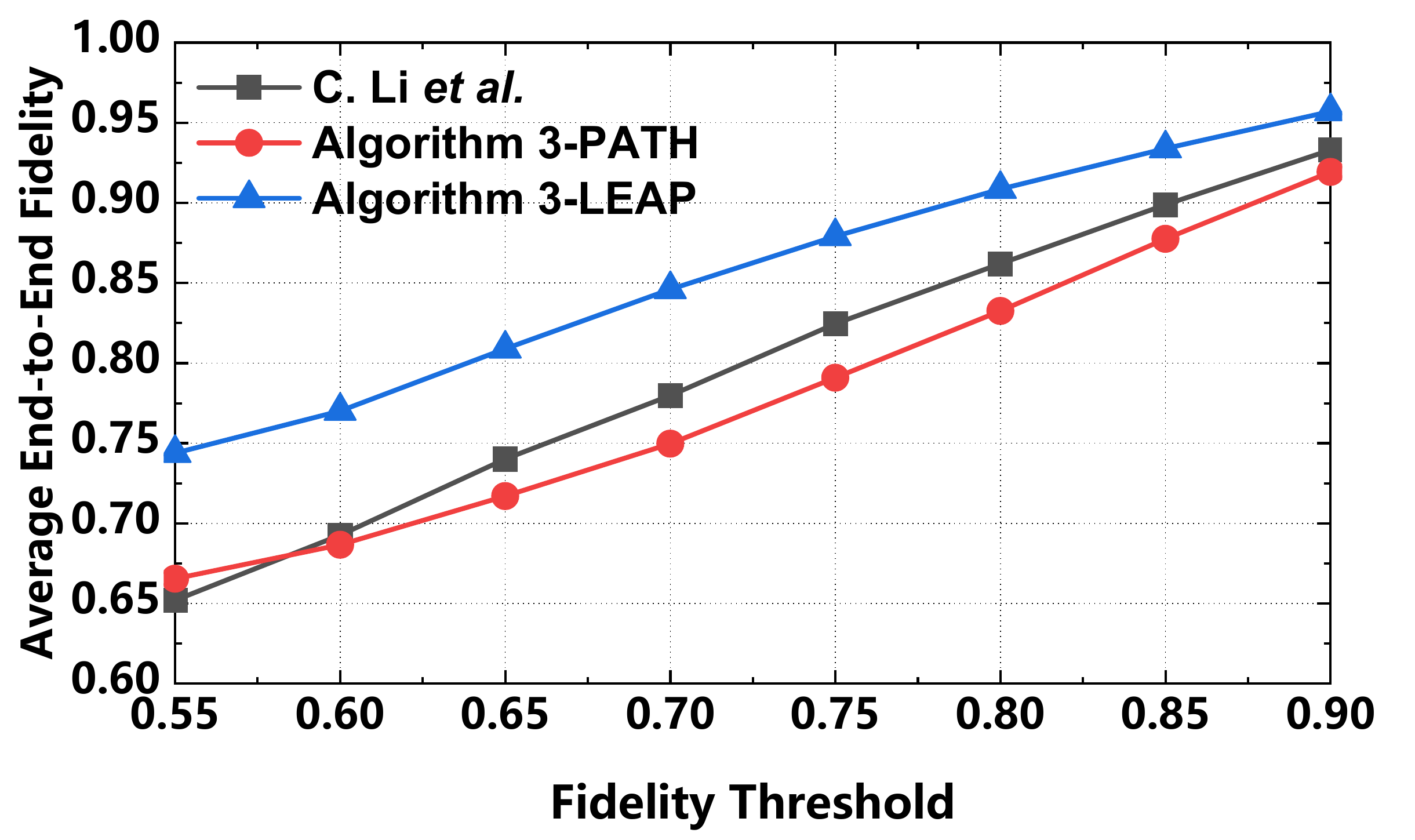}
    }
      \subfigure[Utilization vs. Fidelity Threshold]{
    \label{simulation3-3}
    \includegraphics[width=1.9in]{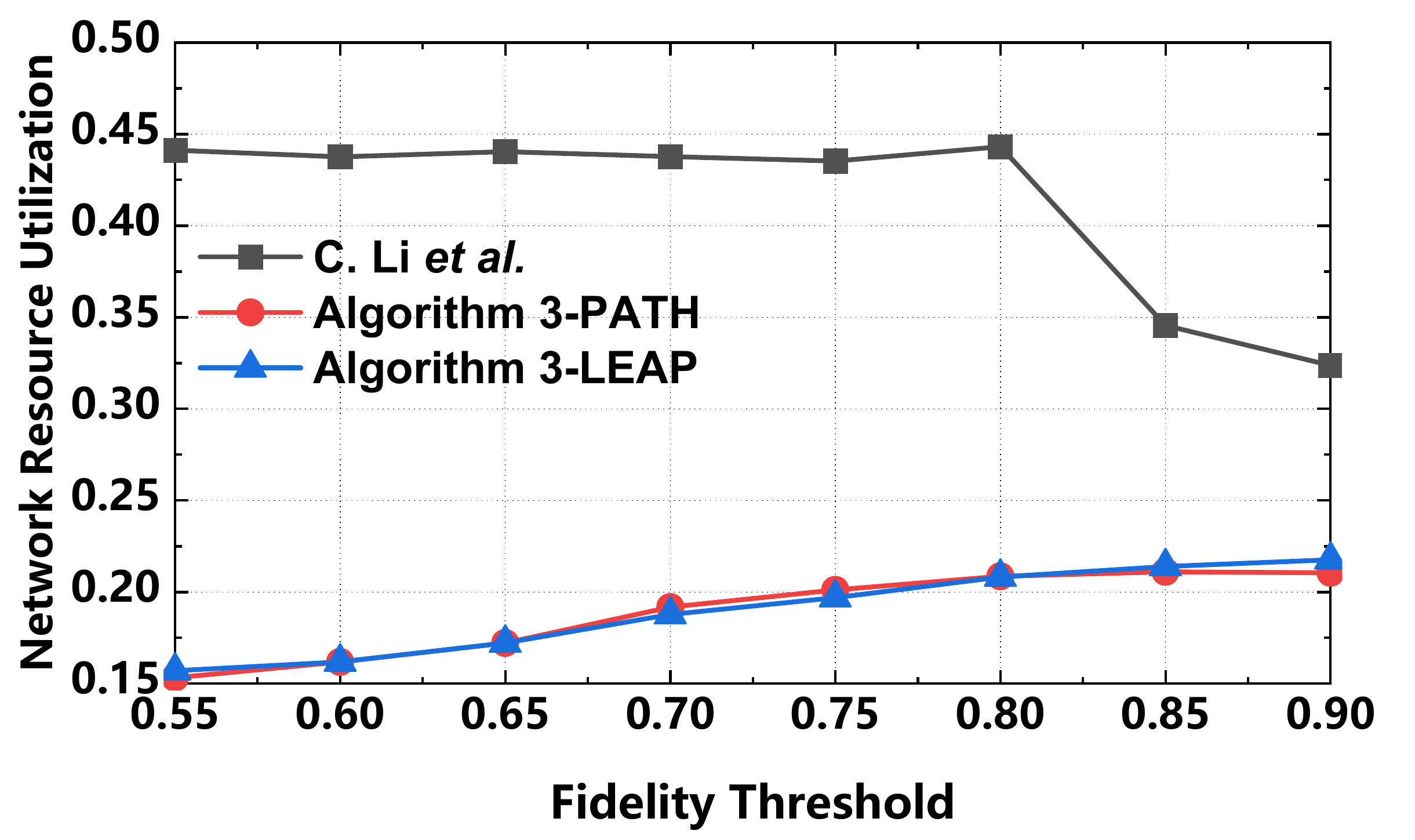}
    }
    \setlength{\abovecaptionskip}{0.1cm}
  \caption{Performance comparison for multiple S-D pairs versus fidelity threshold (channel capacity = 50, SD pairs = 4).}
  \label{simulation3}
\end{figure*}
\begin{figure*}[!ht]
  \centering
  \subfigure[Throughput vs. Channel Capacity.]{
    \label{simulation4-1}
    \includegraphics[width=1.9in]{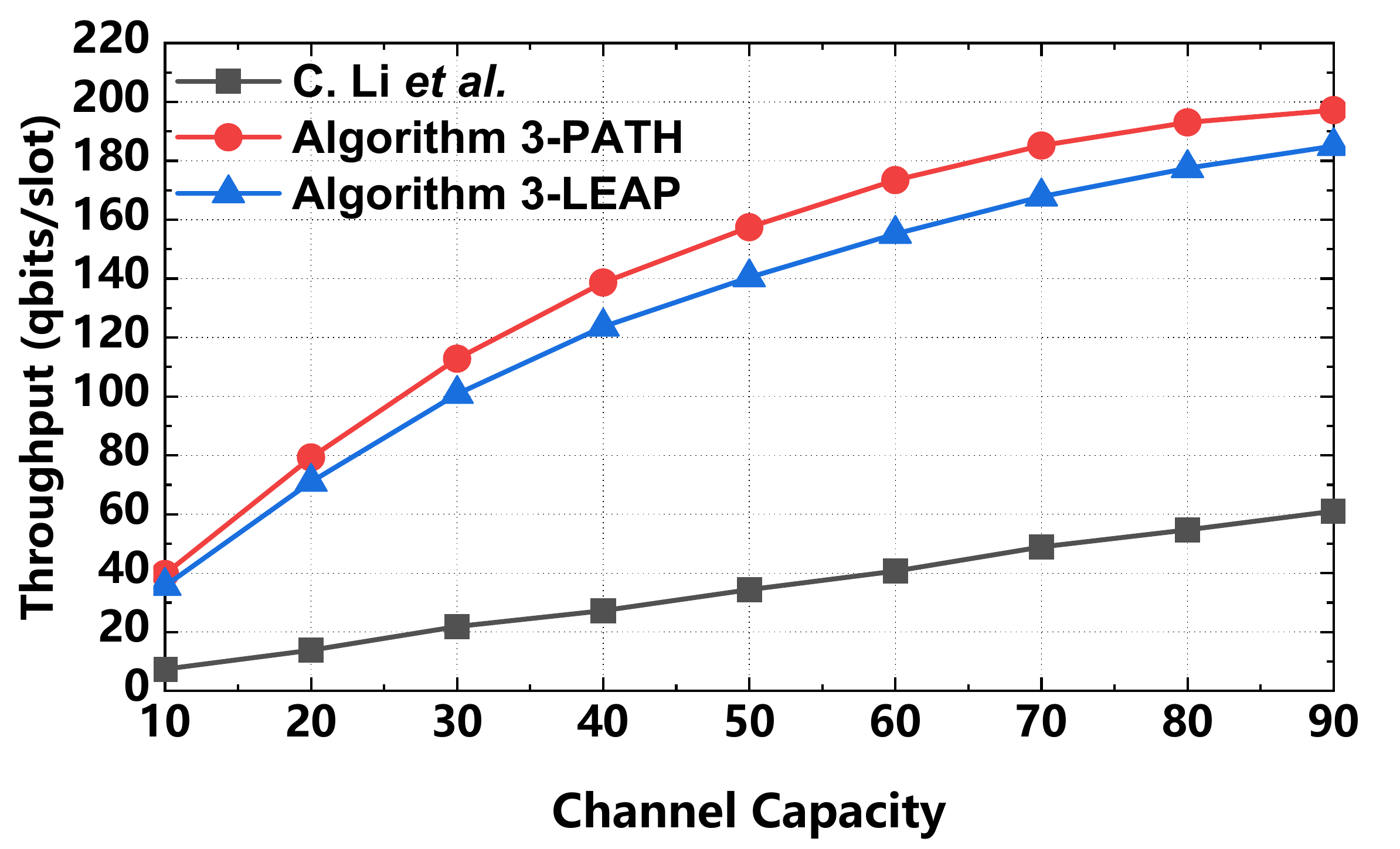}
 }
  \subfigure[Fidelity vs. Channel Capacity.]{
    \label{simulation4-2}
    \includegraphics[width=1.9in]{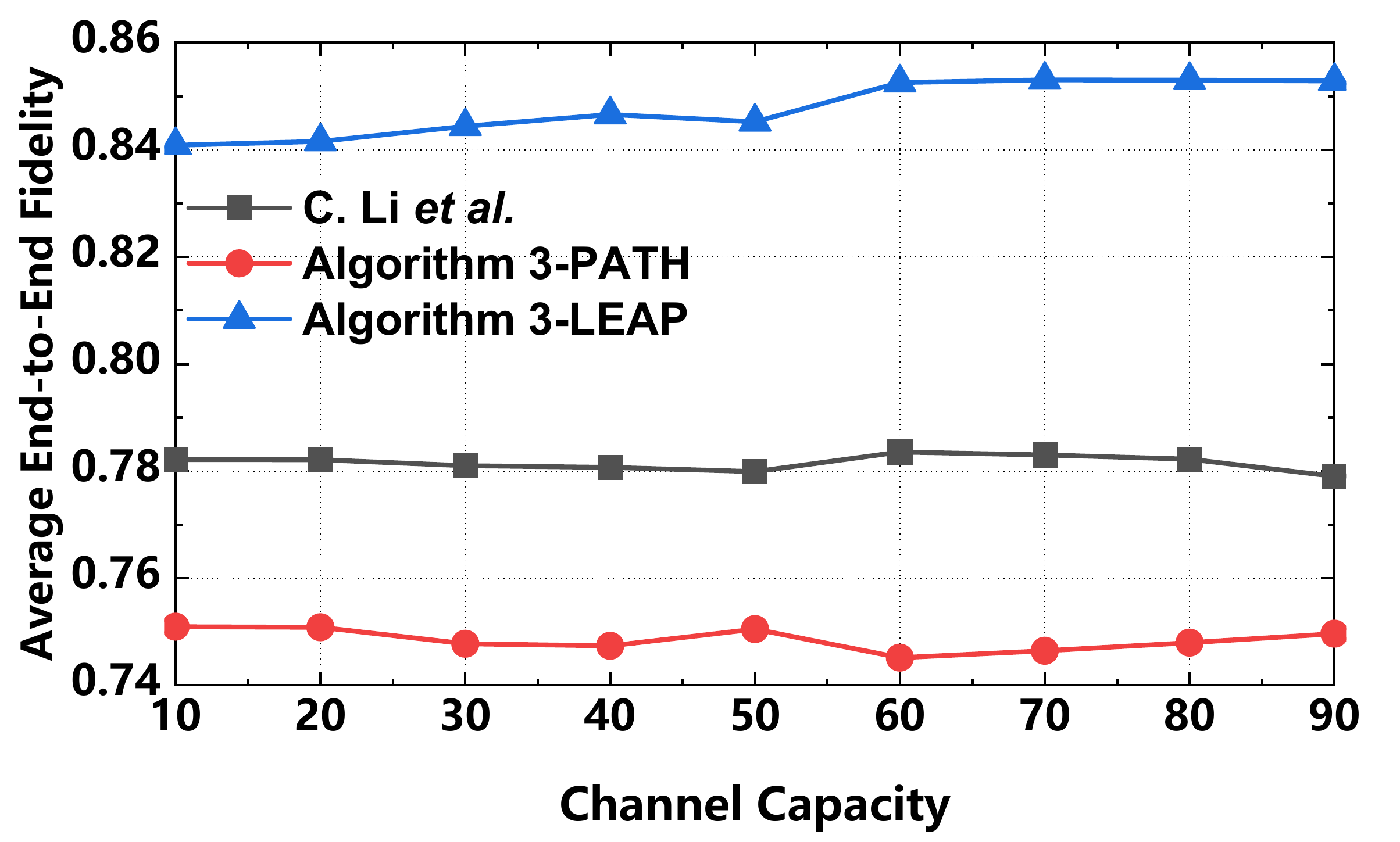}
    }
      \subfigure[Utilization vs. Channel Capacity.]{
    \label{simulation4-3}
    \includegraphics[width=1.9in]{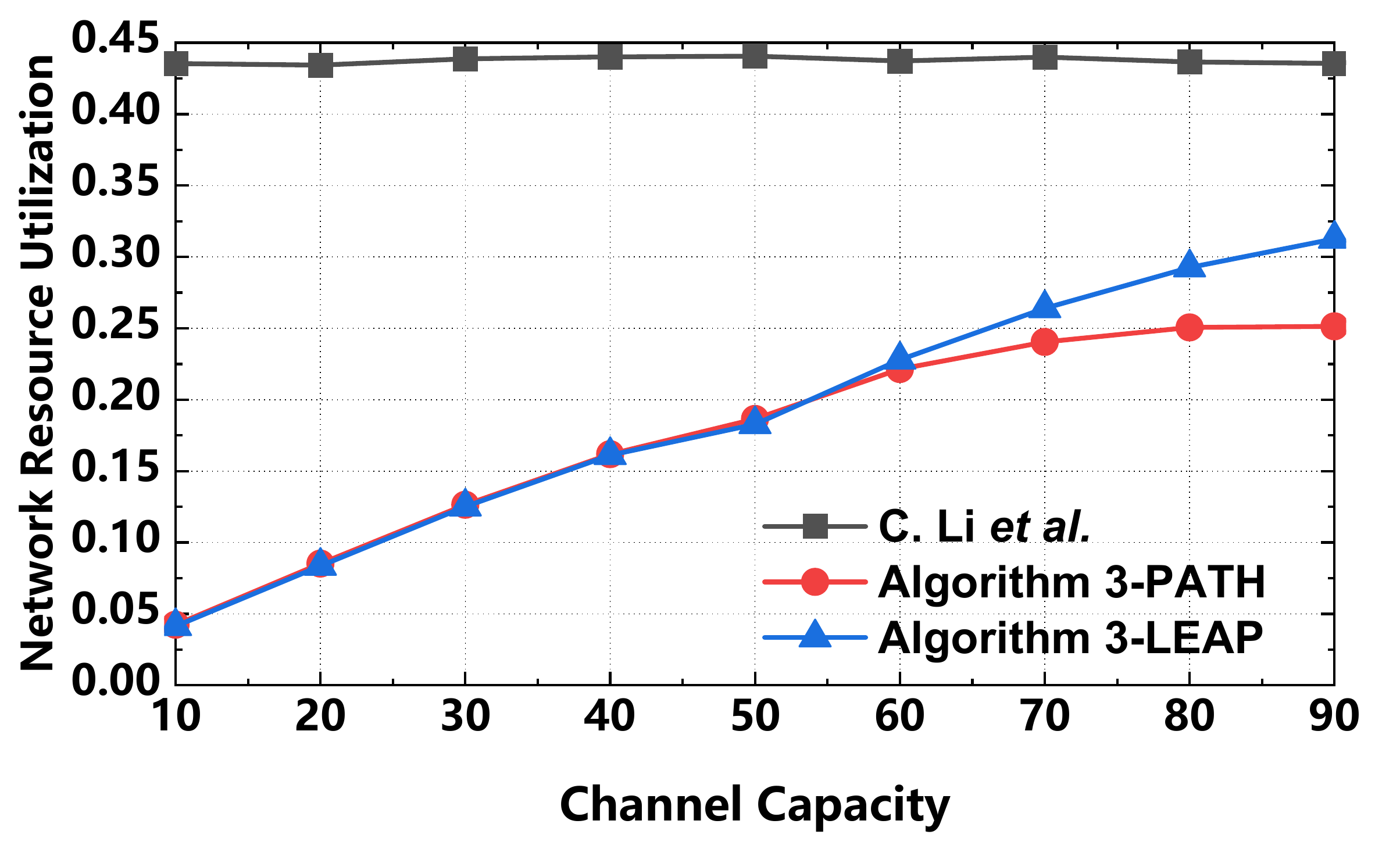}
    }
\setlength{\abovecaptionskip}{0.1cm}
  \caption{Performance comparison for multiple S-D pairs versus channel capacity (fidelity threshold = 0.7, SD pairs = 4).}
  \label{simulation4}
\end{figure*}
\begin{figure*}[!ht]
  \centering
  \subfigure[Throughput vs. S-D Pairs.]{
    \label{simulation5-1}
    \includegraphics[width=1.9in]{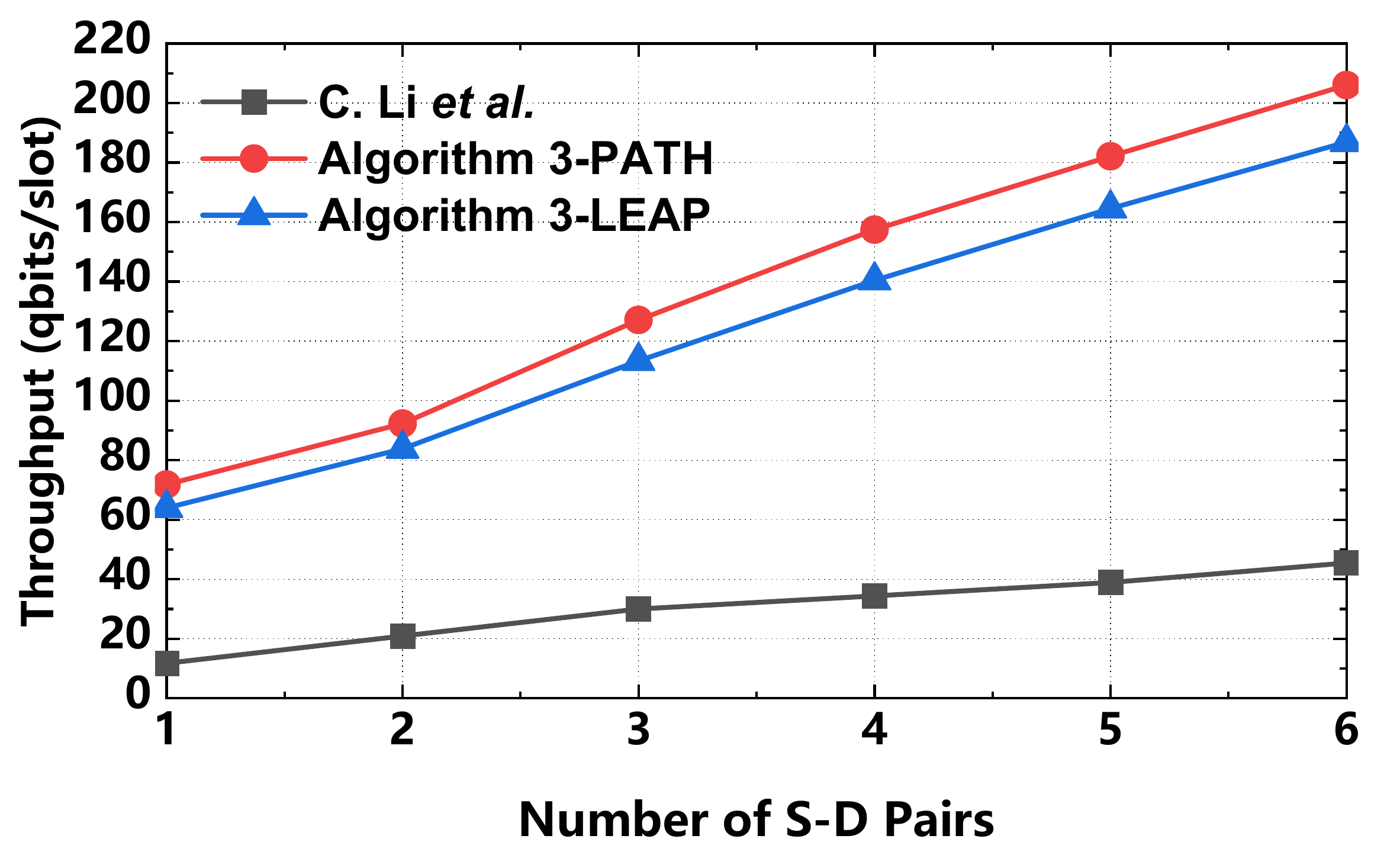}
 }
  \subfigure[Fidelity vs. S-D Pairs.]{
    \label{simulation5-2}
    \includegraphics[width=1.9in]{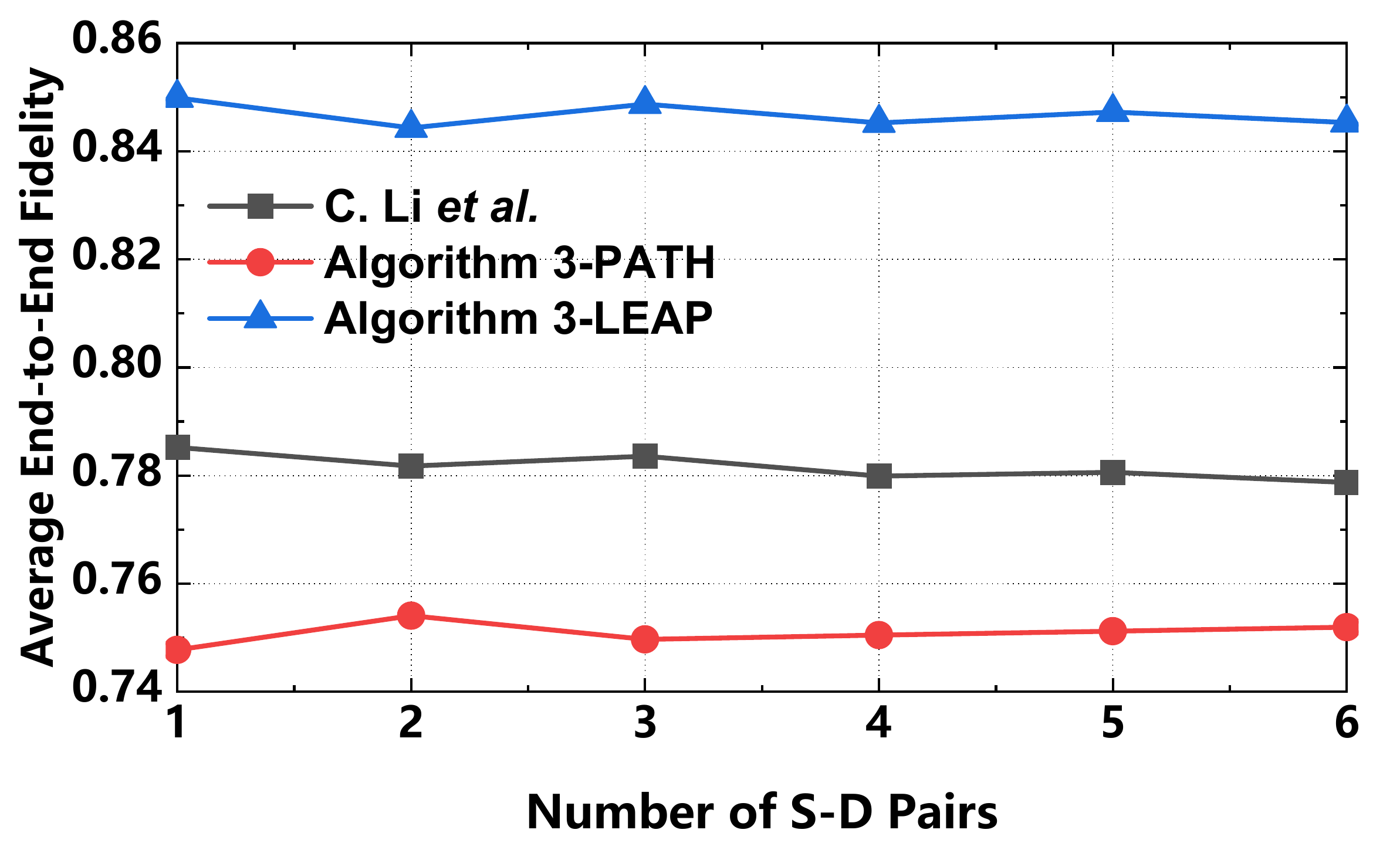}
    }
      \subfigure[Utilization vs. S-D Pairs.]{
    \label{simulation5-3}
    \includegraphics[width=1.92in]{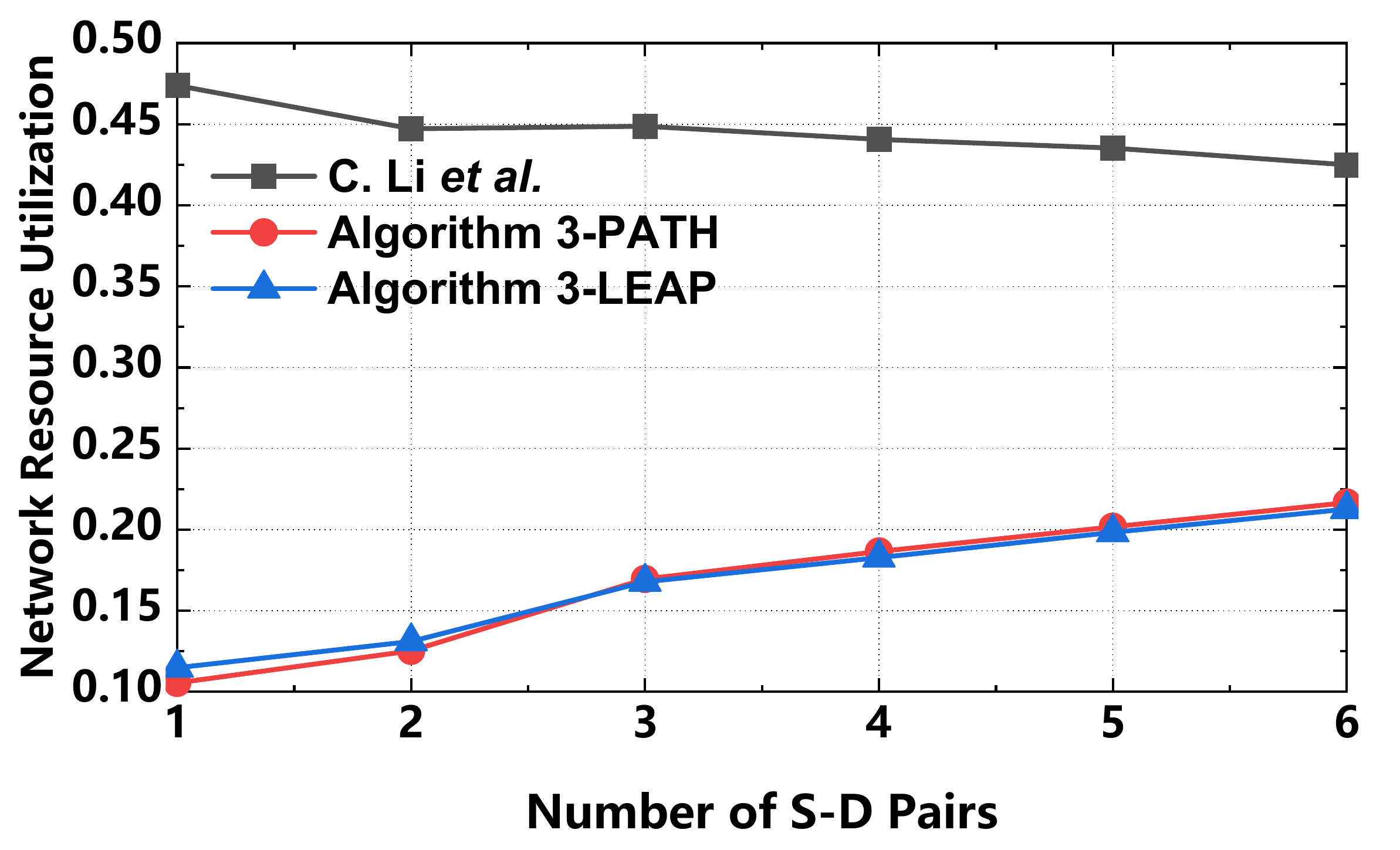}
    }
\setlength{\abovecaptionskip}{0.1cm}
  \caption{Performance comparison for multiple S-D pairs versus the number of S-D pairs (fidelity threshold = 0.7, channel capacity = 50).}
  \label{simulation5}
\end{figure*}

\textbf{\textit{Performance Comparison vs. Channel Capacity:}}
As shown in Fig. \ref{simulation2}, the performance of the proposed Q-PATH and Q-LEAP is evaluated compared with the baseline. In Fig. \ref{simulation2-1}, similarly, \textit{Algorithm} Q-PATH obtain the highest throughput, but the gap between Q-PATH  and the other results is expanding with the increase of channel capacity. This phenomenon shows the superiority of the proposed algorithm, and Q-PATH  can achieve better performance with sufficient entanglement resource in the network.

In Fig. \ref{simulation2-2}, the established end-to-end entanglement connection obtained from Q-PATH always has the lowest fidelity to fidelity threshold. Q-LEAP and the baseline has similar fidelity.
In Fig. \ref{simulation2-3}, due to the advance purification operations of the baseline, it has the highest resource utilization ratio. Similar to the results in Fig. \ref{simulation1-3}, Q-PATH can also achieve the highest throughput but similar resource utilization as Q-LEAP.

%\begin{figure}[t]
%  \centering
%  \includegraphics[width=1.0\linewidth]{algorithm_running_time_comparion}
%  \caption{Gap between the proposed algorithms and brute-force approach.}
%  \label{optimal_simulation_comparison}
%\end{figure}

%\begin{figure}[t]
%  \centering
%  \includegraphics[width=1.0\linewidth]{algorithm_running_time_comparion}
%  \caption{Algorithm running time for single S-D pair.}
%  \label{algorithm_running_time1}
%\end{figure}

\subsection{Results under Multiple S-D Pairs Scenarios}
%\textbf{\textit{Performance comparison vs. Utility Function:}}
As shown in Figs. \ref{simulation3}-\ref{simulation5}, the performance of the proposed \textit{Algorithm} \ref{Greedy_Routing}\footnote{In the simulation, \textit{Algorithm} \ref{Greedy_Routing}-PATH represents Q-PATH is used in the algorithm, \textit{Algorithm} \ref{Greedy_Routing}-LEAP represents Q-LEAP is used in the algorithm.} is evaluated compared with the baseline, the normalization is adopted for weight coefficient $\alpha$ and $\beta$, i.e., $\alpha=\frac{\alpha^*}{2|E|}$ and $\beta=\frac{\beta^*}{|E|C_{channel}}$, where $C_{channel}$ represents channel capacity, and $\alpha^*$ and $\beta^*$ both takes 0.5 in the utility metric.
For fair comparison, we allocate 50 requests for each S-D pair in the simulation.

\textbf{\textit{Performance Comparison vs. Fidelity Threshold:}}
In Fig. \ref{simulation3-1}, \textit{Algorithm} \ref{Greedy_Routing}-PATH obtains the highest throughput, but the gap between \textit{Algorithm} \ref{Greedy_Routing} and the gap between \textit{Algorithm} \ref{Greedy_Routing}-PATH and \textit{Algorithm} \ref{Greedy_Routing}-LEAP is relatively stable, which is inherently caused by the performance gap from Q-PATH and Q-LEAP. %and the other results is shrinking with the increase of fidelity threshold. %The reason can be explained as follows, with the increase of fidelity threshold, multi-round purification operations are required to satisfy end-to-end fidelity requirement, thus, available entangled pairs on each quantum channel are reducing from default 50 pairs to several pairs after necessary purification, which leads to several end-to-end connections in the end.
For the baseline, due to the purification is performed before path selection, end-to-end fidelity can hardly be guaranteed. Thus, a poor performance in terms of throughput is obtained.

In Fig. \ref{simulation3-2}, since \textit{Algorithm} \ref{Greedy_Routing}-PATH always finds the routing solution with minimum entangled pair cost, the fidelity of the obtained solution is also minimum but above the fidelity threshold. Compared to \textit{Algorithm} \ref{Greedy_Routing}-PATH and the baseline, \textit{Algorithm} \ref{Greedy_Routing}-LEAP only selects one path with ``best quality'', which provides a higher fidelity for each obtained routing solution in nature. Thus, \textit{Algorithm} \ref{Greedy_Routing}-LEAP obtains the routing solutions with the highest fidelity. In Fig. \ref{simulation3-3}, \textit{Algorithm} \ref{Greedy_Routing}-PATH and \textit{Algorithm} \ref{Greedy_Routing}-LEAP utilizes similar entangled pair resource to build end-to-end connection. The reason why the resource utilization of the baseline drops when fidelity threshold larger than 0.8 is that, the higher fidelity constraint prevents the successful connection establishments for requests and most of the requests will be denied. Thus, the resource utilization of the baseline significantly drops when fidelity threshold becomes larger.

\textbf{\textit{Performance Comparison vs. Channel Capacity:}}
In Fig. \ref{simulation4-1}, along with the increase of channel capacity, three routing schemes obtain significant improvement in terms of throughput, \textit{Algorithm} \ref{Greedy_Routing}-PATH always obtains the highest throughput, and the gap between \textit{Algorithm} \ref{Greedy_Routing}-PATH and \textit{Algorithm} \ref{Greedy_Routing}-LEAP reaches the peak at channel capacity of 50. This expected phenomenon can be explained as follows. Since the solution space of purification operation is related to channel capacity, \textit{Algorithm} \ref{Greedy_Routing}-PATH can find superior solutions with higher channel capacity compared with \textit{Algorithm} \ref{Greedy_Routing}-LEAP and the baseline. Due to the limited requests of each S-D pair, i.e., 50 requests, the gap between two methods of \textit{Algorithm} \ref{Greedy_Routing} becomes smaller when the channel capacity further increases. The gap between \textit{Algorithm} \ref{Greedy_Routing}-PATH and \textit{Algorithm} \ref{Greedy_Routing}-LEAP is 11.4\%-6.5\% with the increase of channel capacity.

In Fig. \ref{simulation4-2}, due to the same setting of fidelity threshold, the fidelity of established end-to-end entanglement connections obtained by three routing schemes is relatively stable. \textit{Algorithm} \ref{Greedy_Routing}-PATH obtains the lowest fidelity that satisfies threshold, and has the lowest resource consumption as shown in Fig. \ref{simulation4-3}, which indicates that it can obtain better routing decisions and purification decisions with minimum entangled pairs cost. Along with the increase of channel capacity, the gap between \textit{Algorithm} \ref{Greedy_Routing}-PATH and \textit{Algorithm} \ref{Greedy_Routing}-LEAP is also expanding. This phenomenon is similar to the one in single S-D pair scenario, but significantly magnified by multiple S-D pairs.

\textbf{\textit{Performance Comparison vs. S-D Pairs:}}
In Fig. \ref{simulation5-1}, the throughput of three routing schemes is rising along with the increase of the number of S-D pairs. As the routing scheme with best performance, \textit{Algorithm} \ref{Greedy_Routing}-PATH obtain 12.3\%-10.2\% and 510.7\%-353.2\% improvement compared with \textit{Algorithm} \ref{Greedy_Routing}-LEAP and the baseline, respectively.

In Fig. \ref{simulation5-2}, due to the same setting of fidelity threshold, the fidelity of established end-to-end entanglement connections obtained by three routing schemes is relatively stable. In Fig. \ref{simulation5-3}, since the purification decisions of the baseline are made in advance, most of the resource on each quantum channel has been exhausted to generate entangled pairs with higher fidelity. Hence, along with the increase of the number of S-D pairs, the baseline can hardly satisfy the requests from multiple S-D pairs, and lots of requests are denied, which causes the slight decrease of resource utilization ratio. For the proposed routing schemes, the purification decisions are made according to the requests of various S-D pairs, thus the resource utilization ratio of \textit{Algorithm} \ref{Greedy_Routing}-PATH and \ref{Greedy_Routing}-LEAP is rising with the increase of the number of S-D pairs.

\begin{figure}[t]
  \centering
  \includegraphics[width=0.8\linewidth]{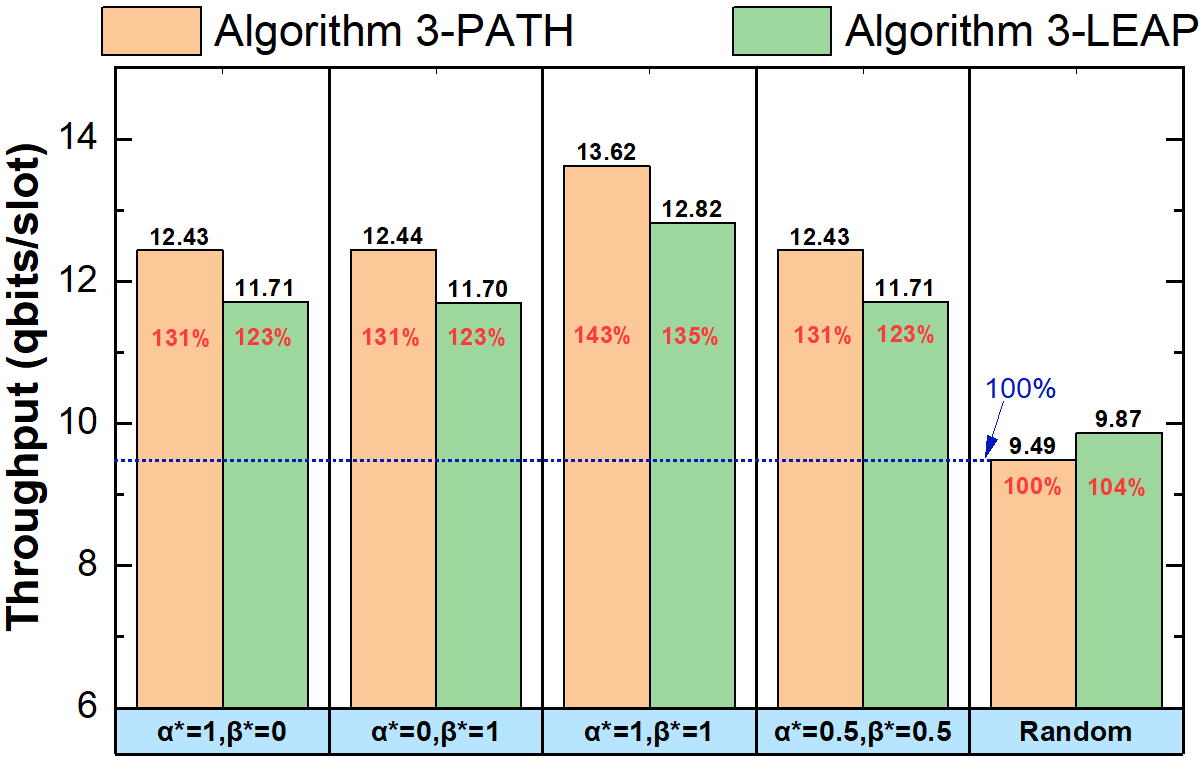}
  \caption{Performance comparison for multiple S-D pairs versus weight coefficient (fidelity threshold = 0.7, channel capacity = 2, number of S-D pairs = 10).}
  \label{simulation6}
\end{figure}

\textbf{\textit{Performance comparison vs. Weight coefficient:}}
In Fig. \ref{simulation6}, the relationship between the setting of weight coefficient and the performance of the proposed algorithms is evaluated. The normalization is adopted for weight  coefficient $\alpha$ and $\beta$, and $C_{channel}$ and $|E|=122$ represent channel capacity and the number of edges, respectively. At first, to prove the effectiveness of the proposed resource allocation method based on the utility metric in Eq. (\ref{utility}), we adopt ``Random'' resource allocation method and set it as performance benchmark ($100\%$ in the figure). As a comparison, the proposed resource allocation method has significant superiority with $23\%-43\%$ performance improvement. Second, the performance comparison under different weight coefficient settings is also conducted. For the given network topology, i.e., US backbone network, both two factors considered in the utility function, i.e., degree of freedom and resource consumption, can provide performance improvement compared with ``Random'' resource allocation method. If we only consider one factor by setting $\alpha^*=1$, $\beta^*=0$ or $\alpha^*=0$, $\beta^*=1$. Furthermore, if we both consider two factors by setting $\alpha^*=1$, $\beta^*=1$ or $\alpha^*=0.5$, $\beta^*=0.5$, the performance can be further improved $12\%$ in terms of throughput. Due to the significant influence of the proposed two factors in the Eq. (\ref{utility}), it can lead to better routing performance by jointly considering degree of freedom and resource consumption.

%\begin{figure}[t]
%  \centering
%  \subfigure[Throughput vs. S-D Pairs.]{
%    \label{simulation6-1}
%    \includegraphics[width=1.5in]{alpha_beta}
% }
%  \caption{Performance comparison for multiple S-D pairs versus weight coefficient (fidelity threshold = 0.7, channel capacity = 1, number of S-D pairs = 10).}
%  \label{simulation6}
%\end{figure}

%\textbf{\textit{Computational Complexity Comparison:}}
%The running time of three routing algorithms under multiple S-D pairs scenarios is shown in Table \ref{Algorithm_Running_Time2}. As we can see, \textit{Algorithm} \ref{Greedy_Routing}-A.
%\begin{table}[t]
%  \centering
%  \caption{Algorithm Running Time (fidelity threshold = 0.7, channel capacity = 50, number of S-D pairs = 4)}
%  \label{Algorithm_Running_Time2}
%  \begin{tabular}{|c|c|c|c|}%p{4.8cm}
%    \hline
%     \textbf{Scale} & \textbf{Algorithm 3-A}& \textbf{Algorithm 3-B} & \textbf{Baseline}\\\hline
%     39 & 0.20337 & 14.18052 & 32.32ms\\\hline
%     100 & 0.20337 & 14.18052 & 1142.89ms\\\hline
%     200 & 0.20337 & 14.18052 & 2662.81ms\\\hline
%     300 & 0.20337 & 14.18052 & 5435ms\\\hline
%     400 & 0.20337 & 14.18052 & 15149.21ms\\\hline
%     500 & 0.20337 & 14.18052 & 34487.81ms\\\hline
%  \end{tabular}
%\end{table}

\section{Conclusion}
In this paper, we studied purification-enabled entanglement routing designs to provide end-to-end fidelity guarantee for various quantum applications. Considering difficulty of entanglement routing designs, we started with single S-D pair scenario, and proposed Q-PATH, an iterative entanglement routing algorithm, and proved the optimality of the algorithm. To further reduce the high computational complexity, we proposed Q-LEAP, a low-complexity routing algorithm which considers ``the shortest path'' with minimum fidelity degradation and a simple but effective purification decision method. Based on the routing designs for single S-D pair scenarios, the design of multiple S-D pair scenarios could be regarded as the resource allocation problem among multiple single S-D routing solutions, and a greedy-based routing algorithm was further proposed. To efficiently allocate resource of entangled pairs for different requests and corresponding routing solutions, two allocation metrics were considered, i.e., degree of freedom and resource consumption.

The superiority of the proposed routing designs was proved by the extensive simulations. For the proposed routing algorithms, Q-PATH achieves the optimal performance with relatively high computational complexity, and Q-LEAP achieves near-optimal performance but highly efficiency, and both algorithms provide significant performance improvement compared to the existing purification-enabled routing scheme. In practice, the performance superior of Q-PATH makes it suitable for network optimization and the efficiency of Q-LEAP makes it suitable for a large-scale quantum networks. In the future, we will study the routing problem under \textit{``on-demand generation''} model with fidelity constraint, and also explore the inherent relationship between the value of fidelity and the possibility of \textit{qubit} error.

\begin{appendices}
\section{Proof of Theorem 1}
\begin{proof}
%At first, we prove that the purification operation to the entangled pair with the lowest fidelity can bring the highest improvement when the original fidelity is above 0.54.
%The resulting fidelity after purification operation can be calculated by eq. (\ref{fidelity_after_purification}).
%The second derivative of eq. (\ref{fidelity_after_purification}) can be obtained as
%\begin{equation}
%\frac{d^2f(x)}{dx^2}=\frac{12(32x^3-168x^2+24x+31)}{(8x^2-4x+5)^3}.\notag
%\end{equation}
%Let $\frac{d^2f(x)}{dx^2}=0$, then we can obtain $x\approx0.54$ when $x\in[0,1]$, and we can also obtain that $\frac{d^2f(x)}{dx^2}<0$ and $\frac{df(x)}{dx}>0$ when $x\in[0.54,1]$.
%The specific performance of purification operation is shown in Fig. \ref{fig:purification_performance}.
%In this case, if the original fidelity is above 0.54, then the fidelity improvement becomes smaller if the original fidelity is approaching 1. In other words, among multiple entangled pairs on different edges, the purification operation to the entangled pair with the lowest fidelity can bring the highest improvement when the original fidelity is above 0.54.

At first, we prove that the purification operation to the entangled pair close to $x^*$ can bring the highest improvement.
Let the original fidelity of two entangled pairs in the first-round purification as $x_1,x_1$, and the original fidelity of two entangled pairs in the second-round purification as $x_1,x_2$, where $x_2=f(x_1,x_1)$. Since $x_2>x_1$, we can have $x_1x_2+(1-x_1)(1-x_2)>x_1^2+(1-x_1)^2$. Thus, we can have:
\begin{align}
f&(x_1,x_2)-f(x_1,x_1)\notag\\
%&=\frac{x_1x_2}{x_1x_2+(1-x_1)(1-x_2)}-\frac{x_1^2}{x_1^2+(1-x_1)^2}\notag\\
&=\frac{x_1^2+x_2-x_1x_2-x_1}{A}\mathop{=}^{(a)}\frac{2x_1^2-3x_1+1}{x_1^2+(1-x_1)^2}\cdot A\mathop{<}^{(b)}0.\notag
\end{align}
Here, $A=\left(x_1x_2+\left(1-x_1\right)\left(1-x_2\right)\right)$ $\left(x_1^2+\left(1-x_1\right)^2\right)$ in $(a)$ and we utilize the following conclusion in (b): When $x_1,x_2\in[0.5,1]$, we can obtain that $A>0$, $x_1^2+(1-x_1)^2>0$ and $2x_1^2-3x_1+1<0$, and thus $f(x_1,x_2)-f(x_1,x_1)<0$.
Similar proof can be done when multi-round purification is considered.
In this case, the improvement of any multi-round purification cannot be higher than the first-round purification.
Thus, let $x_1=x_2$, the resulting fidelity after purification can be calculated by Eq. (\ref{fidelity_after_purification}),
and the derivative of Eq. (\ref{fidelity_after_purification}) can be obtained as:
%\begin{equation}
%\frac{df(x)}{dx}=\frac{(-24x^2+84x-6)}{(8x^2-4x+5)^2}.\notag
%\end{equation}
\begin{equation}
\frac{df(x)}{dx}=\frac{-2x^2+2x}{4x^4-8x^3+8x^2-4x+1}.\notag
\end{equation}
%Let $\frac{df(x)}{dx}=1$, then we can obtain $x^*\approx0.77081$ when $x\in[0.5,1]$, and we can also obtain that $\frac{df(x)}{dx}>0$ and $\frac{d^2f(x)}{dx^2}<0$ when $x\in[0.54,1]$.
Let $\frac{df(x)}{dx}=1$, we can obtain $x^*\approx0.743$ when $x\in[0.5,1]$, and $\frac{df(x)}{dx}>0$ and $\frac{d^2f(x)}{dx^2}<0$ remain when $x\in[0.5,1]$.
In this case, for one-round purification, if the original fidelity equals to $x^*$, the highest fidelity improvement can be obtained. In other words, among multiple entangled pairs on different edges, the purification operation to the entangled pair with the lowest fidelity can bring the highest improvement when the original fidelity is above $x^*$.

Next, we prove that the greedy approach can lead to the optimal purification decision with minimum entangled pair cost.
For a given path $P_{i,j}(s_i,d_i)$, we assume the original fidelity of the entanglement connection following routing path $P_{i,j}(s_i,d_i)$ is lower than the fidelity threshold but higher than $x^*$, i.e., $x^*<F_{i,j}(s_i,d_i)<F^{th}_i$. We assume that $\widehat{D}^{pur}_{i,j}$ is the optimal purification decision with minimum $N^*=\sum\limits_{(u_1,u_2)\in P_{i,j}(s_i,d_i)}N^{pur}_{i,j}(u_1,u_2)$ purification operations.
When $N^*=1$, if the fidelity of entangled pairs on edge $(v_1,v_2)$ is the minimum fidelity, then we have purification decision $D^{pur}_{i,j}$ with $N^{pur}_{i,j}(v_1,v_2)=1$. According to the monotonicity obtained from the first and second derivative when the original fidelity belongs to $[x^*,1]$, we can obtain that:
\begin{align}
\label{proof1-1}
&\widehat{F}_{i,j}(s_i,d_i)-F_{i,j}(s_i,d_i)\notag\\
%&=F^{pur}_{i,j}(u_1,u_2,1)F^{pur}_{i,j}(v_1,v_2,0)-F^{pur}_{i,j}(u_1,u_2,0)F^{pur}_{i,j}(v_1,v_2,1)\mathop{\leq}^{(c)}0,\\
&=\left(y_1+\delta_1\right)y_2-y_1\left(y_2+\delta_2\right)=\delta_1 y_2-\delta_2 y_1\mathop{\leq}^{(c)}0,
\end{align}
where $\widehat{F}_{i,j}(s_i,d_i)$ is the fidelity of entanglement connection under the optimal purification decision $\widehat{D}^{pur}_{i,j}$, $y_1=F^{pur}_{i,j}(u_1,u_2,0)$, $y_2=F^{pur}_{i,j}(v_1,v_2,0)$, $\delta_1=F^{pur}_{i,j}(u_1,u_2,1)-F^{pur}_{i,j}(u_1,u_2,0)$, and $\delta_2=F^{pur}_{i,j}(v_1,v_2,1)-F^{pur}_{i,j}(v_1,v_2,0)$. (c) holds due to $y_1>y_2$ and $\delta_1<\delta_2$. Note that $F^{pur}_{i,j}(u_1,u_2,N^*)$ is the resulting fidelity after $N^*$ round purification on edge $(u_1,u_2)$ in Eq. (\ref{purification_calculation}). Thus, the equality in Ineq. (\ref{proof1-1}) can be only obtained when $\widehat{D}^{pur}_{i,j}=D^{pur}_{i,j}$, which implies the purification decision $D^{pur}_{i,j}$ based on greedy approach is the optimal one when $N^*=1$.
%and $F^{pur}_{i,j}(u_1,u_2,1)$ is the resulting fidelity after one round purification on edge $(u_1,u_2)$. \textcolor{red}{Since edge $(v_1,v_2)$ has the minimum fidelity, thus the improvement after purification $F^{pur}_{i,j}(v_1,v_2,1)-F^{pur}_{i,j}(v_1,v_2,0)$ must be the highest, and ineq. (\ref{proof1-1}) must be satisfied. }

After that, when $N^*=2$, we can assume $\widehat{F}_{i,j}(s_i,d_i)=F_{i,j}(s_i,d_i)$ after the first purification operation. Then the situation goes back to the one when $N^*=1$. Thus, the rest proof when $N^*>2$ can be done in the same manner. This completes the proof.

%After that, when $N^*=2$, we can assume $\widehat{F}_{i,j}(s_i,d_i)=F_{i,j}(s_i,d_i)$ after the first purification operation, then two situations might happen: 1) if the edge $(v_3,v_4)$ with lowest fidelity before next purification decision satisfies $N^{pur}_{i,j}(v_1,v_2)=0$, then the situation goes back to the one when $N^*=1$. 2) if the edge $(u,v)$ with the lowest fidelity before next purification decision satisfies $N^{pur}_{i,j}(v_1,v_2)=1$, i.e., $(u,v)=(v_1,v_2)$, then two purification decisions with the same entangled pair cost can be made, i.e., decision with $N^{pur}_{i,j}(v_1,v_2)=2$ and decision with $N^{pur}_{i,j}(v_3,v_4)=1$ and $N^{pur}_{i,j}(v_5,v_6)=1$, where $(v_3,v_4)$ and $(v_5,v_6)$ are edges with the lowest fidelity except $(v_1,v_2)$. Thus, the fidelity improvement of this two decisions should be compared as described in line \ref{fidelity_improvement_comparison} of Algorithm \ref{Purification_decision}. The one with higher fidelity improvement leads to the optimal purification decision when $N^*=2$.
%Thus, the rest proof when $N^*\geq2$ can be done in the same manner. This completes the proof.
\end{proof}

\section{Proof of Theorem 2}

\begin{proof}
Here, we prove this theorem by contradiction.
For a S-D pair $<s,d>$, we assume $P^*(s,d)$ is the routing path that satisfies fidelity constraint $F^{th}$ with minimum entangled pair cost $cost^*$ for $<s,d>$, and $P'(s,d)$ is the routing path found by Q-PATH with $cost'>cost^*$. Then, the mathematical induction is used in the following:

1. At the beginning, the path with minimum hops $H^{min}$ on graph $G$ is found by using Breadth-First-Search (BFS). Thus, $cost^*\geq H^{min}$ should be satisfied. For the first iteration in line \ref{mincost_loop}, the algorithm would search the shortest paths $P^{SPF}_{H^{min}}$. To satisfy the fidelity constraint $F^{th}$, the cost of each path $P_{i,j}(s,d)\in P^{SPF}_{H^{min}}$ is not less than $H^{min}$. If the cost of the path $P'(s,d)\in P^{SPF}_{H^{min}}$ equals to $H^{min}$ after \textit{\textbf{Step 3}}, and $cost^*!=H^{min}$, then $P^*(s,d)$ must be a shorter path than the one found by BFS, it reaches a contradiction from the assumption that $H^{min}$ is minimum hops for $<s,d>$.

2. For the second iteration in line \ref{mincost_loop}, we have $min\_cost=H^{min}+1$. We assume that the path with $min\_cost=H^{min}$ is not found in the first iteration. If the algorithm finds a path that satisfies condition in line \ref{mincost_satisfied}, and $cost^*!=H^{min}+1$, then $cost^*$ must equal $H^{min}$. Path $P^*(s,d)$ and $P'(s,d)$ must have relationship:
\begin{equation}
\sum_{(u,v)\in P^*(s,d)}cost(u,v)=\sum_{(u',v')\in P'(s,d)}cost(u',v')-1.\notag
\end{equation}
Since the shortest paths $P^{SPF}_{H^{min}}$ have already been checked in the first iteration. If any path $P_{i,j}(s,d)\in P^{SPF}_{H^{min}}$ satisfies the fidelity constraint, condition in line \ref{mincost_satisfied} should be satisfied. Thus, we obtain a contradiction from the assumption that a path with $min\_cost=H^{min}$ is not found in the first iteration.
%the length of these paths is between $H^{min}$ and $min\_cost$.

3. After $k$ iterations in line \ref{mincost_loop}, we have $min\_cost=H^{min}+k$. We assume that the algorithm first finds a path that satisfies condition in line \ref{mincost_satisfied}. If $cost^*<H^{min}+k$, there must exist at least one routing path that is not stored in priority queue $Q$, then path $P^*(s,d)$ and $P'(s,d)$ must have relationship:
\begin{equation}
\sum_{(u,v)\in P^*(s,d)}cost(u,v)<\sum_{(u',v')\in P'(s,d)}cost(u',v').\notag
\end{equation}
Similarly, we obtain a contradiction that a path with $min\_cost<H^{min}+k$ is not found in the first $k-1$ iterations. This completes the proof.
\end{proof}

\end{appendices}

\bibliographystyle{IEEEtran}
%\footnotesize
\small
\bibliography{QuantumReference}

\begin{IEEEbiography}[{\includegraphics[width=1in,height=1.25in,clip,keepaspectratio]{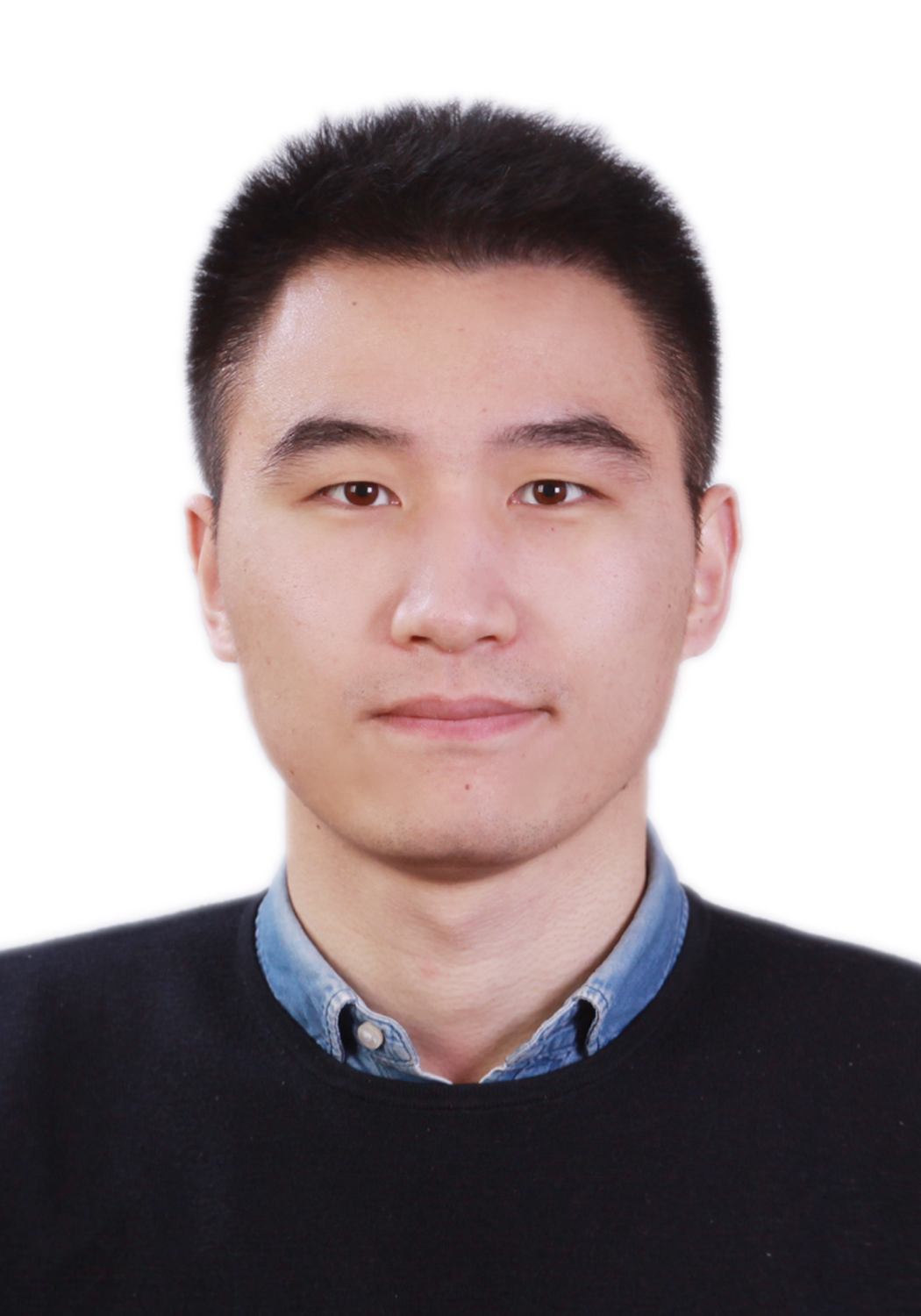}}]{Jian Li} (M'20) received his B.S. degree from the Department of Electronics and Information Engineering, Anhui University, in 2015, and received Ph.D degree from the Department of Electronic Engineering and Information Science (EEIS), University of Science and Technology of China (USTC), in 2020. From Nov. 2019 to Nov. 2020, he was a visiting scholar with the Department of Electronic and Computer Engineering, University of Florida. He is currently a Post-Doctoral fellow with the Department of EEIS, USTC. His research interests include wireless communications, satellite-terrestrial integrated networks, and quantum networking.
\end{IEEEbiography}
\begin{IEEEbiography}[{\includegraphics[width=1in,height=1.25in,clip,keepaspectratio]{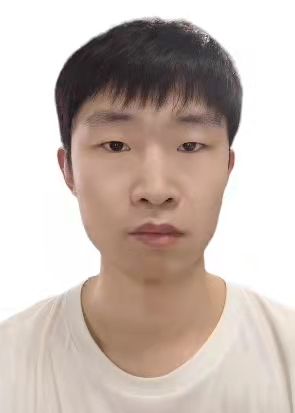}}]{Mingjun Wang} received his bachelor's degree from the School of Information Science and Technology, University of Science and Technology of China(USTC) in 2015. He is now a graduate student in the School of Cyber Science and Technology, USTC. His research interest is quantum networks.
\end{IEEEbiography}

\begin{IEEEbiography}[{\includegraphics[width=1in,height=1.25in,clip,keepaspectratio]{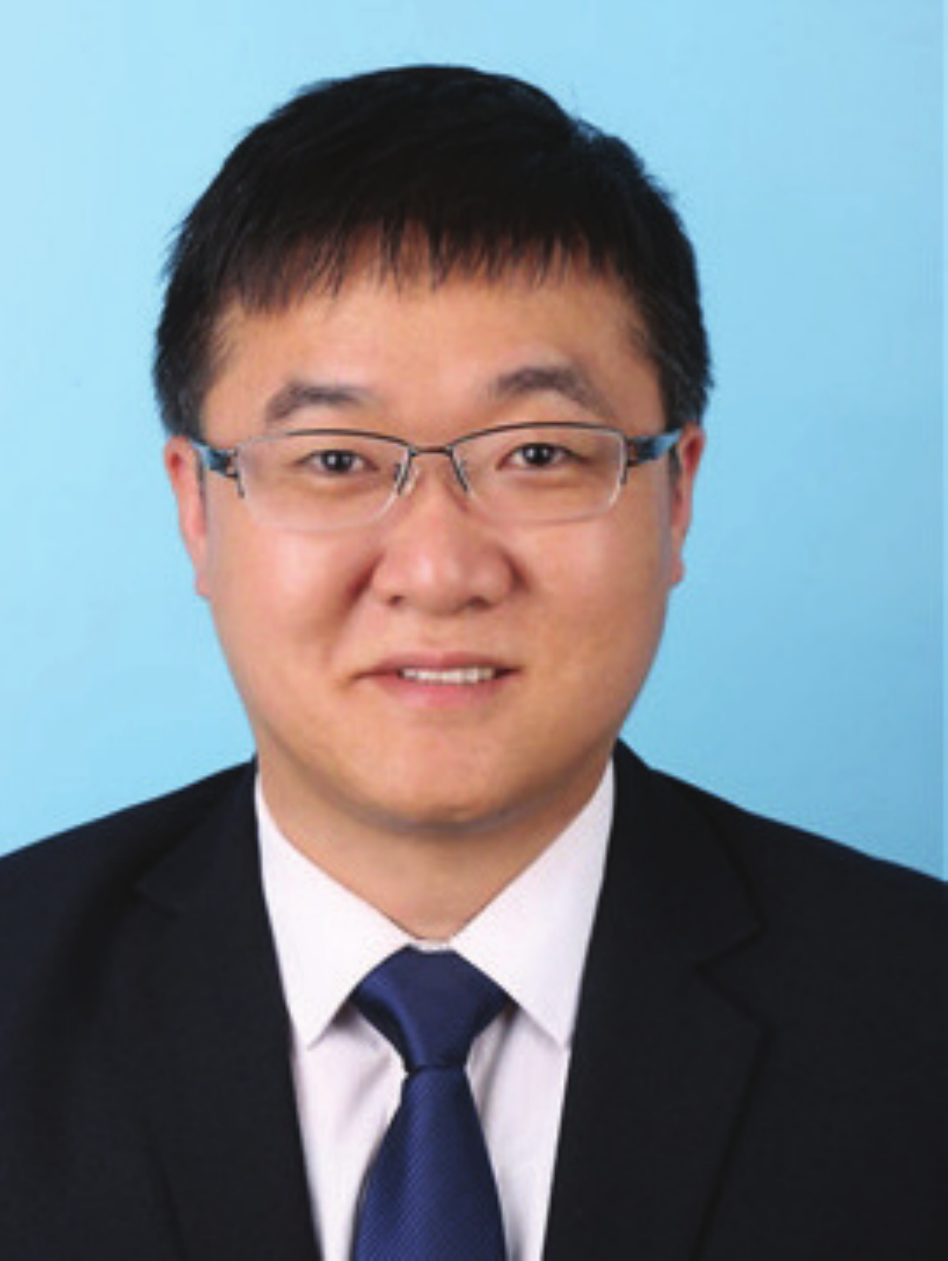}}]{Kaiping Xue} (M'09-SM'15) received his bachelor's degree from the Department of Information Security, University of Science and Technology of China (USTC), in 2003 and received his Ph.D. degree from the Department of Electronic Engineering and Information Science (EEIS), USTC, in 2007. From May 2012 to May 2013, he was a postdoctoral researcher with Department of Electrical and Computer Engineering, University of Florida. Currently, he is a Professor in the School of Cyber Security and the Department of EEIS, USTC. His research interests include next-generation Internet, distributed networks and network security. Dr. Xue has authored and co-authored more than 80 technical papers in the areas of communication networks and network security. He serves on the Editorial Board of several journals, including the IEEE Transactions on Wireless Communications (TWC), the IEEE Transactions on Network and Service Management (TNSM), and Ad Hoc Networks. He is an IET Fellow and an IEEE Senior Member.
\end{IEEEbiography}

	\begin{IEEEbiography}[{\includegraphics[width=1in,height=1.25in,clip,keepaspectratio]{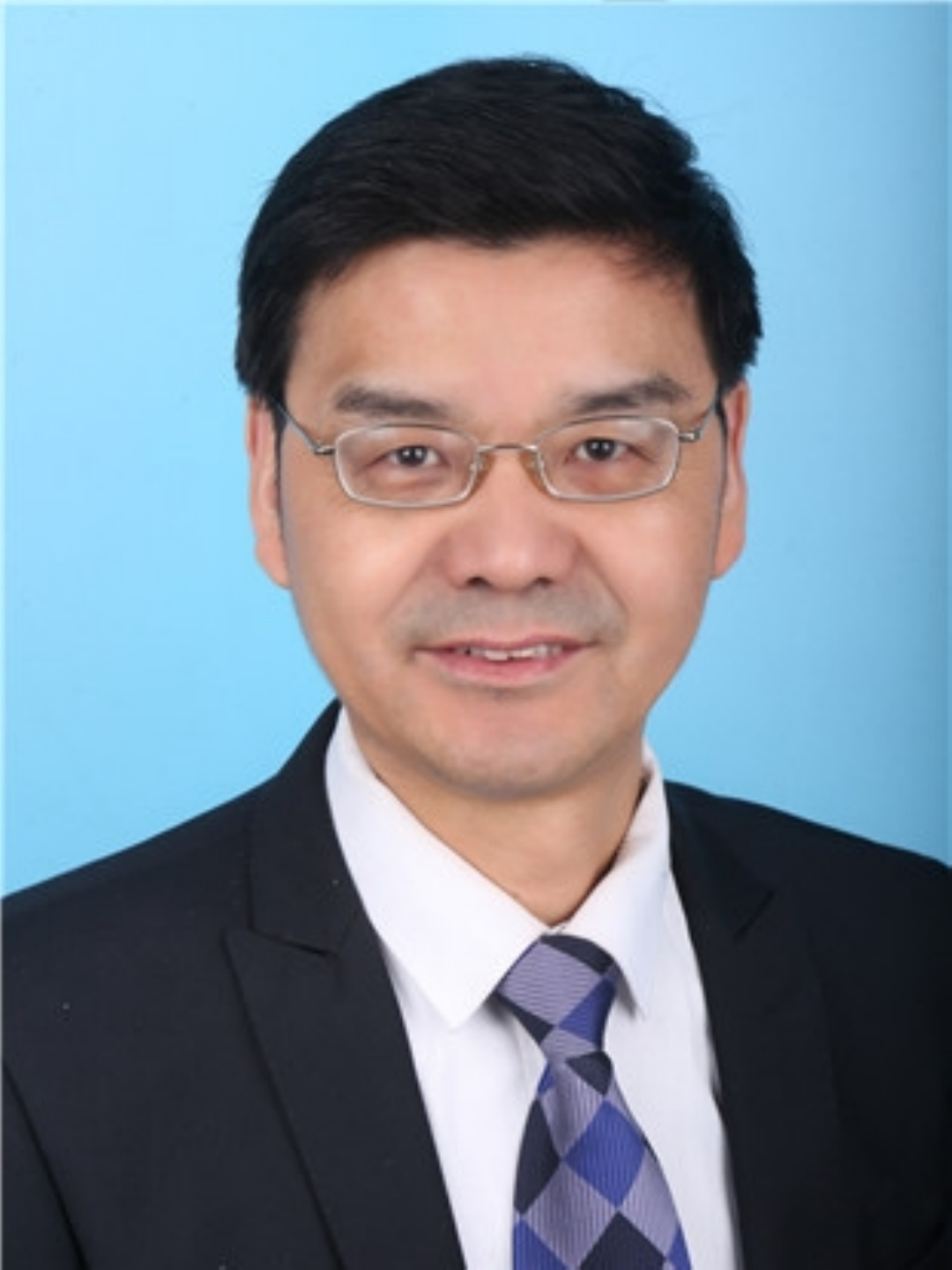}}]{Nenghai Yu} received the B.S. degree from the Nanjing University of Posts and Telecommunications, Nanjing, China, in 1987, the M.E. degree from Tsinghua University, Beijing, China, in 1992, and the Ph.D. degree from the Department of Electronic Engineering and Information Science (EEIS), University of Science and Technology of China (USTC), Hefei, China, in 2004. Currently, he is a Professor in the School of Cyber Security and the School of Information Science and Technology, USTC. He is the Executive Dean of the School of Cyber Security, USTC, and the Director of the Information Processing Center, USTC. He has authored or co-authored more than 130 papers in journals and international conferences. His research interests include multimedia security, multimedia information retrieval, video processing, and information hiding.
\end{IEEEbiography}

\begin{IEEEbiography}[{\includegraphics[width=1in,height=1.25in,clip,keepaspectratio]{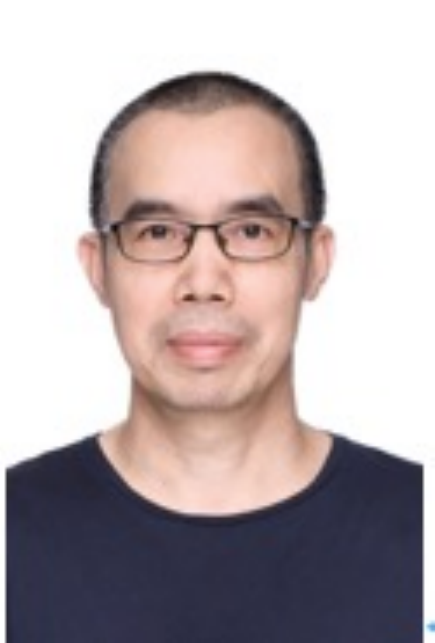}}]
{Qibin Sun} (F'11) received the Ph.D. degree from the Department of Electronic Engineering and Information Science (EEIS), University of Science and Technology of China (USTC), in 1997.  From 1996 to 2007,  he was with the Institute for Infocomm Research,  Singapore, where he was responsible for industrial  as well as academic research projects in the area of  media security, image and video analysis, etc. He  was the Head of Delegates of Singapore in ISO/IEC SC29 WG1(JPEG). He worked at Columbia University during 2000–2001 as a Research Scientist. Currently, he is a professor in the School of Cyber Security, USTC. His research interests include multimedia security, network intelligence and security and so on. He led the effort to successfully bring the robust image authentication technology into ISO JPEG2000 standard Part 8 (Security). He has published more than 120 papers in international journals and conferences. He is an IEEE Fellow.
\end{IEEEbiography}

\begin{IEEEbiography}[{\includegraphics[width=1in,height=1.25in,clip,keepaspectratio]{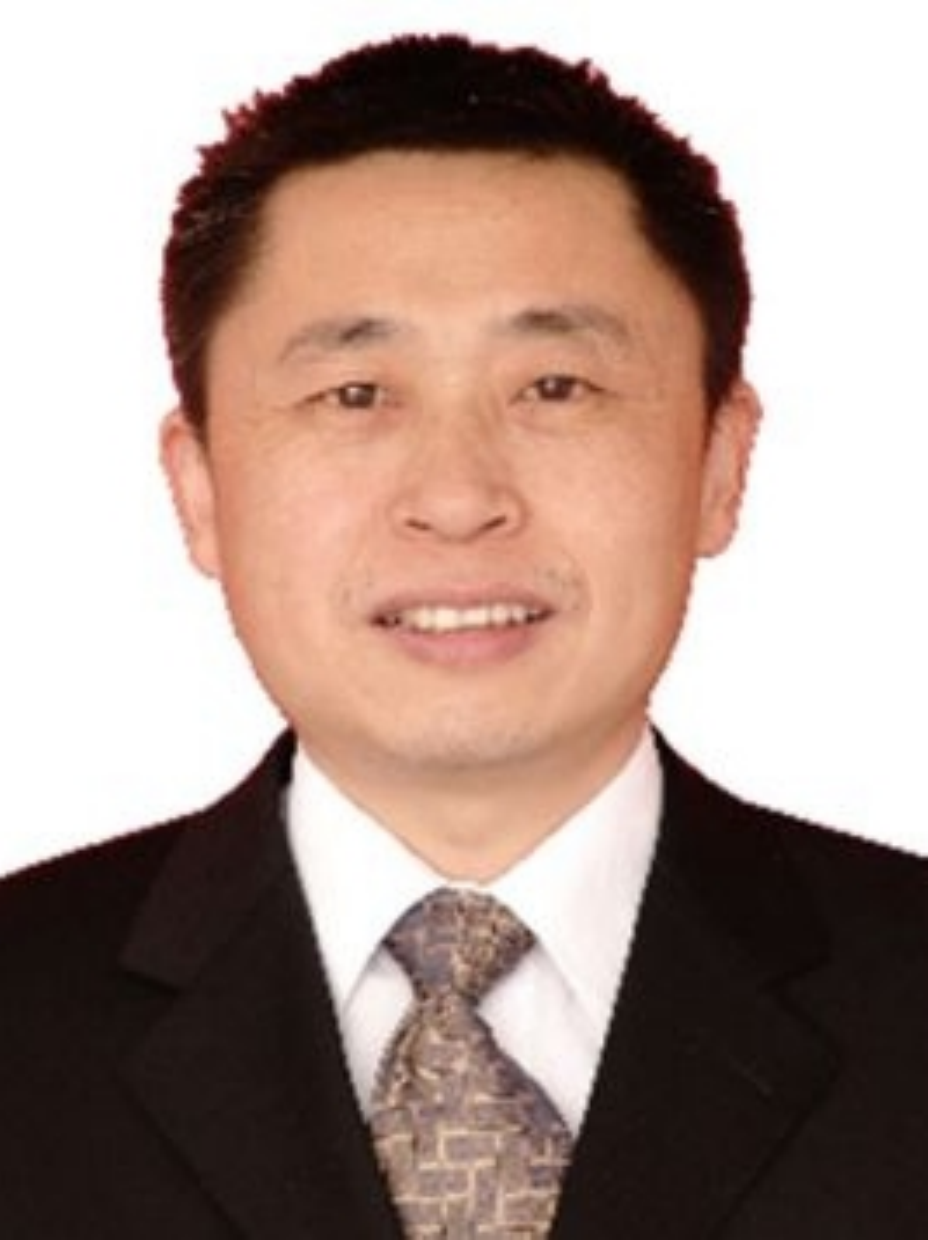}}]{Jun Lu} received his bachelor's degree from southeast university in 1985 and his master's degree from the Department of Electronic Engineering and Information Science (EEIS), University of Science and Technology of China (USTC), in 1988. Currently, he is a professor in the Department of EEIS, USTC. His research interests include theoretical research and system development in the field of integrated electronic information systems. He is an Academician of the Chinese Academy of Engineering (CAE).
\end{IEEEbiography}
\end{document}